\documentclass{aastex631}

\graphicspath{{./}{Figures/Paper 2/}}
\usepackage{xspace} 
\newcommand{\TESS}{TESS\xspace}
\newcommand{\Kepler}{\textit{Kepler}\xspace}

\received{Published in The Astronomical Journal, 167, 203, 2024}

\begin{document}

\title{DIAmante \TESS AutoRegressive Planet Search (DTARPS): II. Hundreds of New \TESS Candidate Exoplanets }

\author{Elizabeth J. Melton}
\affiliation{Department of Astronomy \& Astrophysics, Pennsylvania State University, University Park, PA 16802, USA}
\affiliation{Center for Exoplanets and Habitable Worlds, Pennsylvania State University, University Park, PA 16802, USA}
\affiliation{Department of Physics and Optical Engineering, Rose-Hulman Institute of Technology, 5500 Wabash Avenue, Terre Haute, IN 47803, USA}

\author{Eric D. Feigelson}
\affiliation{Department of Astronomy \& Astrophysics, Pennsylvania State University, University Park, PA 16802, USA}
\affiliation{Center for Exoplanets and Habitable Worlds, Pennsylvania State University, University Park, PA 16802, USA}
\affiliation{Center for Astrostatistics, Pennsylvania State University, University Park, PA 16802, USA}

\author{Marco Montalto}
\affiliation{INAF - Osservatorio Astrofisico di Catania, Via S. Sofia 78, I-95123 Catania, Italy}

\author{Gabriel A. Caceres}
\affiliation{EY-Parthenon, 1540 Broadway, New York, NY 10036, USA}

\author{Andrew W. Rosenswie}
\affiliation{Department of Astronomy \& Astrophysics, Pennsylvania State 
University, University Park, PA 16802, USA}
\affiliation{Institut f\"{u}r Physik und Astronomie, Universit\"{a}t 
Potsdam, D-14476 Potsdam, Germany}
\affiliation{Leibniz-Institut f\"ur Astrophysik Potsdam (AIP), An der 
Sternwarte 16, D-14482 Potsdam, Germany}

\author{Cullen S. Abelson}
\affiliation{Department of Astronomy \& Astrophysics, Pennsylvania State University, University Park, PA 16802, USA}
\affiliation{Department of Physics and Astronomy, University of Pittsburgh, 100 Allen Hall, 
3941 O'Hara St.,
Pittsburgh, PA 15260, USA}

\begin{abstract}

The DIAmante TESS AutoRegressive Planet Search (DTARPS) project seeks to identify photometric transiting planets from 976,814 southern hemisphere stars observed in Year 1 of the TESS mission.  This paper follows the methodology developed by Melton et al.\ 2024 (Paper I) using light curves extracted and pre-processed by the DIAmante project Montalto et al.\ (2020).  Paper I emerged with a list of 7,377 light curves with statistical properties characteristic of transiting planets but dominated by False Alarms and False Positives.  Here a multistage vetting procedure is applied including: centroid motion and crowding metrics, False Alarm and False Positive reduction, photometric binary elimination, and ephemeris match removal.   The vetting produces a catalog of 462 DTARPS-S Candidates across the southern ecliptic hemisphere  and 310 objects in a spatially incomplete Galactic Plane list. Fifty-eight percent were not previously identified as transiting systems. Candidates are flagged for possible blending from nearby stars based on Zwicky Transient Facility data and for possible radial velocity variations based on $Gaia$ satellite data. Orbital periods and planetary radii are refined using astrophysical modeling; the resulting parameters closely match published values for Confirmed Planets. The DTARPS-S population and astrophysical properties are discussed in Paper III.  
\end{abstract}

\keywords{exoplanet catalogs -- exoplanet detection methods -- light curve classification --
period search --
time domain astronomy -- transits}

\section{Introduction}

The search for transiting exoplanets from wide-field space-based surveys like NASA's \Kepler and \TESS missions faces a variety of methodological challenges.  Some are statistical in nature.  The light curves must be extracted from images, detrended to reduce stellar and instrumental variations, and searched for periodic variations resembling planetary transits. Methods often involve spline fits or wavelet filtering followed by construction of a specially designed periodogram. A machine learning classifier trained on confirmed or simulated transits is then applied.  This is typically a Random Forest classifier based on decision trees or a deep neural network.

In \citet[][Paper I]{Melton22a}, we develop such a procedure and apply it to $\sim 0.9$ million light curves from the DIAmante projecct's study of the \TESS Year 1 southern sky survey \cite{Montalto20}.  Our methods are based on \citet{Caceres19b} who successfully applied them to the \Kepler 4-year light curves \citep{Caceres19a}.  The statistical procedures are unusual for exoplanetary discovery: differencing operation for trend reduction, low-dimensional autoregressive modeling to remove remaining autocorrelation, a novel Transit Comb Filter periodogram, and a carefully tuned Random Forest classifier applied to 37 scalar properties of each star.  This project is called the DIAmante \TESS AutoRegressive Planet Search for the southern ecliptic hemispehre (DTARPS-S); in Paper I we emerge with a DTARPS-S Analysis List of 7,377 possible exoplanets.   

But the lists of potential exoplanets are quite unreliable at this stage. Efforts to detect small planets also allow False Alarms with unconvincing periodic behaviors.  Astronomical False Positives are common, sometimes overwhelming, at this stage particularly due to eclipsing binaries from stars in the vicinity of the target that are blended in the large detector pixels. 

A multifaceted vetting procedure by skilled humans, sometimes assisted by a machine learning automated procedure, is needed to produce a catalog of candidate planets with sufficient reliability for spectroscopic followup observations and science analysis.  Exoplanet candidate lists from large-scale efforts of transit detection and vetting include the \Kepler Objects of Interest \citep{Jenkins10, Jenkins20, Batalha10, McCauliff15, Coughlin16},  NGTS \citep{Armstrong18}, AstroNet \citep{Yu19}, \TESS Objects of Interest \citep[TOIs,][]{Guerrero21}, and DIAmante \TESS candidates \citep{Montalto20}. A neural network method is described by \citet{Shallue18} and others.  Additional recent vetting efforts include \citet{Zink21}, \citet{Dvash22}, \citet{Ofman22}, \citet{Kunimoto22}, \citet{Kostov22}, \citet{Cacciapuoti22}, \citet{Mantovan22}, and \citet{Magliano23}.

Here we describe the vetting process applied to the DTARPS-S Analysis List produced in Paper I.  The goal is to refine the list by removing, as much as feasible, False Alarms (FAs) and False Positives (FPs). The sample of 7,377 light curves in the DTARPS-S Analysis List has many contaminants: the classifier has an FP rate of 0.43\% with respect to the training sample that predicts about half of the 7,743 passing the classifier are false. But there will also be FPs with behaviors that were not put in the injected training sample, and FAs where the periodic signal itself is spurious or unconvincing. A strong vetting effort is thus needed to produce a catalog of candidate planets that sacrifices recall rate but enhances the reliability. 

This vetting is mostly based on astronomical rather than statistical considerations, and requires subjective judgments by the vetters. We have chosen here to have very strict criteria.  The vetting greatly reduces the sample size and therefore sensitivity (recall or True Positive rate) but also substantially decreases the False Positive rate of the resulting sample. 

We emerge here with a sample of 462 light curves that satisfy all of our vetting tests, and an additional 310 light curves from Galactic Plane targets that satisfy most of the vetting tests.  We report the 462 stars as the {\it DTARPS-S Candidate Catalog} and the 310 stars as the {\it DTARPS-S Galactic Plane list}.  Of these 772 stars, 457 have not been reported previously as possible host for transiting exoplanets. These stars are ready for spectroscopic followup for astronomical characterization. \citet[][Paper III]{Melton22c}  proceeds with astronomical analysis of these samples including results on the Neptune desert, ultra-short period planets, and planet occurrence rates. 

The present paper is organized as follows. Section \ref{sec:vet} describes the vetting process applied to the DTARPS-S Analysis List obtained in Paper I. The main results are given in Section \ref{sec:cands} with electronic tables for the DTARPS-S Candidates Catalog and DTARPS-S Galactic Plane List with  a Figure Set providing graphical displays of light curve properties from the ARPS analysis.  Candidate properties such as planet radius are refined and other catalog improvements are described in \S\S\ref{sec:cand_fit}-\ref{sec:flag}.  The DTARPS-S Galactic Plane list is discussed further in \S\ref{sec:eval_ca}.  Appendices give notes on individual planet candidates (Appendices \ref{sec:app_DTARPS}-\ref{sec:app_DTARPS-GP}), and comparison between the DTARPS-S results and other surveys (Appendix \ref{sec:app_other}).

\section{Vetting Procedures \label{sec:vet}}

\begin{figure}[t]
    \centering
    \includegraphics[width=0.45\textwidth]{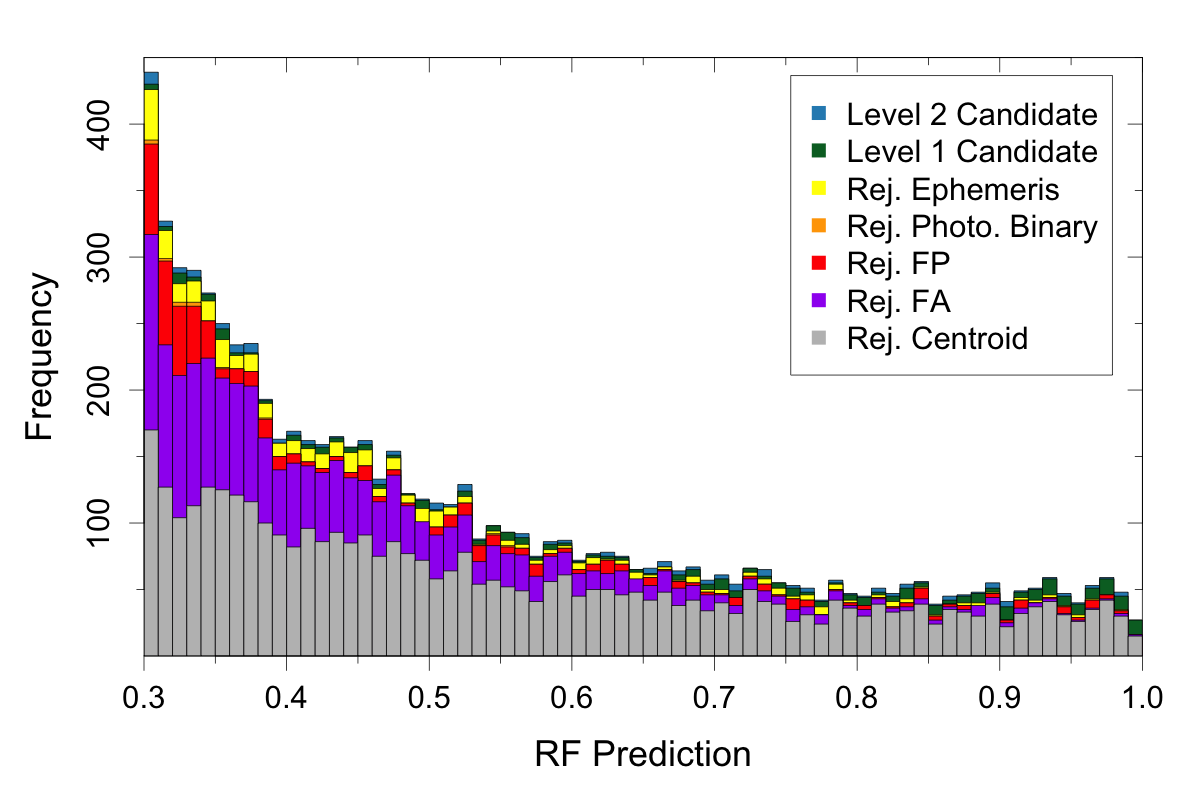}
    \includegraphics[width=0.45\textwidth]{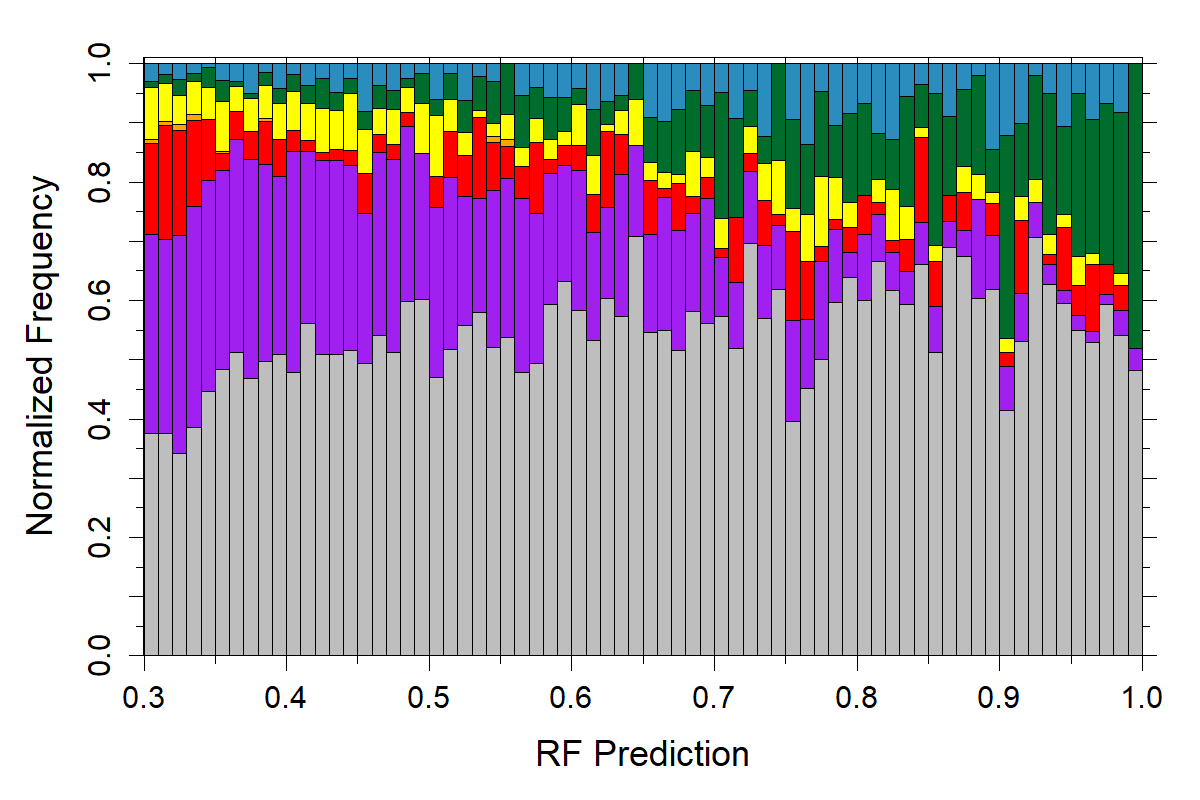}
    \caption{Number of light curves rejected for each step in the vetting process as a function of Random Forest prediction value. Level 1 objects satisfied all vetting criteria, while Level 2 objects passed a vetting test only marginally.  FP = False Positives, FA = False Alarm.}
    \label{fig:vet_breakdown}
\end{figure}

The DTARPS vetting process applied to the 7,377 objects in the DTARPS-S Analysis List (Paper I) contains five steps for vetting the candidates above the RF threshold: centroid and crowding analysis based on FFI image analysis; False Alarm elimination; False Positive determination; potential photometric binary identification based on the \emph{Gaia} color-magnitude diagram; and removal of ephemeris matches arising from light leakage in the \TESS telescope. Centroid-crowding analysis and ephemeris match removal are automated analysis procedures adopted from \citet[M20]{Montalto20}  and \citet{Coughlin14}, respectively. False Alarms (FAs) are cases where, despite a high probability from the Random Forest (RF) classifier, no convincing periodicity is present.  False Positives (FPs) are cases here a periodicity is clearly present but is not convincingly attributable to a planetary transit; typically these are blended eclipsing binaries.  FA elimination, FP determination, and potential photometric binary identification are performed by independent human vetters. Objects with conflicting dispositions from human vetters are analyzed collaboratively until a final disposition is agreed upon.

The distribution of the reasons light curves are rejected during vetting is shown in Figure \ref{fig:vet_breakdown}.  The biggest reduction came from the centroid-crowding analysis (\S \ref{sec:centroid}); this effectively excluded candidates near the Galactic Plane. Most of the visual vetting steps (FA elimination, FP determination, and potential photometric binary identification) removed a higher fraction of the DTARPS-S Analysis List sources with RF prediction values near the 0.30 threshold chosen in Paper I.  This is expected; light curves with very high RF probability are more likely to be True Positives that those near the threshold.

Vetting is conducted by the authors using visual examination of graphical and scalar properties. During the vetting process, DTARPS-S Candidates are sorted into two categories referred to as Level 1 candidates and Level 2 candidates. Level 1 candidates are the strongest exoplanet candidates for which no concerns were raised during the vetting process. All evidence is consistent with a true transiting planet.  Level 2 candidates are exoplanet candidates that passed the vetting process, but passed one or more vetting test marginally. Any objects for which significant concerns were raised during the vetting process were rejected. {\it We choose to be very conservative in our vetting in order to present a list of transiting exoplanet candidates with the highest possible purity.}

\subsection{Centroid Motion and Crowding Metrics \label{sec:centroid}}

Unlike the DTARPS analysis in Paper I  based on \TESS light curves, these initial vetting steps are based on analysis of the \TESS images. The procedure is only outlined here; details are provided in \S11 of M20. Centroid and crowding metrics for estimating FP probabilities also appear in other transit analysis software packages such as \texttt{vespa} \citep{Morton16}, \texttt{robovetter} \citep{Thompson18}, and \texttt{TRICERATOPS} \citep{triceratops21}.  

Centroid motion analysis measures the motion of the center of light in an aperture applied to the \TESS FFIs during the proposed transit event. If the center of light moves substantially from a target during a transit event, it indicates that the transit event detected is most likely due to a blended eclipsing binary whose transit depth is being diluted by the target star making it appear like a planetary transit. Centroid analysis is especially necessary for \TESS because its large 21\arcsec\/ pixels increase the likelihood of blended background eclipsing binaries, particularly in the crowded Galactic Plane.  

In the procedure of M20, flux-weighted centroid in each FFI are calculated in four circular apertures centered on the DTARPS-S analysis list objects. Orbital parameters from the best TCF periodogram peak are used to identify the in-transit and out-of-transit times for centroid measurements. Smoothing splines are fitted to the out-of-transit centroid measurements to correct for low-frequency variations in the centroid motion.  Principal Components Analysis is applied to the in-transit centroid measurements to a identify bi-variate Gaussian distribution of the centroid for each aperture.  The transit event is considered to be associated with the target star if the target had the smallest multivariate (Mahalanobis) distance of all \emph{Gaia} objects within 3\arcmin\/ of the target in at least three of the four apertures. 

In addition to centroid wobble, M20 calculates two probabilities that the transit signals is associated with the target star based on the crowding around the target (equation 19, M20) and the distribution of the centroid distributions for all four apertures (equation 22). We adopt the same probability cutoffs as M20 and rejected objects from the DTARPS-S analysis list for which the probability density was less than 30\% or the probability that the centroid distribution is associated with the target is less than 50\%. 

Only 27\% of the objects in the DTARPS-S Analysis List survive this centroid motion and crowding analysis (Figure \ref{fig:vet_breakdown}). In particular, nearly all candidates near the Galactic Plane are excluded. As these are probabilistic metrics that do not reflect measured properties of the stars themselves, in \S \ref{sec:cands} below we relax these stringent exclusion criteria for a portion of Galactic longitudes to capture some DTARPS-S candidates at low Galactic latitudes.

\begin{figure}[b]
    \epsscale{0.8}
    \plotone{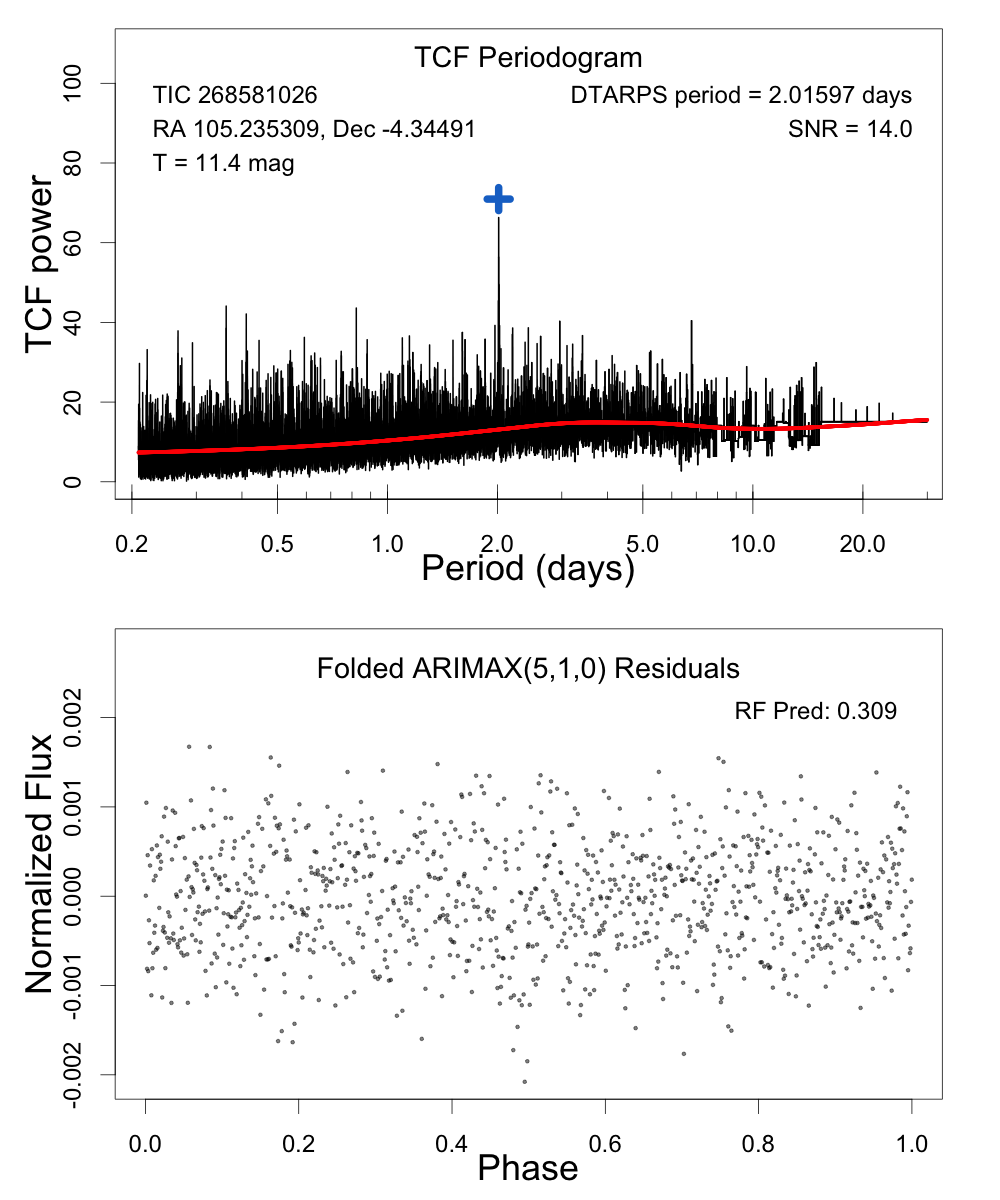}
    \caption{Example of a False Alarm light curve. The top panel shows the Transit Comb Filter periodogram with highest-SNR period marked.  The bottom panel shows no evident transit in the folded light curve after autoregressive noise is removed.  This case had a low TCF signal-to-noise ratio and a Random Forest prediction value only marginally above the threshold of 0.30.}
    \label{fig:FA_example}
\end{figure}

\subsection{False Alarm Elimination}
\label{sec:centroid_FA}

The DIAmante light curve, TCF periodogram, phase-folded ARIMA residuals, and phase-folded ARIMAX residuals\footnote{
{Full descriptions of these ARPS analysis stages are given in Paper I (\S2).}
} are examined for evidence of a periodic signal. The phase-folded light curves are folded using the TCF peak parameters for period, phase, and duration.  The goal is to remove objects from the 7,377 cases with unconvincing periodicities that passed the RF classifier threshold based on 37 features. The FA label is assigned where the periodic signal in the the phase-folded light curves or the DIAmante light curve is either unconvincing or entirely absent. FAs are also recognized by their noisy TCF periodograms with no strong isolated peaks, as illustrated in Figure \ref{fig:FA_example}. Some FAs arise when sharp features not removed by the ARIMA fit, such as stellar flares or instrumental features (Paper I, \S2.5), align to produce a spurious peak in the periodogram. FAs often have marginal TCF periodogram peak signal-to-noise ratios, $SNR \simeq 12-17$.  

FAs are expected to be present with any automated classification procedure that seeks to identify smaller planets. Of the 7,377 stars in the DTARPS-S Analysis List, 1,006 (66\%) that passed centroid and crowding analysis (\S\ref{sec:centroid}) were labeled as FAs.

\subsection{False Positive Identification}
\label{sec:centroid_FP}

The phase-folded light curve and the TCF periodogram are carefully scrutinized for indicators that the transit signal may come from an eclipsing binary or other variable star rather than a planetary transit.  We recall that the RF classifier is trained $against$ simulated eclipsing binaries as well as $towards$ simulated transiting planets (Paper I, \S4.2) so many blended eclipsing binary light curves likely receive low RF probabilities and do not enter the DTARPS-S Analysis List.  

\begin{figure}
    \centering
    \includegraphics[width=0.48\textwidth]{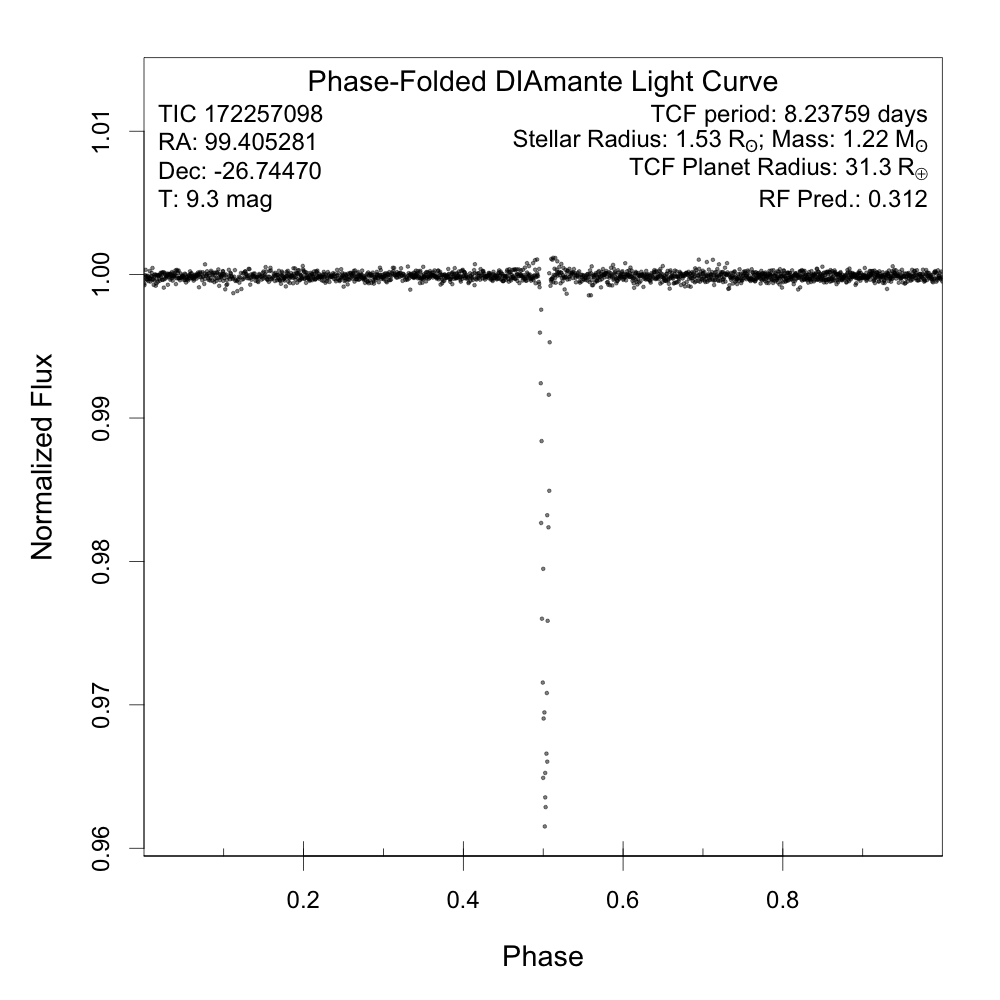}
    \includegraphics[width=0.48\textwidth]{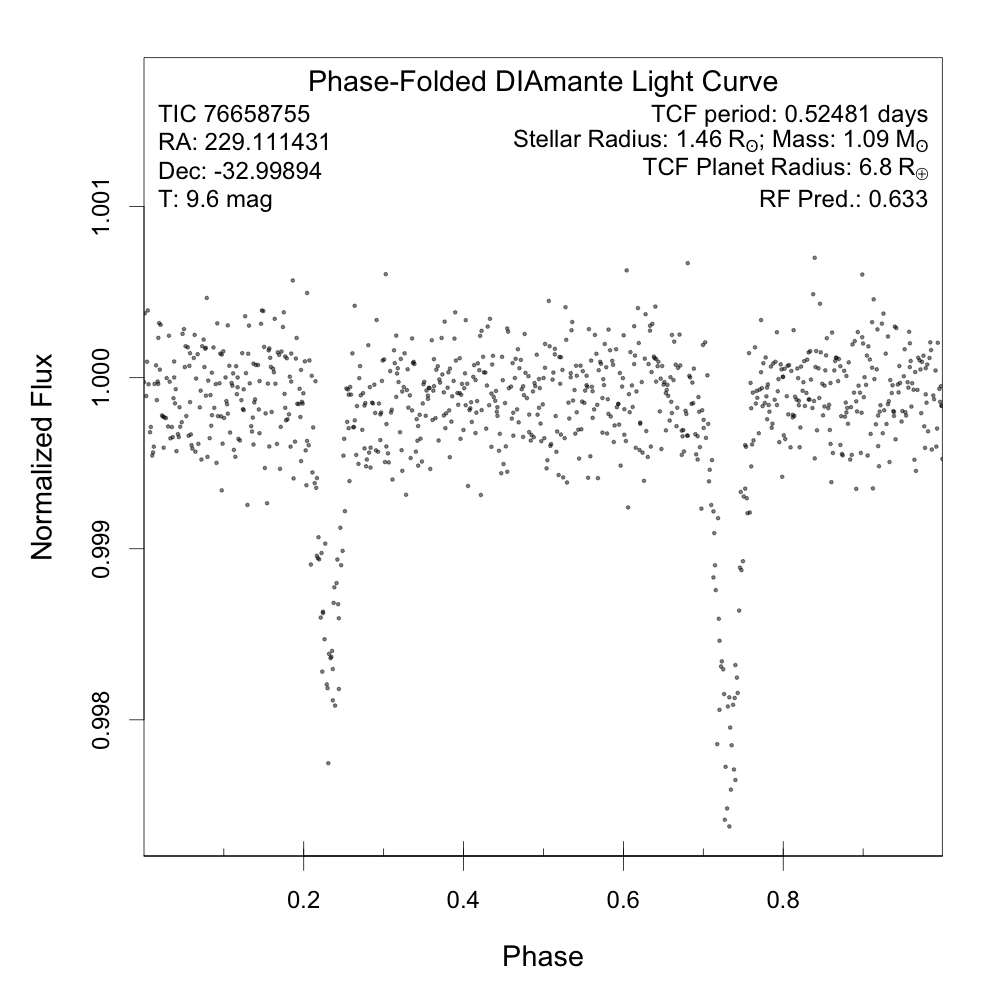} \\
    \includegraphics[width=0.48\textwidth]{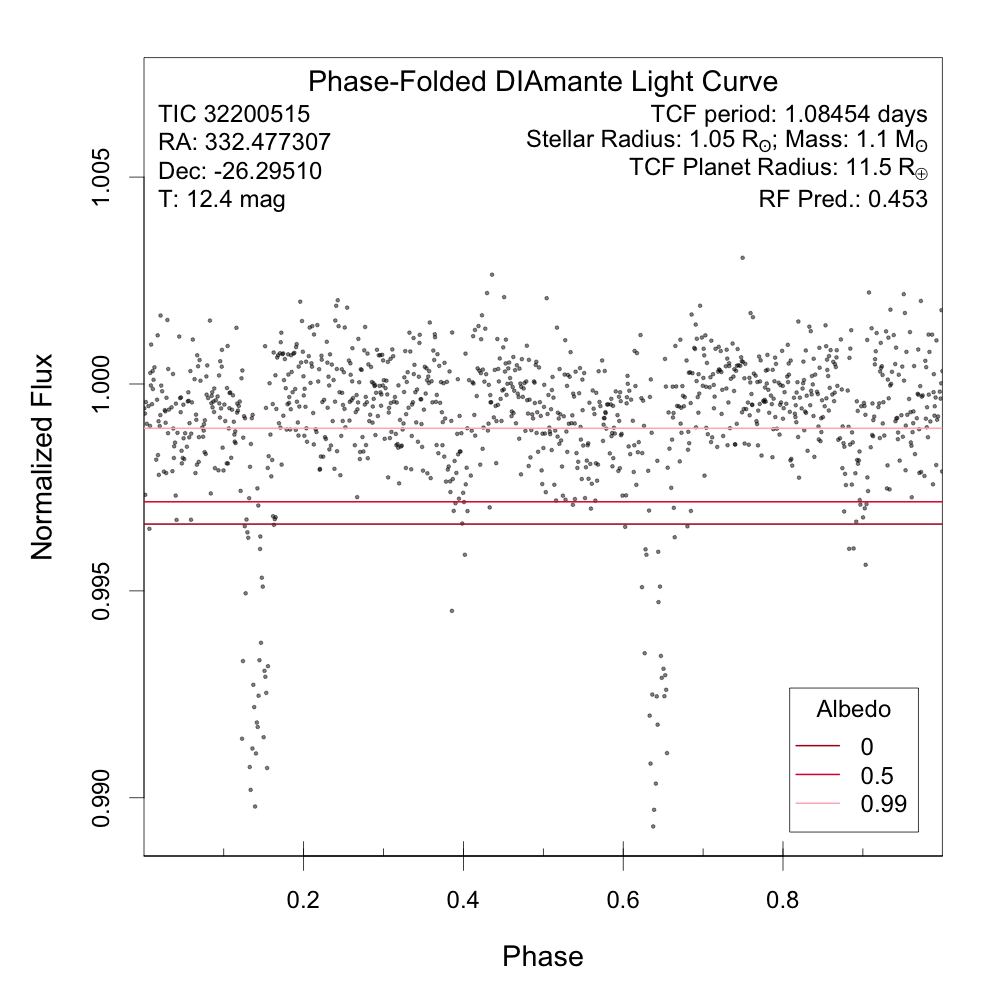}
    \includegraphics[width=0.48\textwidth]{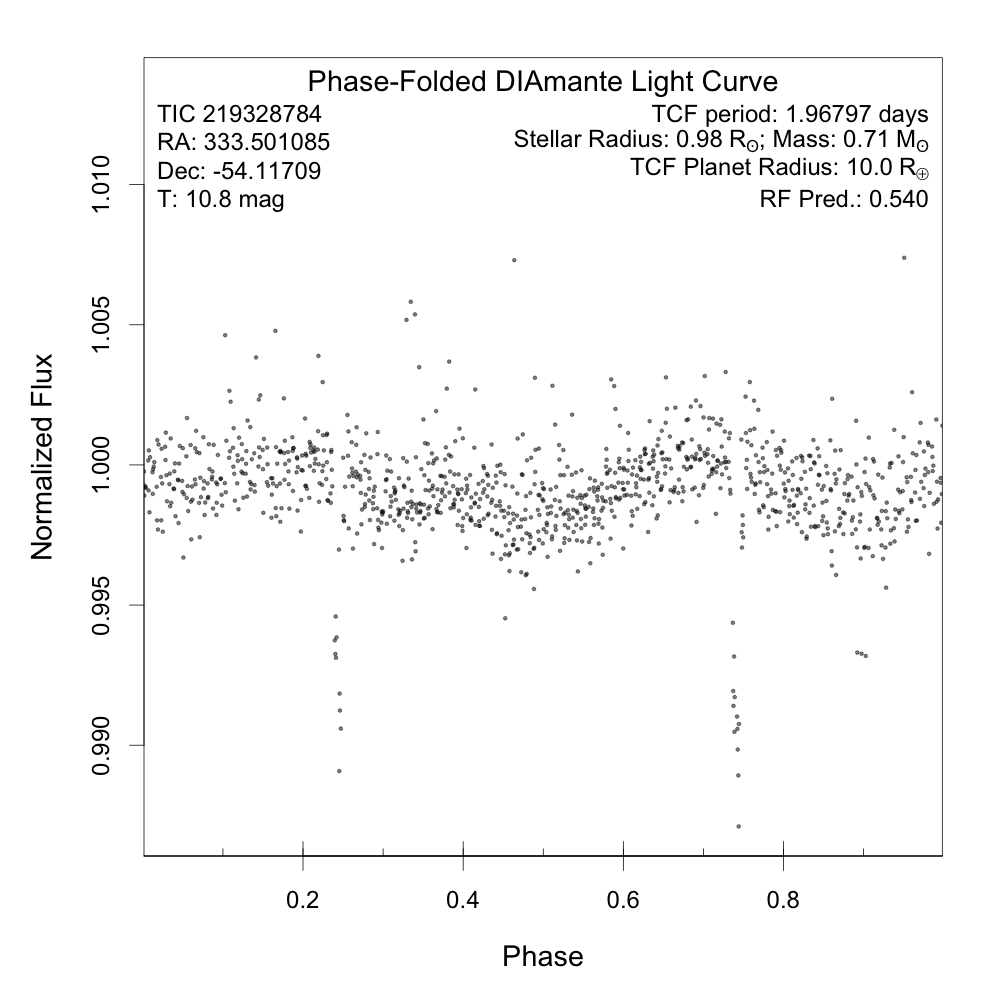}
    \caption{Examples of False Positive light curves.  These are the light curves from the DIAmante survey prior to ARPS analysis, folded with the TCF peak period.  {\it Top left:} The depth of the transit from the best TCF periodogram peak is too deep. {\it Top right:}  The even and odd transits have different depths. {\it Bottom left:}  The secondary transit is too deep. {\it Bottom right:}  Photometric curvature is present in the phase-folded light curve.}
    \label{fig:FP_examples}
\end{figure}

First, an object is labeled FP if the transit depth from the best TCF periodogram peak orbital parameters indicated a `planet' with a radius larger than 20 $R_{\oplus}$. An example is shown in the top left panel of Figure \ref{fig:FP_examples}.  We set this limit low because of the known tendency for the TCF periodogram peak to underestimate the planet radius (Paper I, \S8.1).  \citet{Hou22} placed an upper limit on the radius of hot Jupiters at 2.2 $R_J$ or $\sim$ 25 $R_{\oplus}$, due to tidal heating.  

Second, differences in even- $vs.$ odd-transits are examined.  Measures of even-odd differences are included in the RF classifier but might not dominate other positive measures; the light curve might then exceed the $P_{RF}=0.300$ threshold. DIAmante and ARIMAX residual light curves are phase-folded using the double harmonic of the period from the orbital parameters from the TCF periodogram peak. The depths of the even and odd transits are visually compared (Figure \ref{fig:FP_examples}, top right). Different even and odd transit depths indicate that the signal is strong evidence for an eclipsing binary system whose stars have different luminosities. Due to the paucity of points during transits for \TESS single sector light curves,  this vetting step is often not sensitive to small differences in stellar luminosities.  Only objects whose even and odd transit depths differ by more than the spread of the out-of-transit flux values around the median of the light curve are labeled FPs.

Third, the DIAmante and ARIMAX residual phase-folded light curves are examined for a strong secondary transit.  If present, the depth of the secondary transit is compared with the theoretical secondary transit depth given by 
\begin{equation}
    \delta_{2} = \delta_1 \times \sqrt{\frac{R_{\star}}{2} \times \left(\frac{4 \pi^2}{G M_{\star} P^2}\right)^{1/3} \times \left(1-A_b\right)^{1/4}}
\end{equation}
where $\delta_2$ is the theoretical secondary transit depth, $\delta_1$ is the primary transit depth (measured from the folded light curve), $R_{\star}$ is the stellar host radius from the TIC (in $R_{\sun}$), $G$ is Newton's gravitational constant, $M_{\star}$ is the stellar host mass from the TIC (in $M_{\sun}$), $P$ is the period of the primary transit depth (in days), and $A_b$ is the albedo of the planet. If the observed secondary transit depth is deeper than the theoretical secondary eclipse depth with an albedo of 0.5, than the object is labeled FP. An example is shown in the bottom left panel of Figure \ref{fig:FP_examples}.  The position of the secondary transit in phase is also considered.  Objects with secondary transits that did not occur half-way between primary transit were labeled as FPs as a possible indicator of an eccentric eclipsing binary system.  

Fourth, residual curvature present in the DIAmante and ARIMAX residual phase-folded light curves (Figure \ref{fig:FP_examples} bottom right) may indicate the object is a mutually illuminating and/or tidally distorted binary. Many examples are found in the ensemble of \Kepler and \TESS light curves \citep{Prsa11, Prsa22}.  These objects are removed as likely FPs. 

From the list of objects that passed the centroid-motion analysis and False Alarm vetting stages, 333 are labeled as False Positives. Most, but not all, of these light curves have Random Forest probabilities close to the 0.30 threshold (Figure \ref{fig:vet_breakdown}).   

\subsection{Photometric Binary Identification}
\label{sec:photom_bin}

Photometric properties of the host star for the DTARPS-S Analysis List is also vetted based on the \emph{Gaia} photometric catalog.  The DIAmante data set included subgiants and dwarf FGKM stars identified from color-magnitude cuts (M20).  We additionally remove objects potentially involving equal-mass photometric binary stars as they may possibly be eclipsing.  Potential photometric binaries were identified by comparing the position of the star on a \emph{Gaia} color-magnitude diagram \citep{GaiaCollab_DR2}, and on a plot of the stellar TIC radius as a function of the TIC effective temperature \citep{Stassun19}, to a random sample of stars from the full DIAmante data set. Objects whose stars were brighter than the bulk majority of the DIAmante stars with the same \emph{Gaia} color, or stars whose TIC stellar radius was larger than the bulk majority of the DIAmante stars with the same TIC effective temperature, were rejected as possible FPs. 

Sixteen DTARPS-S Analysis List objects that passed centroid-motion analysis, False Alarm elimination and False Positive Determination are rejected using this criterion.

\subsection{Ephemeris Match Removal}
\label{sec:ephem}

Ephemeris contamination arises in several ways: nearby bright stars contribute to the measured flux for an object; multiple reflections on telescope optical planes give stray light contamination from more distant very bright stars; and cross-talk and column bleed can occur in the CCD detector array \citep{Coughlin14}. In an exoplanet search, ephemeris contamination causes multiple sources to exhibit photometric transit signals with the same period and transit epoch.

\begin{figure}[t]
    \centering
    \epsscale{0.8}
    \plotone{ 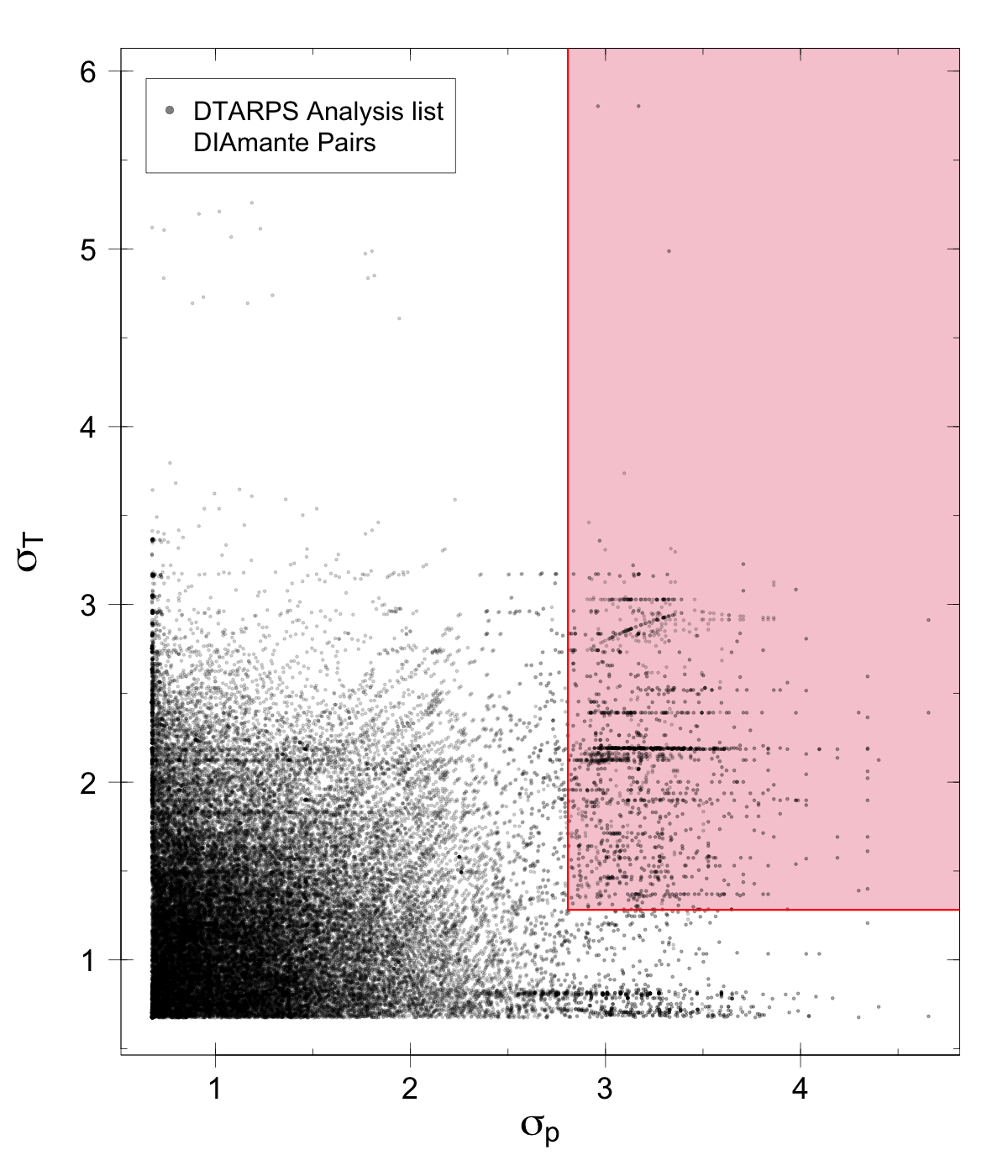}
    \caption{Plot of $\sigma_P$ and $\sigma_T$ for matches between the 7,377 DTARPS-S Analysis List objects and the full DIAmante data set of $\sim 1$ million stars. The red shaded region represents the pairs of points removed as likely ephemeris matches. The limit for $\sigma_P$ is set to 2.807 or a 0.5\% difference or less in the periods between two sources.  The limit for $\sigma_T$ is set to 1.281 or a 20\% difference or less in the epoch of first transits of two sources.}
    \label{fig:ephem_all_points}
\end{figure}

We identify and remove ephemeris contamination from our candidate list following the procedure developed by \citet{Coughlin14}. Sources from the full DIAmante data set within a distance, $d_{max}$ of the remaining DTARPS-S list objects were paired for ephemeris analysis where 
\begin{equation}
    d_{max} = 12.5 \times ps_T \times  \sqrt{10^6 \times 10^{-0.4 \times m_{T}} + 1}, \label{eq:d_max}
\end{equation}
$ps_T$ is the 21$\arcsec$ pixel scale of \TESS, and $m_T$ is the \TESS T magnitude of the DTARPS-S Candidates. The constant of 12.5 was the same constant from \citet{Coughlin14} to search a wide area around each potential candidate for ephemeris matches.  For each of the paired sources we calculated a modified version of $\sigma_P$ and $\sigma_T$ from Coughlin et al. For each pair of sources A and B,
\begin{equation}
    \sigma_P = \sqrt{2} \cdot \textrm{erfcinv}\left(\left|\frac{P_A - P_B}{P_A}\right|\right) \label{eq:sigma_p}
\end{equation}
and 
\begin{equation}
    \sigma_T = \sqrt{2} \cdot \textrm{erfcinv}\left(\left|\frac{T_A - T_B}{P_A} - round\left(\frac{T_A - T_B}{P_A}\right)\right|\right) \label{eq:sigma_t}
\end{equation}
where $P_A$ and $P_B$ are the periods of source A and B respectively, $T_A$ and $T_B$ are the epoch first transit for source A and B respectively, and erfcinv$()$ refers to the inverse complementary error function. We focus on ephemeris problems with a source A period to source B period ratio of 1:1, 2:1, or 1:2 by calculating $\sigma_P$ and $\sigma_T$ for each of the paired sources for each permutation of source A and B with a period ratio of 1:1, 2:1 and 1:2.  

Figure \ref{fig:ephem_all_points} shows the scatter plot of all values for $\sigma_P$ and $\sigma_T$ from the source pairs of the DTARPS-S Analysis List objects and objects from the full DIAmante data set. The minimum value for $\sigma$ is 0.674 and is equivalent to the worst match between the two sources \citep{Coughlin14}.  The phase from the TCF periodogram peak for a DTARPS-S Analysis List object is given in integer time steps which constrains the $\sigma_T$ values into stratified values. The red shaded region in Figure \ref{fig:ephem_all_points} shows the limits we apply to $\sigma_T$ and $\sigma_P$ values.  The limit for $\sigma_P$, 2.807, corresponds to a match within 0.5\% between the periods of the two sources (or between the period of one source and the double or half period of another source). The limit for $\sigma_T$, 1.282, corresponds to a match within 20\% of a period of one source between the epoch of first transits for the two sources.

The objects in the DTARPS-S Analysis List were also examined for ephemeris matches to the eclipsing binary catalog presented by \citet{Prsa22}.  There were only three ephemeris matches with the eclipsing binary catalog which are all rejected in the analysis above. Of the remaining DTARPS objects, 170 lie in the shaded region and were rejected as being likely contaminated by ephemeris matching.

\section{DTARPS-S Candidates Catalog} \label{sec:cands}

Of the 7,377 DTARPS-S Analysis List objects that exceeded the Random Forest classifier threshold in Paper I, the 462 cases that pass the vetting tests in \S \ref{sec:vet} are listed in Table \ref{tab:DTARPS_cands}. These are designated DTARPS-S Candidates and represent the principal result of this paper. Notes on previous published information on individual objects are presented in Appendix~\ref{sec:app_DTARPS}. 

The table has 26 columns giving the DTARPS-S Candidate catalog number and disposition from the DTARPS vetting process (Levels 1 and 2); designations and disposition from other studies \citep[][TOI list, cTOI list, M20 or NASA Exoplanet Archive]{TOI, Montalto20, NEA}; information about the host star from the {\it Gaia} catalog; information about the DIAmante light curve and ARPS processing; the best TCF period, depth, planet radius and other orbital information; the RF classifier probability; and an indicator of notes based on previous research in Appendix~\ref{sec:app_DTARPS}. 
 
For the first five DTARPS-S Candidates shown here, the NASA Exoplanet Archive currently labels two as `Ambiguous Planetary Candidate', one as a spectroscopically `Confirmed Planet', and one as a TOI `Planetary Candidate'. The appearance of independently identified exoplanet candidates in the DTARPS-S Candidate catalog lends support for these candidates as true planets. DTARPS-S 5 is newly reported here, but the Appendix~\ref{sec:app_DTARPS} note tells of a possible blended variable star contaminant; further investigation is needed.  

\begin{figure}[ht]
    \centering
    \includegraphics[width=0.99\textwidth]{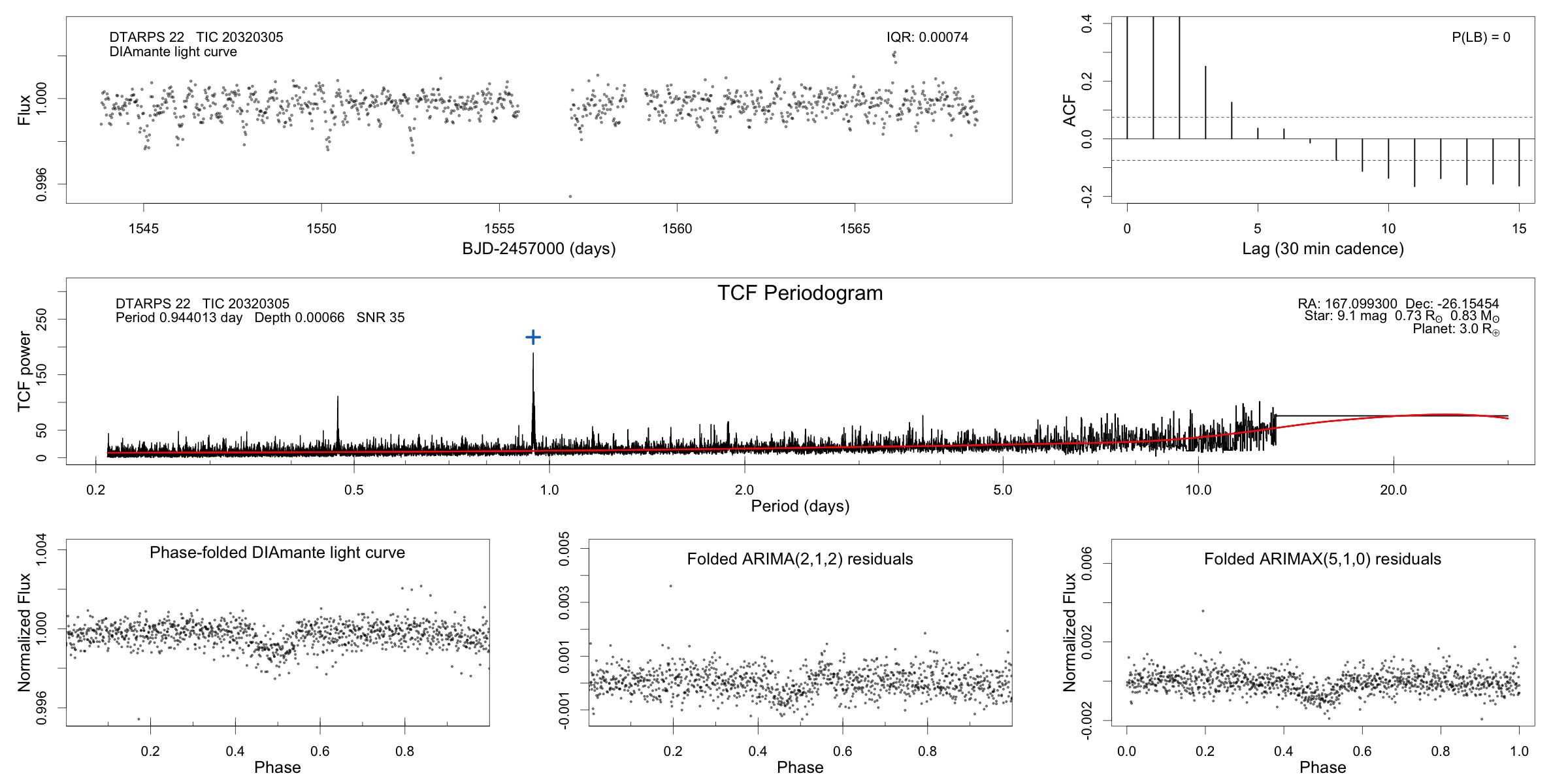} \\ \vspace*{0.6in}
    \includegraphics[width=0.99\textwidth]{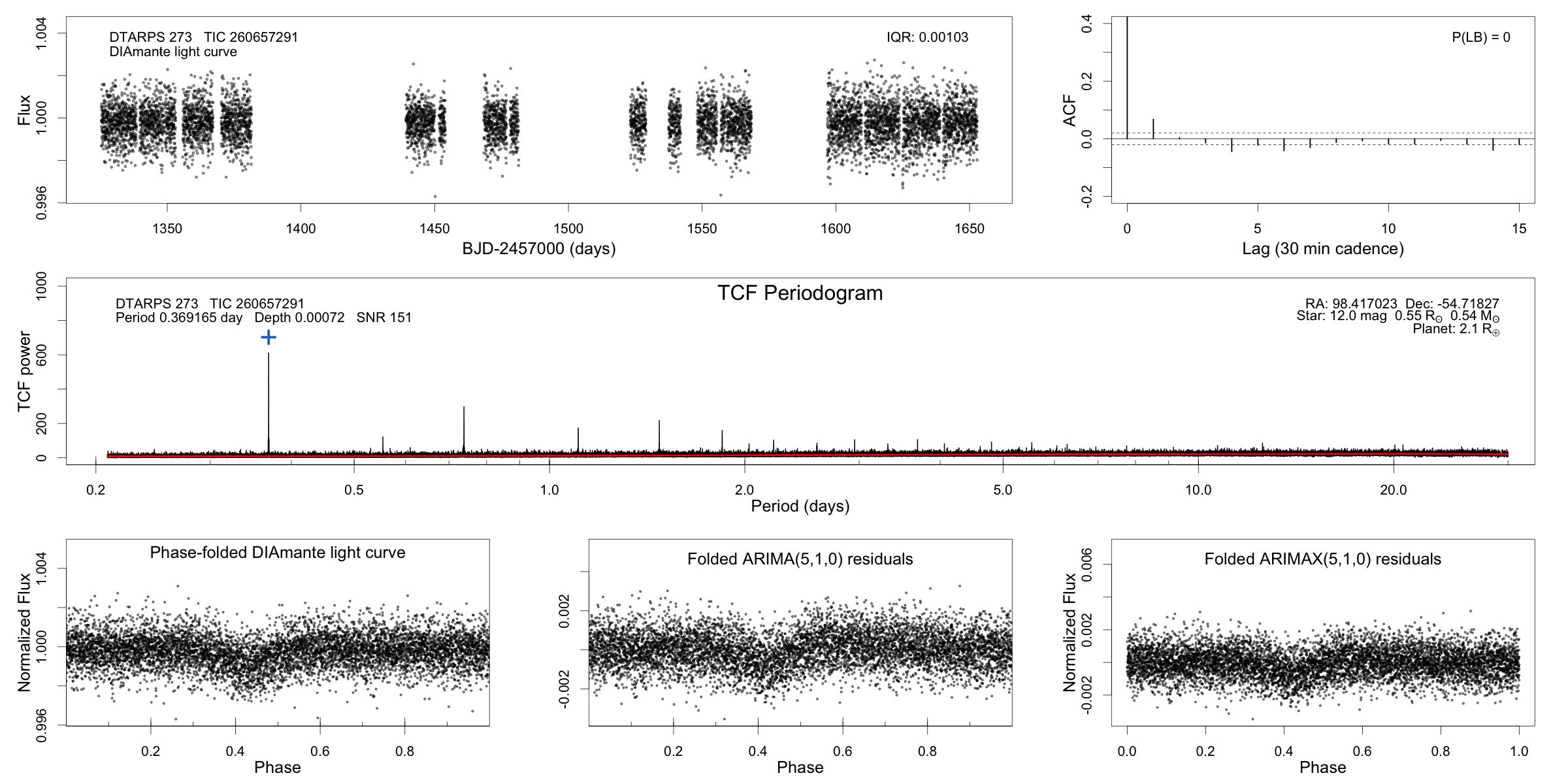}
    \caption{Figure Set graphics for two newly reported DTARPS-S Candidates with small radii: DTARPS-S 22 and 273.  See text for descriptions.  The complete Figure Set for 462 DTARPS-S Candidates is available in the electronic version of the paper.
    }
    \label{fig:cand_example}
\end{figure}

A Figure Set of six plots on one page is available for each  DTARPS-S Candidate; two examples are shown in Figure~\ref{fig:cand_example}. The top panels give the DIAmante extracted and preprocessed light curve and autocorrelation function. Annotations give the DTARPS-S and TIC identifiers, noise level (IQR = InterQuartile Range), and the probability of the Ljung-Box test that the time series is uncorrelated (Paper I, \S2.4). The middle panel shows the TCF periodogram where the best TCF period (highest signal-to-noise ratio after subtraction of the smooth LOESS curve shown in red) is marked with a blue cross (Paper I, \S3.2). Annotations give TCF results (best period, depth, and SNR of the peak) and stellar metadata (celestial location, $T$ magnitude, mass and radius from the TIC) \citep[v8, ][]{Stassun19}. The bottom panels show the light curve folded modulo the best period at three stages: the DIAmante light curve shown in the top panel; the ARIMA residuals (where a transit signal appears as a double spike); and the ARIMAX residuals (where a transit signal appears as a box-shaped dip; Paper I, \S2.2). 

The two cases illustrated in Figure~\ref{fig:cand_example} represent newly reported smaller planets.  The graphics, together with scalar values from Table~\ref{tab:DTARPS_cands} and information from Appendix A, can be described as follows:
\begin{description}

\item[DTARPS-S 22] This is CD -25$^\circ$8457, a bright T=9.1 mag K3 star at distance d = 52 pc. The single-sector \TESS light curve extracted by the DIAmante project shows autocorrelated variations with amplitude $\sim 0.2$\% likely arising from magnetic activity (top panels).  These are removed by a ARIMA(2,1,2) model revealing a signal with SNR=35 at P=0.94401 days in the TCF periodogram (middle panel).  The folded light curve shows a transit-like dip with depth 0.14\% corresponding to an UltraShort Period sub-Neptune with radius 3.0~R$_\oplus$ (lower panels).  

\item[DTARPS-S 273]  This is a T=12.0 mag M0 star at distance d = 83 pc that  \citet{Kaltenegger21} places in the {\it TESS Habitable Zone Star Catalog}.  Lying near the southern ecliptic pole, it has several sectors of \TESS observation giving a greater ability to reveal small planets than with stars covered by a single sector.  The DIAmante light curve shows little variability so the ARIMA model had little effect.  The TCF periodogram has a strong peak with SNR=151 at period P=0.369165 day with several harmonics visible.  The folded light curve shows a transit-like dip with depth 0.12\% corresponding to an extreme UltraShort Period\footnote{
The term `extreme UltraShort Period' exoplanet is presented in Paper III for DTARPS-S exoplanets with periods in the range $0.2 < P < 0.5$ days.
} 
sub-Neptune with radius 2.1~R$_\oplus$.  The periodogram shows a possible peak around 0.55 day that is not an alias of the principal spectral peak; this might represent a second planet.  

\end{description}

\begin{deluxetable}{rrrrrrrrrrrrrr}[hb]  \label{tab:DTARPS_cands}
\tablecaption{DTARPS-S Candidates Catalog}     
\tabletypesize{\small}
\tablewidth{0pt}
\centering
\tablehead{ \\
\multicolumn{6}{c}{Designations} && \multicolumn{7}{c}{Star} \\ \cline{1-6} \cline{8-14} 
\colhead{DTARPS-S} & \colhead{Disp} & \colhead{TIC} & \colhead{TOI} & \colhead{NEA} & \colhead{DIAm} &&  \colhead{R.A.} & \colhead{Dec} &   \colhead{T} &  \colhead{T$_{eff}$} & \colhead{R$_\star$} & \colhead{M$_\star$} & \colhead{Dist} \\  
\colhead{(1)} & \colhead{(2)} & \colhead{(3)} & \colhead{(4)} & \colhead{(5)} & \colhead{(6)} && \colhead{(7)} & \colhead{(8)} & \colhead{(9)} & \colhead{(10)} & \colhead{(11)} & \colhead{(12)} & \colhead{(13)}   }
\startdata
  1~~~~~~ & 1~~~  &       17361  & 3127 &  APC  & T~~~~ &&  219.33632  & -24.95848  & 11.3  & 5775  & 1.26  & 1.03  & 309  \\  
  2~~~~~~ &  1~~~   &   1003831  & 564 &   CP &   T~~~~  && 130.29516  & -16.03628  & 10.7  & 5550  & 1.12  & 0.98  & 200 \\  
  3~~~~~~ &  1~~~   &   1449640  & 446  &  APC &  F~~~~  &&  75.17161  & -35.20261  & 11.9 &  6411 &  1.67  & 1.29  & 674 \\
  4~~~~~~ &  1~~~   &   1525480  & 2402 &   PC &  T~~~~  &&     156.52780 & -17.26265  & 12.3 &  5519 &  0.97  & 0.97  & 345 \\
  5~~~~~~ &  1~~~   &   1599403 &\nodata&\nodata& F~~~~ &&  131.11166 & -2.47356  & 11.7 & 6508 & 1.26  & 1.33  & 453 \\
\enddata
\tablecomments{The full table of 462 DTARPS-S planet candidates is available in the electronic version of the paper in file DTARPS.MRT. Column definitions: \\
(1) DTARPS-S: Sequence number of DTARPS-S planet candidates\\
(2) Disp: DTARPS-S disposition based on vetting procedures.  1 = very high confidence.  2 = high confidence\\
(3) TIC: TESS Input Catalog identifier\\
(4) TOI: TESS Object of Interest \citep{ExoFop-TOI} \\
(5) NEA Disp: NASA Exoplanet Archive (NEA) Disposition.  APC = Ambiguous Planet Candidate. CP = Confirmed Planet.  FP = False Positive. KP = Known Planet.  ... = previously unidentified by NEA \citep{NEA-CP}\\
(6) DIAm: Boolean indicator of DIAmante planet candidate (Montalto et al. 2020)\\
(7-8) R.A., Dec = Right Ascension and Declination (J2000) from Gaia DR2 catalog \citep[accessed using VizieR X-match,][]{vizier_gaiaDR2}\\
(9) T = TESS band magnitude\\
(10-13): T$_{eff}$, R$_\star$, M$_\star$, Dist = Stellar effective temperature (K), radius (R$_\odot$), mass (M$_\odot$), and distance (pc) from Gaia DR2 catalog
}
\end{deluxetable}

\begin{deluxetable}{rrrrrrrr}
\tabletypesize{\small}
\centering
\tablehead{
\multicolumn{3}{c}{Lightcurve} && \multicolumn{2}{c}{ ARIMA} && \colhead{TCF} \\ \cline{1-3} \cline{5-6} \cline{8-8}
\colhead{N$_{lc}$} & \colhead{IQR$_{lc}$} & \colhead{P$_{LB,lc}$} && \colhead{IQR$_{AR}$} & \colhead{P$_{LB,AR}$} && \colhead{SNR} \\
\colhead{(14)} & \colhead{(15)} & \colhead{(16)} && \colhead{(17)} & \colhead{(18)} && \colhead{(19)} 
 }
\startdata
 849 &  0.0009 & -6.0~~~  &&  0.0012 & -0.1~~~~  &&  83~~ \\ 
 610 &  0.0005 &  -6.0~~~ &&  0.0006 &  -0.6~~~~ &&  34~~ \\ 
1147 &  0.0011 &  -6.0~~~ &&  0.0015 &   0.0~~~~ &&  52~~ \\
1079 &  0.0014 &  -6.0~~~ &&  0.0016 &  -0.2~~~~ &&  66~~ \\
 590 &  0.0009 &  -6.0~~~ &&  0.0010 &  -0.1~~~~ &&  29~~ \\
\enddata
\tablecomments{Column definitions: \\
(14) N$_{lc}$:  Number of measurements in TESS FFI lightcurve \\
(15) IQR$_{lc}$: InterQuartile Range of normalized lightcurve fluxes\\
(16) P$_{LB,lc}$: Log probability of autocorrelation in lightcurve from the Ljung-Box test. Values above -2 are consistent with white noise while values below -4 are not.\\
(17) IQR$_{AR}$: InterQuartile Range of ARIMA residuals\\
(18) P$_{LB,AR}$: Log probability of autocorrelation of ARIMA residuals from the Ljung-Box test\\
(19) SNR: Signal-to-noise of peak power in Transit Comb Filter periodogram }
\end{deluxetable}

\vspace*{-0.5in}
\begin{deluxetable}{ccccccccc}
\tabletypesize{\small}
\tablewidth{0pt}
\centering
\tablehead{
\multicolumn{5}{c}{Planet properties} && \colhead{Classifier} && Notes\\ \cline{1-5} 
\colhead{Period} & \colhead{Depth} &  \colhead{R$_{pl}$} & \colhead{T$_0$} & \colhead{b} && \colhead{P$_{RF}$} && \\
\colhead{(20)} & \colhead{(21)} & \colhead{(22)} & \colhead{(23)} & \colhead{(24)} && \colhead{(25)} && \colhead{(26)}
 }
\startdata
 3.60667$\pm$0.00038 & 0.0111$\pm$0.0014 & 16.1$\pm$2.6 & 1600.570$\pm$0.001 & 0.9$\pm$0.0 &&0.94 &&T~  \\
 1.65092$\pm$0.00047 & 0.0034$\pm$0.0009 & 8.9$\pm$2.1 & 1518.205$\pm$0.003 & 0.9$\pm$0.0 && 0.76 && T~ \\ 
 3.50173$\pm$0.00023 & 0.0165$\pm$0.0002 & 22.1$\pm$1.3 & 1440.432$\pm$0.001 & 0.6$\pm$0.1 && 0.47 && T~ \\
 5.50845$\pm$0.00121 & 0.0082$\pm$0.0003 &  8.6$\pm$0.6 & 1548.866$\pm$0.002 & 0.4$\pm$0.2 && 1.00 && T~ \\
 1.26251 & 0.0018 & 5.9 & 1473.641 & \nodata && 0.67 && T~ 
\enddata
\tablecomments{
Period, Depth, Planet Radius and T$_0$ are from the ARPS analysis if their associated errors are missing.  Otherwise they are from the best-fit 'Exoplanet' model.  Column definitions: \\
(20) Period of transit (day) \\
(21) Depth of transit (fraction of normalized flux)\\
(22) Planet radius (in Earth radii)\\
(23) Time of first transit (day)\\
(24) Impact parameter (fraction of stellar radius)\\
(25) P$_{RF}$: Probability of planet classification from Random Forest classifier\\
(26) Notes on individual objects appear in Appendix~\ref{sec:app_DTARPS}. 
}
\end{deluxetable}

\clearpage
\newpage

\clearpage
\newpage

\section{Refining Candidate Properties \label{sec:cand_fit}}

The ARPS methodology \citep{Caceres19b} is designed for sensitive transit detection, not for accurate charactrization of transit properties. It is therefore unsurprising that the properties emerging from ARPS, such as planet radius and orbital period, suffer biases and/or inaccuracies.  Paper I (\S8.1) showed that the transit depth obtained from the TCF algorithm tends to underestimate the planetary radius for injected planets. This is due to the combined effects of ARIMA model overfitting the stellar variations and the TCF underfitting the double-spike pattern.  It was also noted during the vetting process that, in some cases, an inaccurate TCF period causes a slight drift in phase from the earlier to later transits.  This section describes our efforts to ameliorate these problems. Table \ref{tab:DTARPS_cands} incorporates the improved radius and period values.

\subsection{Improved Radius with \texttt{exoplanet} Fits} \label{sec:exoplanet}

Limb darkening transit models \citep{Mandel02} are fitted to the DIAmante light curve for each of the DTARPS-S Candidates using the \texttt{exoplanet} package \citep{exoplanet21} based on Bayesian inference, Gaussian Processes regression, and the \texttt{PyMC3} algorithm  \citep{exoplanet:pymc316}. 
The light curve models were averaged into 30-minute bins with an oversampling factor of 15 for each of the bins. 
 
\begin{table}[b]
\fontsize{9}{11}\selectfont
  \centering
    \caption{Assumed Priors for \texttt{exoplanet} Fits}     \label{tab:fit_prior}
    \begin{tabular}{llll}
        \\\hline \hline
        Parameter & Description & Prior Distribution & Units \\\hline
         $P$ & Orbital period & U$P_\textrm{TCF}-0.5$, $P_\textrm{TCF}$+0.5) & days\\
$T_0$ & Epoch of first transit  & N($T_{0,TCF}$, 1) & BJD - 2457000 \\
$\delta$ & Transit depth & log N($\delta_{TCF}$, 2) & \\
$b$ & Impact parameter & U(0,1) & \\
$M_{\star}$ & Stellar mass & Bounded N($M_{\star}$, $\epsilon_{M_{\star}}$, 0, 3) & $M_{\sun}$\\
$R_{\star}$ & Stellar radius & Bounded N($R_{\star}$, $\epsilon_{R_{\star}}$, 0, 3) & $R_{\sun}$\\
$u_0$, $u_1$ & Quadratic limb-darkening coefficients & Adopted from \citet{Kipping13}\\\hline
    \end{tabular}
\end{table}

The free parameters for these transit models are: the period of the candidate transit ($P$), the epoch of the first transit ($T_0$), the transit depth ($\delta$), the impact parameter ($b$), the stellar mass ($M_{{\star}}$), the stellar radius ($R_{{\star}}$), and the quadratic limb-darkening coefficients ($u_0$, $u_1$). Assumed prior distributions are given in Table \ref{tab:fit_prior}. We use a narrow distribution for the orbital period and epoch of first transit because these are strongly constrained by the TCF and Random Forest classifier. But as TCF transit depths tended to be underestimated, a wider prior is used for depth. Priors for the stellar mass and radius parameters are based on the tabulated errors from the TIC \citep{Stassun19} with additional wide bounds. A wide uniform prior is assumed for the impact parameter. Prior distributions for the quadratic limb-darkening coefficients are adopted from \citet{Kipping13}. 

The posteriors were sampled with the No U-Turns step method \citep{NUTS14} with \texttt{PyMC3} using two chains with 2,000 tuning steps and 1,000 draws each. The target acceptance rate was set to 0.99. The posterior for the planet radius was drawn directly from the \texttt{starry} transit model for each candidate. Median and standard deviation posterior values are calculated from the 2,000 samples for each parameter. The transit model using the posterior parameters was plotted over the DTARPS-S candidate light curve and manually inspected.  

Well-fitting transit model solutions were obtained for 53\% of the DTARPS-S Candidates.  The failures can be attributed to the sparse nature of \TESS single-sector data where, in many cases, only a few observations are available to define the transit shape.

\subsection{Improved Period and Radius by Visual Inspection \label{sec:man_fit}}

When \texttt{exoplanet} failed to find a good transit model fit, the transit period and depth associated with the best TCF periodogram peak are adjusted visually to best fit the candidate transit signal. These DTARPS-S Candidates are identified in Table \ref{tab:DTARPS_cands} by the absence of errors for the transit parameters.

The period of the candidate transit was adjusted if a drift in the transit midpoint was seen over multiple transits. Due to the sparse nature of the light curves (with time steps of 30 minutes or 0.021 days), the period could not be visually adjusted with high precision. These adjusted periods have estimated errors of $\pm$0.001 day or $\sim$1 minute. TCF transit depths were visually adjusted so the depth matched the median of the light curve points during the transit event in the folded light curve. The error for the manual adjustment of the candidate transit depth is approximately estimated to be  $\pm$0.02\%.

\subsection{Reliability and Accuracy of Planet Properties} \label{sec:accuracy}

In Paper I (\S8.1), we found that the TCF periodogram tended to identify spuriously short periods for shallow injected transits and tended to underestimate the radius for deep injected transits.  The latter effect is understood: the ARIMA model incorporates some of transit dip and thus the signal in the ARIMA residuals indicate a smaller planet. The former effect is not understood. We now examine whether these problems are alleviated by the corrections described above. 

\begin{figure}[ht]
    \includegraphics[width=\textwidth]{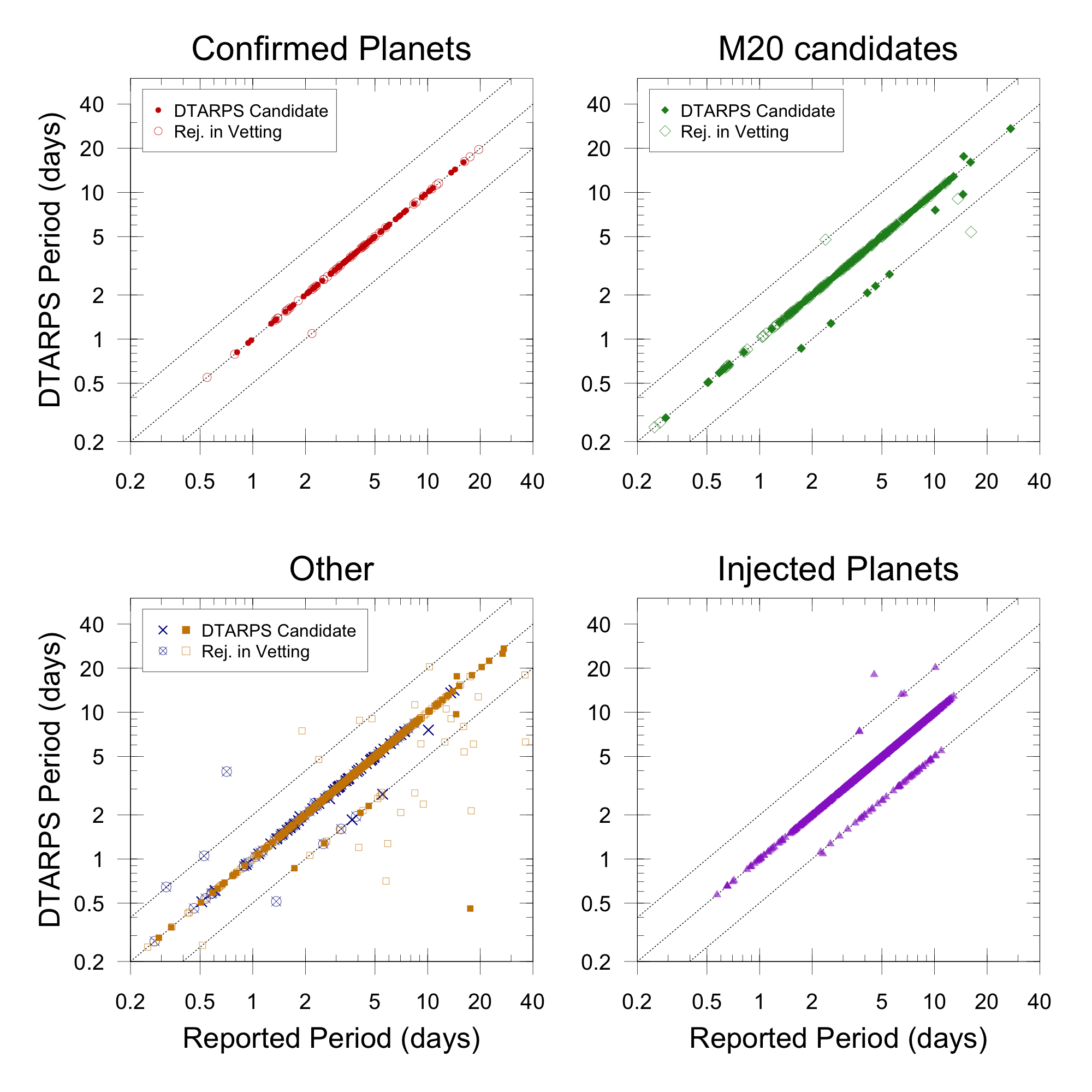}
    \caption{Comparison between the refined DTARPS period and the reported periods for different groups of objects: NASA/TOI Confirmed Planets (upper left); M20 candidates (upper right); candidates from other surveys (lower left); and our synthetic injected planets (Paper I).  The bottom left panel contains both planet candidates (gold squares) and previously identified False Positives (blue x's). Open symbols represent objects rejected by vetting. The dashed lines indicate the 2:1, 1:1, and 1:2 ratios between the reported and DTARPS period.  \label{fig:period_comparison}}
\end{figure}

Figure \ref{fig:period_comparison} compares the DTARPS periods with those reported for previously identified Confirmed Planets, M20 candidates, unconfirmed Planet Candidates, and planetary injections that were recovered by DTARPS from Paper I (\S8.1).  Each plot shows stars that were accepted and rejected by our vetting procedure.

The results are excellent.  All (100\%) of the Confirmed Planets in the DTARPS-S Candidates list have a DTARPS fitted period that matches the reported period in the NASA Exoplanet Archive or TOI list over the range $1 < P < 20$ days with a standard deviation of 0.0016 day or 2.3 minutes (top left panel Figure \ref{fig:period_comparison}). There is no case of DTARPS preferring a 1/2- or 2-times harmonic.  Similarly, 98\% of the DTARPS-S Candidates in the candidate lists of M20 and other surveys agree.  Most discrepant cases have DTARPS periods equal to 1/2 of the reported period; it is generally not known which is correct. One discrepant period is present: TIC 64071894 has a DTARPS $P = 0.45917$ day while TOI 458.01 has a reported period $P = 17.52883$ day. These may be different planets within the same system.  

It is gratifying that the erroneous short periods obtained when the DTARPS analysis is applied to planetary injections (Paper I, \S8.1) were not found when true Confirmed Planets were examined.  We note from Figure~\ref{fig:period_comparison} (upper left) that our vetting was somewhat too stringent, as it rejected some Confirmed Planets even though ARPS found the correct period.

\begin{figure}[ht]
    \includegraphics[width=\textwidth]{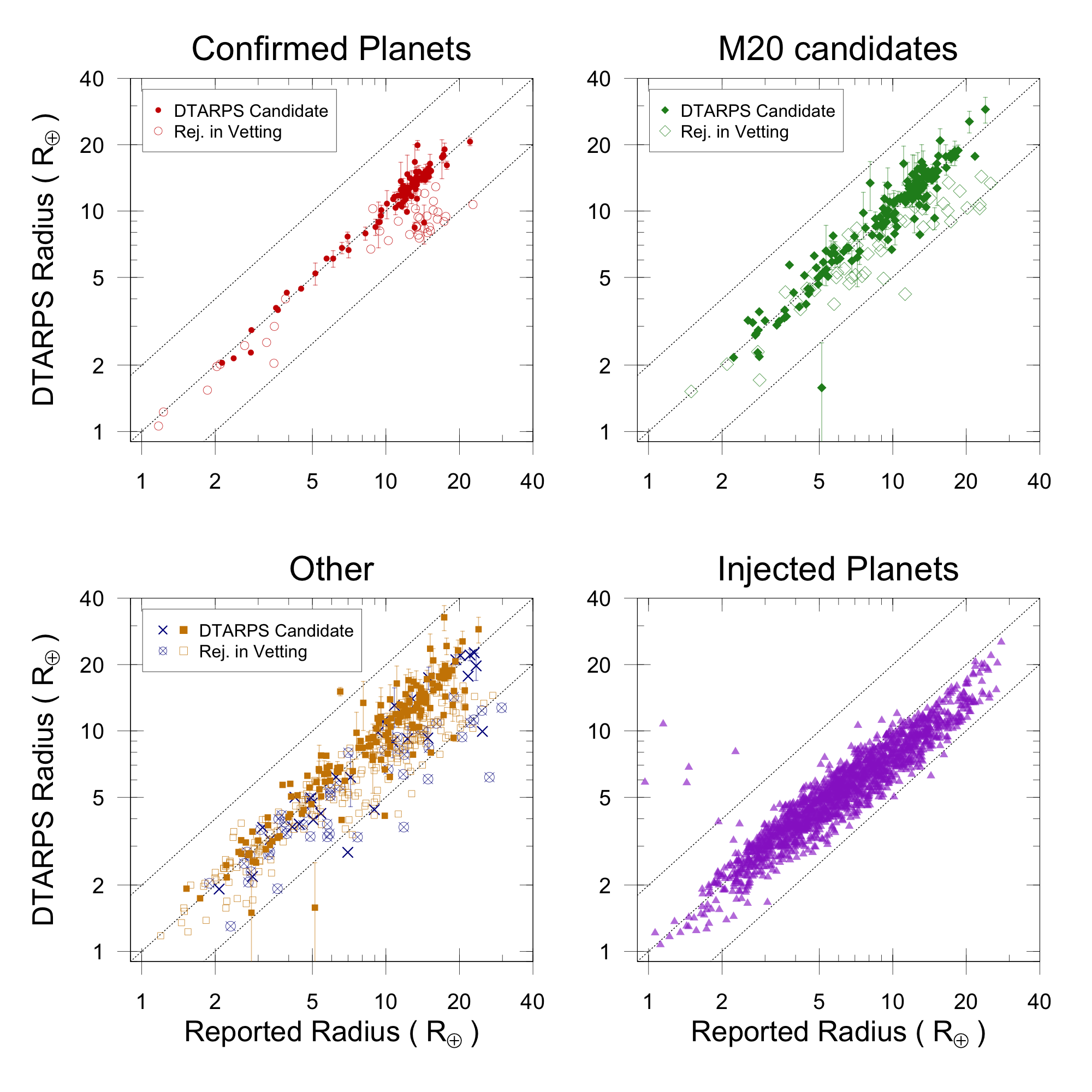}
    \caption{Comparison between the refined DTARPS radius and the reported radii for different groups of objects with reported radii and the planetary injections from the positive training set of the RF classifier. The bottom left panel contains both planet candidates (in gold squares) and previously identified False Positives (in blue x's). The bottom right panel is from Paper I.  The open points are plotted by the radius from their TCF periodogram peak. The dashed lines indicate the 2:1, 1:1, and 1:2 ratios between the reported radius and the DTARPS radius. \label{fig:rad_comparison}}
\end{figure}

Figure \ref{fig:rad_comparison} and Table \ref{tab:radii_diff} show a similar comparison between planet radii obtained by DTARPS and other surveys. The astrophysical \texttt{exoplanet} or manually corrected radii are used (\S\ref{sec:exoplanet}-\ref{sec:man_fit}) except in the lower right panel where the underestimation of large radii is evident. Error bars from the \texttt{exoplanet} astrophysical fits are plotted when available. The agreement between DTARPS and published radii is good over the range $2 < R < 20$~$R_\oplus$ with $70-80$\% of radii agreeing within $\pm 15$\%.  The underestimation of radii seen in Paper I with the injections is not apparent in the comparison to NASA and TOI Confirmed Planets.  The radius agreement is worse for objects that were rejected by DTARPS vetting (open points in Figure \ref{fig:rad_comparison}), but this has no impact on the quality of the DTARPS-S Candidate catalog.

\begin{deluxetable}{rcccc}[hb]
\centering
\caption{Differences between Published and DTARPS-S Planet Radii for Confirmed Planets  \label{tab:radii_diff}}
\tablehead{
          \colhead{Population} & \colhead{$<$ 85\%} & \colhead{100\%\tablenotemark{a}} & \colhead{$> 115\%$} & \colhead{N} \\
           &  \colhead{(1)} &  \colhead{(2)} & \colhead{(3)} & \colhead{(4)}
}
\startdata
        Super-Earth ($<$ 1.8 $R_{\oplus}$) & 67\% & 33\% & \nodata & 3\\
        Sub-Neptune (1.8 to 4 $R_{\oplus}$) & 7\% & 78\% & 15\% & 27\\
        Super-Neptune (4 to 8 $R_{\oplus}$) & 12\% & 72\% & 21\% & 58\\
        Jovian ($>$8 $R_{\oplus}$) & 9\% & 87\% & 15\% & 183\\
\enddata
\tablenotetext{a}{Objects whose DTARPS radius error bars fall within 15\% of the reported radii.}
\tablecomments{Columns $1-3$ give the percent of the Confirmed Planets and previously identified planet candidates whose DTARPS fitted radius are in the stated range of agreement.  Column 4 gives the number of DTARPS-S planets with Confirmed Planet counterparts in each radius stratum.}
\end{deluxetable}

\clearpage\newpage 

\clearpage\newpage

\section{Flagging Potential Problems \label{sec:flag}}

The analysis in this section is used to identify potential problems in DTARPS-S Candidates but does not remove objects in the candidate list. The DTARPS-S disposition of any Level 1 candidate flagged here is changed to Level 2.

\subsection{Evidence for Blending and Dilution from the ZTF survey}

The Zwicky Transient Facility \citep[ZTF;][]{Bellm19, Masci19} conducts a wide-field survey that scans the northern hemisphere sky every $\sim$2 days from the Palomar Observatory in the $g$ and $r$ optical bands with median depths of 20.8~mag and 20.6~mag, respectively. Despite the wide field of view of the telescope (47 deg$^2$), ZTF is able to achieve a 2\farcs0 full-width-half-maximum point spread function in images. The project releases both photometric measurements and light curves for sources to aid in the identification and characterization of astrophysical transients. ZTF light curves are examined here for eclipsing binaries near DTARPS-S Candidates that would be blended in the 21\arcsec\/ \TESS pixel and the photometric ZTF catalog is searched for nearby bright stars that would dilute a true transit signal. Only $\sim 1/4$ of DTARPS-S Candidates are covered by the ZTF survey due to their southern declinations.  Our study used the ZTF DR9 catalog hosted by the NASA/IPAC Infrared Science Archive. 

We examine ZTF light curves from stars within 15\arcsec\/ around each DTARPS-S candidate within the ZTF coverage area to identify potential blended eclipsing binaries.  Nearby stars with $> 50$ observations and $r<18$ were considered as potential contaminants. Fifty-eight DTARPS-S Candidates had blended sources with light curve data that fit these parameters. If a contaminant light curve had a $\chi^2$ value greater than 1.5, indicating that the light curve was not well fit with a constant magnitude stellar model, then the light curve was examined for evidence of periodic dips. Six DTARPS-S Candidates (10\% of the accessible sample) with nearby objects were found to have variable contaminants and were assigned Level 2 dispositions.   Details are provided in Appendix~\ref{sec:app_DTARPS}. 

Another 15\arcsec\/ search around each DTARPS-S candidate is performed to identify nearby bright stars that may dilute the depth of a true transit signal. Out of 94 candidates with ZTF observations, 18 had a nearby star with a $r<16$ that may cause transit dilution and thus had their disposition set to Level 2.  Details appear in Appendix~\ref{sec:app_DTARPS}. Most of the identified possible dilution contaminants are faint (e.g., a 15 magnitude star near an 11$-$12 magnitude candidate), and only $\sim$5 of the possible sources of dilution are likely to substantially affect our depth measurements. Overall, about 5\% of DTARPS-S Candidates may have blended field stars that diluted the DTARPS-S candidate star by 20-80\% in the \TESS pixel, reducing the inferred planet radius by 10-30\%. In such cases, a Neptune with tabulated radius of 7 $R_{\oplus}$ might actually have radius 8-9 $R_{\oplus}$. Individual cases are described in Appendix \ref{sec:app_DTARPS}.

\subsection{Gaia Radial Velocity Constraints}

\emph{Gaia} DR2 includes radial velocities (RVs) with goodness-of-fit measures of the astrometric solution for 7,224,631 sources \citep{GaiaCollab_Mission, GaiaCollab_DR2}. These quantities can be used to flag candidates with suspected stellar companions, although the relation between the astrometry measurements, RV measurements, and the candidate transit signal is unknown. Therefore if a candidate's \emph{Gaia} DR2 astrometric solution has poor fitting parameters or if the standard deviation of the RV measurement is too large, the disposition of the DTARPS-S Candidate is set to Level 2 indicating a possible stellar companion.  

\begin{figure}[ht]
    \includegraphics[width=0.48\textwidth]{ 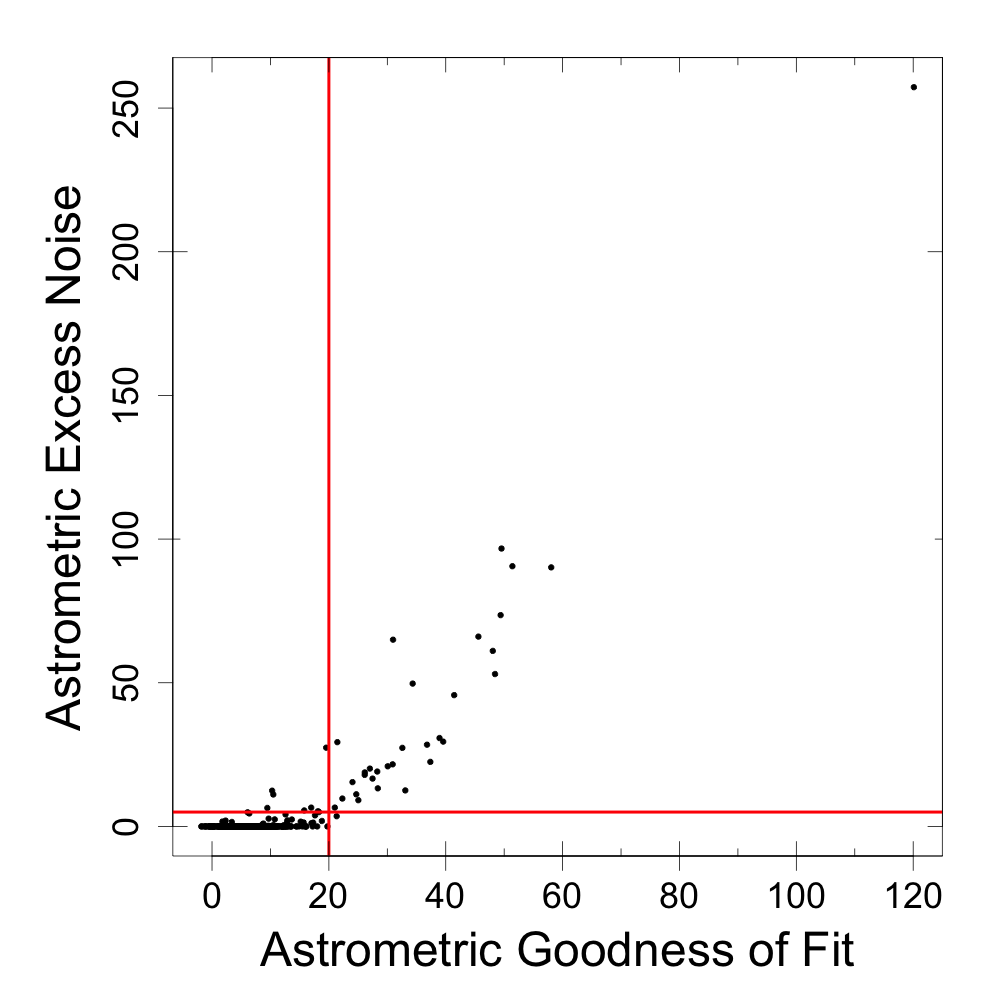}
    \includegraphics[width=0.48\textwidth]{ 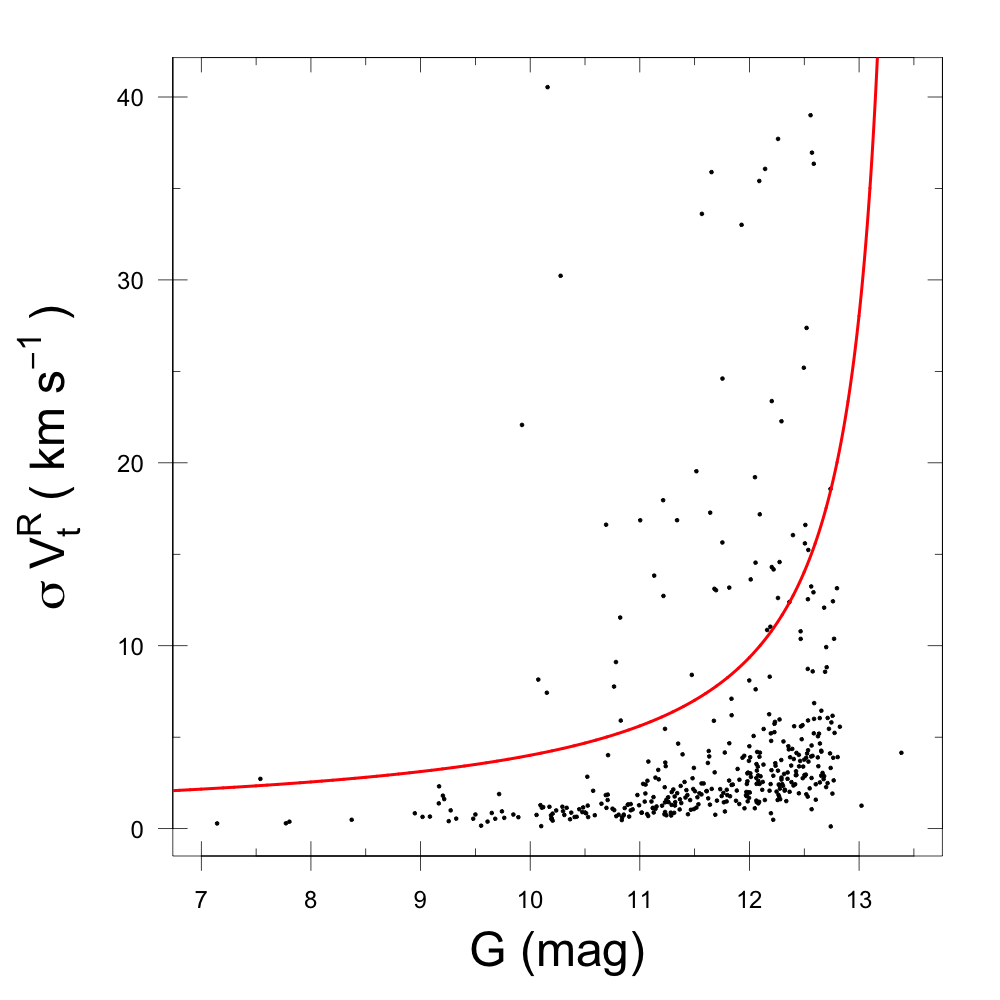}
    \caption{Astrometric analysis of DTARPS-S Candidates from \emph{Gaia} DR2 observations. The red lines in each plot show the upper boundary for a Level 1 candidate. The red boundaries are from \citet{Evans18}. (Left) Significance of the astrometric excess noise and astrometric goodness-of-fit in the along-scan direction for DTARPS-S Candidates. (Right) Standard deviation of the radial velocity measurements. The red boundary is from \citet{Montalto20}.}
    \label{fig:gaia_limits}
\end{figure}

\citet{Evans18} found that an astrometric excess noise significance \citep[][astrometric\_excess\_noise]{vizier_gaiaDR2} of 5 and an astrometric goodness-of-fit in the along-scan direction \citep[][astrometric\_gof\_al]{vizier_gaiaDR2} of 20 separated planetary systems orbiting close/unresolved binaries from planetary systems orbiting single stars. Thirty-nine DTARPS-S Candidates had either a significance of astrometric excess noise or an astrometric goodness-of-fit in the along-scan direction value greater than these limits (Figure \ref{fig:gaia_limits}, left panel).  

Additional thresholds for flagging DTARPS-S Candidates based on \emph{Gaia} RV measurements were adopted from M20.  The standard deviation of the RV measurements,
\begin{equation}
    \sigma_{V^t_R} = \sqrt{\frac{2N}{\pi}\left(\epsilon_{V_R}^2 - 0.11^2\right)},
\end{equation}
was obtained by inverting equation 1 in \citet{Katz19} where $\epsilon_{V_R}$ is the reported error of the RV \citep[][radial\_velocity\_error]{vizier_gaiaDR2} and $N$ is the number of radial velocity measurements \citep[][rv\_nb\_transits]{vizier_gaiaDR2}. The 5$\sigma$ boundary (red line) shown in the right panel of Figure \ref{fig:gaia_limits} was calculated from a fit to RV standard deviation measurements for known planet hosts based on the \emph{Gaia} $G$ magnitude. Fifty-three DTARPS-S Candidates have a RV standard deviation in excess of this 5$\sigma$ criterion. Seven of these stars exceeded both thresholds for a poor astrometric solution and RV standard deviation. Dispositions were changed from Level 1 to Level 2 for these cases.

\subsection{Upper Planetary Radii Limit}

The radii of the candidates were re-examined after many DTARPS-S Candidates radii had been fitted with parametric orbital models (\S \ref{sec:cand_fit}). Recall that during the vetting process, we had removed DTARPS-S Analysis List objects with radii greater than 20 $R_{\oplus}$ (\S\ref{sec:centroid_FP}). The disposition of any DTARPS-S candidate with a corrected planetary radius greater than 28 $R_{\oplus}$ (2.5 $R_{J}$) was set to Level 2. Seven DTARPS-S Candidates have a planetary radius above this threshold.

\section{Galactic Plane Analysis \label{sec:eval_ca}}

The centroid analysis applied to the 7,377 DTARPS-S Analysis List objects at the beginning of our vetting procedure measures possible spatial jitter of light during the photometric transit is combined with a crowding analysis measuring the density of brighter \emph{Gaia} DR2 stars in the immediate vicinity (\S\ref{sec:centroid}, M20). These criteria were perhaps too restrictive since 73\% of the stars were removed, including most objects in the Galactic Plane. Centroid-crowding analysis also was responsible for most of the rejections of Confirmed Planets. These problems may arise because the centroid-crowding bias combines the flux-weighted centroid in four apertures in the \TESS FFI image;  this may not be effective in crowded fields. 

\begin{figure}[hb]
    \centering
    \includegraphics[width=0.99\textwidth]{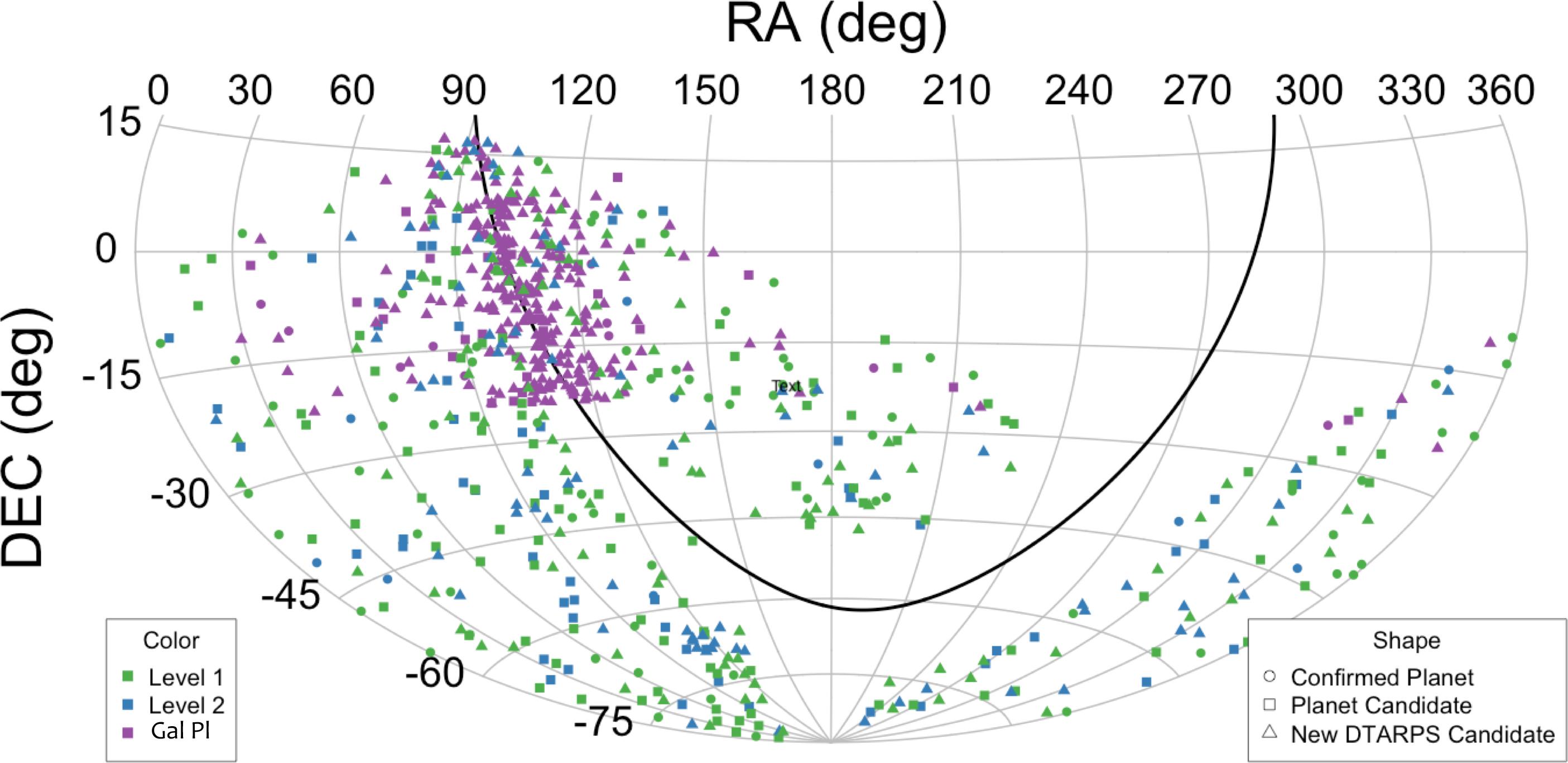}
    \caption{Spatial distribution of DTARPS-S Candidates (Tables \ref{tab:DTARPS_cands} and 2). Objects are differentiated by DTARPS-S disposition (DTARPS-S Level 1 and 2, DTARPS\_GP) and status with respect to previous surveys (previously known Confirmed Planet, previously known Planet Candidates, and new DTARPS-S Candidates).  The black curve indicates the approximate plane of the galaxy.}
    \label{fig:cands_ra_dec}
\end{figure}

In order to evaluate these restrictions on the DTARPS-S Candidate sample, we processed a sample of $\sim$1,800 DTARPS-S Analysis List objects through every stage of the vetting process except centroid-crowding analysis. This sample has Galactic longitudes from $193-236$ and latitudes mostly within $\pm 15^\circ$ of the Galactic Plane. The result was an additional 310 DTARPS-S Candidates that we call the {\it DTARPS-S Galactic Plane list}. The sky distribution of these additional DTARPS-S Candidates is shown in Figure~\ref{fig:cands_ra_dec}; most lie close to the Galactic Plane where the crowding criterion was violated.  These objects are listed in Table~\ref{tab:DTARPS-GP} using nearly the same format at Table~\ref{tab:DTARPS_cands}. Previously published information on them is summarized in Appendix B.

\begin{figure}
    \centering
    \includegraphics[width=0.99\textwidth]{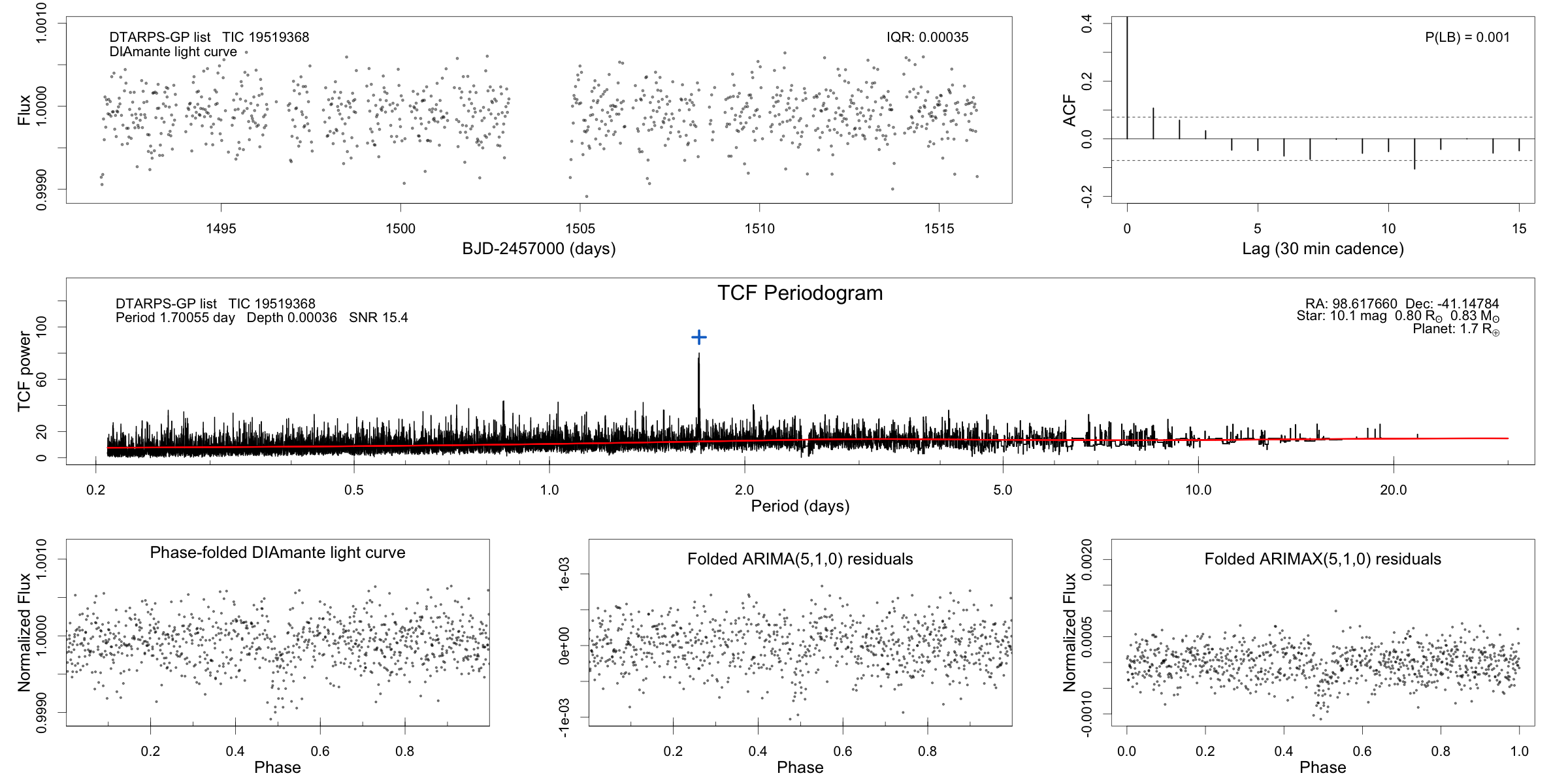} \\
    \vspace*{0.6in}
    \includegraphics[width=0.99\textwidth]{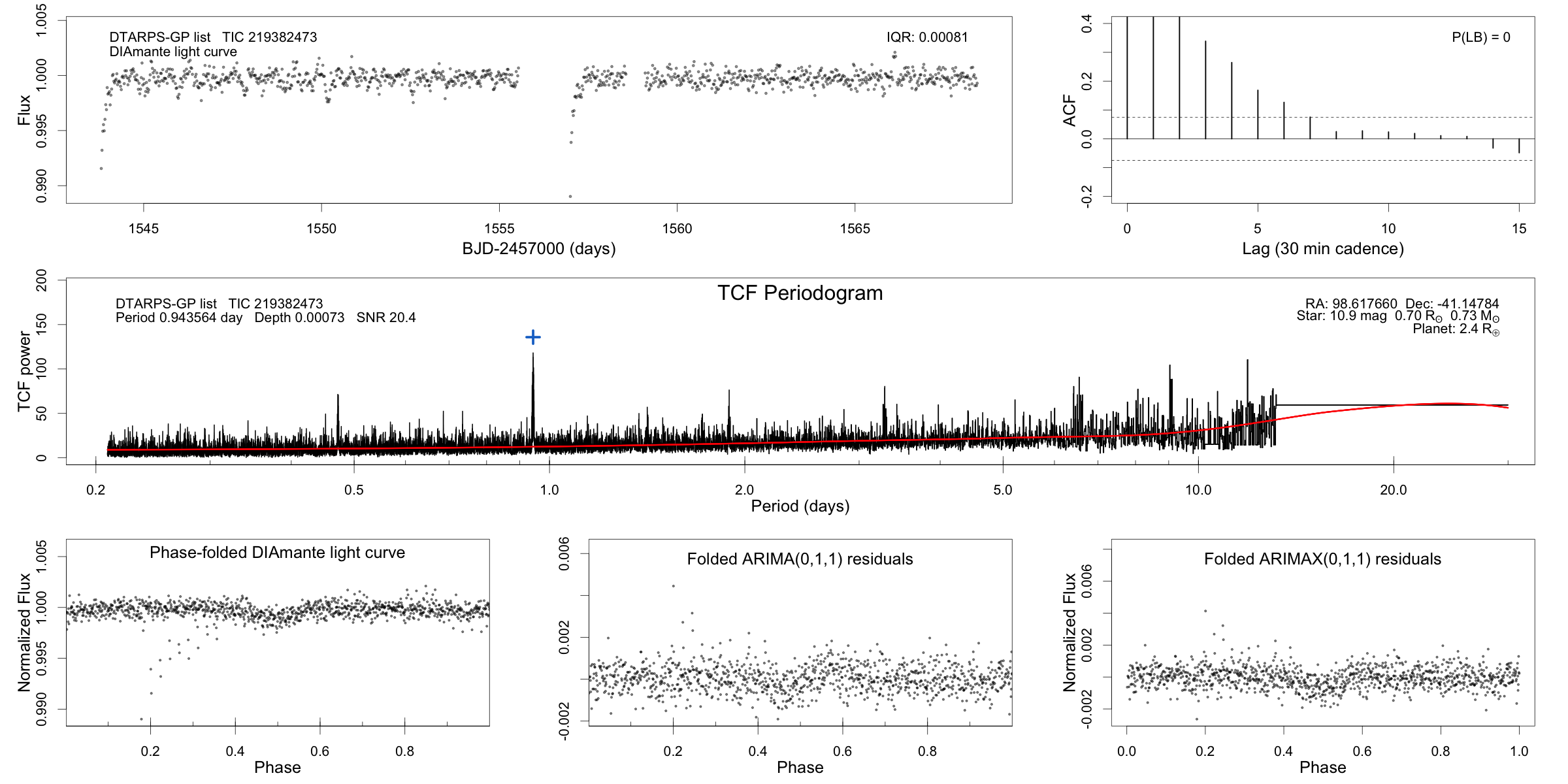} \\
    \caption{Figure Set graphics for two newly reported DTARPS-S Galactic Plane objects with small planetary radii: DTARPS-GP019519368 and DTARPS-GP219382473.  See text for descriptions.  The complete Figure Set for 310 DTARPS\_GP stars is available in the electronic version of the paper.
    }
    \label{fig:gp_example}
\end{figure}

Two cases representing smaller planets newly reported here are illustrated in Figure~\ref{fig:gp_example}, together with scalar values from Table~\ref{tab:DTARPS-GP} and information from Appendix B, can be described as follows:
\begin{description}

\item[DTARPS-GP019519368 = TOI-494] This is a T=10.1~mag K1 star at a distance of 90~pc with TOI designation 'Planetary Candidate'.   It has period $P=1.7005$ day, transit depth 0.04\% and corresponding planet radius 1.7~R$_\oplus$; a super-Earth. These parameters agree with the TOI values.  The Random Forest probability is high at $P_{RF}=0.51$. The Galactic coordinates are $(l,b)= (229,16)$.  

\item[DTARPS-GP219382473] This is a newly reported possible transiting planet orbiting a T=10.9~mag K4 star at d=96~pc.  It has period $P=0.9436$ day, transit depth 0.1\% and corresponding planet radius 2.4~R$_\oplus$.  The Random Forest probability $P_{RF}=0.32$ is only slightly above the 0.30 threshold.  The Galactic coordinates are $(l,b)= (279,35)$.  

\end{description}

\clearpage\newpage

\begin{deluxetable}{rrrrrrrrrrrr}[t]       \label{tab:DTARPS-GP}
\tablenum{4}
\tablecaption{DTARPS-S Galactic Plane List}     
\tabletypesize{\small}
\tablewidth{0pt}
\centering
\tablehead{
\multicolumn{4}{c}{Designations} && \multicolumn{7}{c}{Star} \\ \cline{1-4} \cline{6-12} 
\colhead{TIC} & \colhead{TOI} & \colhead{NEA} & \colhead{DIAm} &&  \colhead{R.A.} & \colhead{Dec} &   \colhead{T} &  \colhead{T$_{eff}$} & \colhead{R$_\star$} & \colhead{M$_\star$} & \colhead{Dist} \\  
\colhead{(1)} & \colhead{(2)} & \colhead{(3)} & \colhead{(4)} && \colhead{(5)} & \colhead{(6)} & \colhead{(7)} & \colhead{(8)} & \colhead{(9)} & \colhead{(10)} & \colhead{(11)}   }
\startdata
   883943  & \nodata & \nodata  & F && 127.02816 & -17.49844 & 10.6 & 5976 & 1.18 & 1.10 & 232 \\
  1129033  & 398     & KP       & F &&  37.15511 &  -7.06066 &  9.6 & 5605 & 0.93 & 0.99 & 105 \\
  1605476  & \nodata & \nodata  & F && 131.26395 &  -4.11884 & 11.6 & 6056 & 1.05 & 1.13 & 329 \\
  4616346  & \nodata & PC       & F && 101.95155 & -22.12333 & 11.6 & 5845 & 1.21 & 1.05 & 350 \\
  4784880  & \nodata & \nodata  & F && 109.66720 & -20.88884 & 11.0 & 5725 & 0.91 & 1.02 & 198  
\enddata
\tablecomments{The full table of 310 DTARPS\_GP objects is available in the electronic version of the paper in file DTARPS\_GP.MRT. Column definitions are the same as for Table~1 except the first two columns are absent.  Notes on individual objects appear in Appendix~\ref{sec:app_DTARPS-GP}.}
\end{deluxetable}

\setcounter{table}{3}
\begin{deluxetable}{rrrrrrrr}
\tabletypesize{\small}
\centering
\tablehead{
\multicolumn{3}{c}{Lightcurve} && \multicolumn{2}{c}{ ARIMA} && \colhead{TCF} \\ \cline{1-3} \cline{5-6} \cline{8-8}
\colhead{N$_{lc}$} & \colhead{IQR$_{lc}$} & \colhead{P$_{LB,lc}$} && \colhead{IQR$_{AR}$} & \colhead{P$_{LB,AR}$} && \colhead{SNR} \\
\colhead{(12)} & \colhead{(13)} & \colhead{(14)} && \colhead{(15)} & \colhead{(16)} && \colhead{(17)} 
 }
\startdata
  610 & 0.0006 & -6.0~~~ && 0.0007 & -0.2~~~ &&  29~~  \\
  799 & 0.0004 & -6.0~~~ && 0.0007 & -0.1~~~ &&  65~~  \\
  590 & 0.0010 & -6.0~~~ && 0.0015 & -0.1~~~ &&  56~~  \\
 1962 & 0.0009 & -2.8~~~ && 0.0010 & -0.3~~~ &&  62~~  \\
 1023 & 0.0009 & -5.2~~~ && 0.0008 & -5.3~~~ &&  63~~  \\
\enddata
\end{deluxetable}

\begin{deluxetable}{ccccccccc}
\tabletypesize{\small}
\tablewidth{0pt}
\centering
\tablehead{
\multicolumn{5}{c}{Planet properties} && \colhead{Classifier} && Notes\\ \cline{1-5} 
\colhead{Period} & \colhead{Depth} &  \colhead{R$_{pl}$} & \colhead{T$_0$} & \colhead{b} && \colhead{P$_{RF}$} && \\
\colhead{(18)} & \colhead{(19)} & \colhead{(20)} & \colhead{(21)} & \colhead{(22)} && \colhead{(23)} && \colhead{(24)}
 }
\startdata
1.47000             & 0.0038            &  8.0         & 1468.454           &  \nodata     && 0.66 && F \\
1.36000$\pm$0.00005 & 0.0203$\pm$0.0003 & 13.7$\pm$0.5 & 1413.705$\pm$0.000 & 0.2$\pm$ 0.1 && 0.52 && T \\
0.42100$\pm$0.00004 & 0.0058$\pm$0.0002 &  7.3$\pm$0.4 & 1517.578$\pm$0.001 & 0.1$\pm$ 0.1 && 0.72 && F \\
0.69047             & 0.0010            &   4.6        & 1468.891           & \nodata      && 0.38 && T \\
2.53600             &  0.0015           &   3.8        & 1520.183           & \nodata      && 0.8  && F \\
\enddata
\end{deluxetable}

\clearpage\newpage

\clearpage\newpage

The breakdown of the vetting labels for the $\sim$1,800 Galactic Plane sample was 53\% False Alarms, 18\% False Positives, 0.2\% possible photometric binaries, 1.5\% ephemeris match rejections, and 27\% planetary candidates.  These proportions are very similar to the vetting label breakdown of the DTARPS-S Candidates after centroid-crowding analysis (\S\ref{sec:centroid_FA}-\ref{sec:ephem}). 

The 310 DTARPS-S Galactic Plane objects were also analyzed for ZTF imaging, \emph{Gaia} dynamical constraints, and large planetary radii (\S\ref{sec:flag}). The ZTF analysis identified 10 stars with possible blended contamination from nearby variable ZTF light curves out of 140 candidates with light curves available; 34 objects may affected by possible dilution from a nearby star with $r<16$ out of 176 candidates. Details appear in Appendix~\ref{sec:app_DTARPS-GP}.  The \emph{Gaia} dynamical constraints found 20 DTARPS-S Galactic Plane objects with poor astrometric fits and 17 candidates with a RV standard deviation greater than the 5$\sigma$ threshold. Twelve have planetary radii greater than 28 $R_{\oplus}$.  These flags appear with similar rates as with DTARPS-S Candidates (\S \ref{sec:flag}).  

It thus appears that applying centroid-crowding criteria significantly reduced the number of  planetary candidates in our DTARPS vetting procedure while exhibiting no reduction in sensitivity to Confirmed Planets and little increase in the False Positives or other worrisome properties. Paper III reports that early results from reconnaissance spectroscopy DTARPS Galactic Plane objects appear to have the same low fraction of variable radial velocities as DTARPS-S candidates. (This result, however, is based on a very small sample.) We find no indication that DTARPS Galactic Plane objects are drawn from a different population than DTARPS-S Candidates.  At present, it appears that the 310 Galactic Plane objects in Table~\ref{tab:DTARPS-GP} are promising for further study including spectroscopic follow-up.   

This experiment on part of the Galactic Plane suggests that the centroid-crowding vetting step designed by M20 is too strict.  Removing it entirely would add $\sim$1,500 candidates to our \TESS southern hemisphere survey, increasing the fraction of candidates in the 7,377 DTARPS-S Analysis List from 6.3\% to $\sim$27\%.  The overall sensitivity for this enlarged DTARPS-S Candidate catalog would be $\sim$71\% for Confirmed Planets and $\sim$42\% for the planet candidates with specificity $\sim$37\%. The number of DTARPS-S discoveries in a combined sample would be similar to the pre-flight prediction of \citet{Barclay18} discussed in Paper I (\S1).  However, searching for periodic centroid wobbling in \TESS images is clearly advantageous for removal of contaminating eclipsing binaries (Appendix~\ref{sec:sens_FP}).  We therefore suggest that future efforts be made to design a vetting step of  intermediate severity that admits more likely candidates in the Galactic Plane region.

\section{Summary}\label{sec:summary}

In Paper I, the AutoRegressive Planet Search (ARPS) methodology was uniformly applied to $\sim 1$~million \TESS Year 1 light curves. This processing had four major steps: removal of trends and instrumental artifacts; removal of best-fit ARIMA models to give nearly-Gaussian white noise residuals; detection of transit-shaped periodic signals with the Transit Comb Filter in ARIMA residuals; and application of an optimized Random Forest classifier applied to 37 scalar features for each star. The classifier was trained towards injected planetary signals and against a combination of random light curves and injected eclipsing binary signals. A threshold to the RF probability was then applied to give the DTARPS-S Analysis List of 7,377 promising cases for transiting planets.

This procedure has advantages of uniformity and objectivity. Once various analysis parameters and thresholds were chosen, the $\sim 1$~million stars are processed in identical fashion without human judgment. The DTARPS-S Analysis List can thus be viewed as well-defined and `complete'; for instance, allowing completeness heat maps as a function of period and radius (Paper I, Figure~17). But the DTARPS-S Analysis List has limited utility for identifying individual exoplanets for further study because it is dominated by contaminants. Thus, a vetting program is needed to cull problematic cases and provide a smaller list of highly probable exoplanet candidates. In the parlance of statistical classification, the DTARPS-S Analysis List produced in Paper I optimizes $recall$ while the DTARPS-S Candidate Catalog emerging here seeks to optimize $precision$ by reducing False Alarms and False Positives through vetting procedures.   

Our vetting procedure described in \S\ref{sec:vet} reduces the list 7,377 promising cases nearly 10-fold, giving lists of 772 stars likely to be hosting short-period exoplanets: 462 in the spatially complete DTARPS-S Candidate catalog (Table~\ref{tab:DTARPS_cands}) and 310 in the spatially incomplete DTARPS Galactic Plane List (Table  4). But the vetting necessarily sacrifices the high standard of uniformity and objectivity achieved with the DTARPS-S Analysis List. The vetting procedures are based on human evaluations of graphical data, and in the case of centroid motion and crowding metrics, application of thresholds that have major effects on the final sample.  

In the DTARPS-S Candidate catalog, we segregate the more subjective judgments by `DTARPS-S disposition' flag. Level 1 dispositions are cases where the object satisfies all tests for a planetary transit and there is no evidence for an alternative classification. Level 2 dispositions are cases where we believe the evidence for a planetary transit is convincing, but there is some marginal vetting confirmation.  The DTARPS Galactic Plane objects have additional uncertainty as they were not subject to the centroid-crowding analysis. The evidence suggests that the Galactic Plane objects have similar recall and precision rates as the DTARPS-S Candidates, but there are reasons to believe that they may include more False Positive eclipsing binaries. 

Once the DTARPS-S Candidates catalog membership is established, we make efforts to improve estimates of the inferred planet properties $-$ orbital period and planet radius $-$ during the vetting process.  These efforts appear to be very successful:   Figures~\ref{fig:period_comparison} and \ref{fig:rad_comparison} show that the DTARPS properties are accurate. 

The DTARPS-S Candidates catalog (Table~1) and DTARPS Galactic Plane list (Table~4) are ready for spectroscopic follow-up to remove remaining False Positives, measure other planetary properties, and further investigate the planetary systems.  Paper III examines various astronomical properties of these samples, discussing exoplanet types such as the hot Neptune population, (extreme) Ultra Short Period planets, planets orbiting low-mass stars, and candidates for atmospheric characterization by transmission spectroscopy. The results are informed by preliminary results of reconnaissance spectroscopy of a small portion of the DTARPS-S Candidate sample. Paper III also returns to the completeness heat maps introduced in Paper I showing that the DTARPS-S Candidate planet occurrence rates are compatible with those obtained with the \Kepler mission.   

\begin{acknowledgements}

The ARPS project is supported at Penn State by NASA grant 80NSSC17K0122 and NSF grant AST-1614690.

The DTARPS project benefits from the vibrant community of Penn State's Center for Exoplanets and Habitable Worlds that is supported by the Pennsylvania State University, the Eberly College of Science, and the Pennsylvania Space Grant Consortium.  This study is also a product of the Center for Astrostatistics supported by the Eberly College of Science. We appreciate comments on the manuscript by members of these Centers $-$ Ian Czekala, Rebekah Dawson, Hyungsuk Tak, Jason Wright $-$ as well as Joel Hartman (Princeton). E.J.M. thanks Jiayin Dong (Penn State) for help executing the \texttt{exoplanet} software.  We also appreciate comments by a thoughtful anonymous referee. 

This paper includes data collected by the \TESS mission. Funding for the \TESS mission is provided by the NASA Explorer Program. This work has made use of data from the European Space Agency (ESA) mission {\it Gaia} (\url{https://www.cosmos.esa.int/gaia}), processed by the {\it Gaia} Data Processing and Analysis Consortium (DPAC, \url{https://www.cosmos.esa.int/web/gaia/dpac/consortium}). Funding for the DPAC has been provided by national institutions, in particular the institutions participating in the {\it Gaia} Multilateral Agreement. This research has made use of the NASA Exoplanet Archive, which is operated by the California Institute of Technology, under contract with the National Aeronautics and Space Administration under the Exoplanet Exploration Program.  It also made use of the SIMBAD database, operated at CDS, Strasbourg, France. The ZTF analysis is based on observations obtained with the Samuel Oschin Telescope 48-inch and the 60-inch Telescope at the Palomar Observatory as part of the Zwicky Transient Facility project. ZTF is supported by the National Science Foundation under Grant No. AST-2034437 and a collaboration including Caltech, IPAC, the Weizmann Institute for Science, the Oskar Klein Center at Stockholm University, the University of Maryland, Deutsches Elektronen-Synchrotron and Humboldt University, the TANGO Consortium of Taiwan, the University of Wisconsin at Milwaukee, Trinity College Dublin, Lawrence Livermore National Laboratories, and IN2P3, France. This research made use of \texttt{exoplanet} and its dependencies  \citep{exoplanet21}. 

\end{acknowledgements}

\facility{\TESS, NASA Exoplanet Archive, \emph{Gaia}}

\software{R core \citep{Rcore}: \emph{forecast} \citep{Hyndman21}; 
Python: \texttt{Jupyter} \citep{Jupyter16}, \texttt{ArviZ} \citep{ArviZ19}, \texttt{Astropy} \citep{astropy13, astropy18}, \texttt{celerite} \citep{celerite17, celerite18}, \texttt{exoplanet} \citep{exoplanet21}, \texttt{NumPy} \citep{numpy11, numpy20}, \texttt{Matplotlib} \citep{matplotlib07, matplotlib16}, \texttt{PyMC3} \citep{exoplanet:pymc316}, \texttt{starry} \citep{starry19}, \texttt{Theano} \citep{Theano16}
}

\bibliography{Paper_II_revised}{}
\bibliographystyle{aasjournal}

\appendix

\section{Notes on DTARPS-S Candidates}
\label{sec:app_DTARPS}

This Appendix gives a brief summary of previously published identifications of the DTARPS-S Candidates catalog, and some details from the DTARPS-S vetting process.  Detailed studies of known planetary systems are not covered. Conflicts in classification are not resolved;  in particular, \citet{Prsa22} reports eclipsing binaries with uncertain reliability because some are spectroscopically confirmed transiting planets. Catalog cross-references provide a sense of the star brightness:  Henry Draper catalog stars include the brightest $\sim 300,000$ stars in the sky, Durchmusterung (BD, CD, CPD) catalog stars include the brightest $\sim 700,000$ stars, and Tycho catalog stars include the brightest $\sim 2,500,000$ stars.  Sources: \\
CTOI: Community \TESS Objects of Interest \\
\citet{Montalto20}:  DIAmante project \\
NEA/TOI: NASA Exoplanet Archive and \TESS Objects of Interest (accessed February 2022 and  December 2023) \\
SIMBAD:  SIMBAD Astronomical Databased, CDS, Strasbourg FR (accessed February 2022 and December 2023)  \\
SPOC: \TESS Science Processing Operations Center, NASA-Ames (Jon Jenkins, private communication, November 2022) \\

DTARPS-S 1 = TIC 17361 = TOI-3127 = CTOI 17361.01. Disposition 'Ambiguous Planetary Candidate' with period P = 3.6071013, transit depth 0.97\% and planet radius 13 R$_\oplus$. [NEA/TOI] Planetary candidate from DIAmante analysis \citep{Montalto20}. [CTOI] 

DTARPS-S 2 = TIC 1003831 = TOI-564.  Spectroscopically 'Confirmed Planet' with period P = 1.6511 d, mass M = 460 M$_\oplus$ and radius R = 1.0 R$_J$ orbiting a F6 IV-V star. [NEA/TOI] Eclipsing binary with P = 1.6511490 d. \citep{Prsa22}

DTARPS-S 3 = TIC 1449640 = TOI-446.  Disposition 'Ambiguous Planetary Candidate' with period P = 3.5018023, transit depth 1.94\% and planet radius 13 R$_\oplus$. [NEA/TOI]  Eclipsing binary with P = 3.5017948 d.  \citep{Prsa22}

DTARPS-S 4 = TIC 1525480 = TOI-2402 = CTOI 1525480.01.  Disposition 'Planetary Candidate' with period P = 5.5094951, transit depth 1.61\% and planet radius 22 R$_\oplus$. [NEA/TOI] DIAmante survey and CTIO entry \citep{Montalto20}. [CTOI]

DTARPS-S 5 = TIC 1599403. ZTF 414209300031141 star 6\arcsec\/ from TIC star at r=16.0 mag with multiple dips to 16.2.  Possible blended contaminant. 

DTARPS-S 9 = TIC 9033144 = TOI-367.  Disposition 'False Positive' with period P = 4.71502, transit depth 0.47\% and equivalent planet radius 18 R$_\oplus$. [NEA/TOI] Eclipsing binary with P = 14.1451608 d. \citep{Prsa22}

DTARPS-S 10 = TIC 9725627 = WASP 30 = TOI-239.  Disposition 'Known Planet' with period P = 4.1567815, transit depth 0.47\% and planet radius 10 R$_\oplus$. [NEA/TOI]

DTARPS-S 11 = TIC 11356662 = CTOI 1356662.01.   Planetary candidate from DIAmante analysis with period P = 3.24887 d, transit depth 0.77 mmag and planet radius 2.8 R$_\oplus$ \citep{Montalto20}. [NEA/CTOI]  Very deep V-shaped transits, likely EB. [SPOC]

DTARPS-S 12 = TIC 11762560.  ZTF 409204100025935 star 6\arcsec\/ from TIC star at r=15.9 and g=17.0 with 0.3 mag dips.  Possible blended contaminant. 

DTARPS-S 13 = TIC 13021029 = WASP 61 = TOI-439.  Spectroscopically 'Confirmed Planet' with period P = 3.8559 d, mass M = 654  M$_\oplus$ and radius R = 1.24 R$_J$. [NEA/TOI] Eclipsing binary with P = 3.8523922 d. \citep{Prsa22}

DTARPS-S 14 = TIC 13349647 = WASP 36 = TOI-567.  Spectroscopically 'Confirmed Planet' with period P = 1.5373653 d, mass M = 730  M$_\oplus$ and radius R = 1.3 R$_J$. [NEA/TOI] Eclipsing binary with P = 1.5370594 d. \citep{Prsa22}

DTARPS-S 15 = TIC 14091633 = TOI-447.  Disposition 'Ambiguous Planetary Candidate' with period P = 55292743, transit depth 1.79\% and planet radius 23 R$_\oplus$. [NEA/TOI]  Very deep V-shaped transits, likely EB. [SPOC] 

DTARPS-S 16 = TIC 14091704 = TOI-445.  Disposition 'Planetary Candidate' with period P = 0.7649571, transit depth 0.19\% and planet radius 13 R$_\oplus$. [NEA/TOI] Planetary candidate with period P = 0.764880 d, transit depth = 0.19\% and width 1.5 h, triage probability 99.5\% and vetting probability 49\% from deep learning analysis  \citep{Yu19}. Eclipsing binary with P = 0.7650591	d. \citep{Prsa22}

DTARPS-S 18 = TIC 16490297 = 1SWASP J074724.19+122830.3.  Double-line spectroscopic eclipsing binary with period P = 2.751483 d, transit depth 2.3\% and width 3\% \citep{Schanche19}. False Positive in the SuperWASP and DIAmante surveys \citep{Montalto20}. [SIMBAD] 

DTARPS-S 21 = TIC 20178111 = TOI-467.  Disposition 'False Positive' with period P = 1.734422, transit depth 0.12\% and planet radius 4 R$_\oplus$. [NEA/TOI] Planetary candidate with period P = 1.734102 d, transit depth = 0.23\% and width 2.2 h, triage probability 99.4\% and vetting probability 41\% from deep learning analysis  \citep{Yu19}.

DTARPS-S 22 = TIC 20320305 = CD-25 8473.  Bright K3 V star at distance d = 52 pc. [SIMBAD]  Blended eclipsing binary 20\arcsec\/ offset from target.  [SPOC]

DTARPS-S 25 = TIC 21639186 = CD-43 7422. Eclipsing binary with P = 4.7297276 d. \citep{Prsa22} 

DTARPS-S 27 = TIC 22221375 = HD 86226 = TOI-652.  Spectroscopically 'Confirmed Planet' HD 86226 c with period P = 3.98442  d, mass M = 7.25  M$_\oplus$ and radius R = 0.193 R$_J$ orbiting a G1 V star.  HD 86226 b has period P = 1695 d.. [NEA/TOI]

DTARPS-S 28 = TIC 22529346 = WASP-121 = TOI-495.  Spectroscopically 'Confirmed Planet' with period P = 1.2749255  d, mass M = 375  M$_\oplus$ and radius R = 1.865 R$_J$ orbiting a F6 V star. [NEA/TOI] 

DTARPS-S 30 = TIC 22767596.  Transit depth differs in two TESS  sectors.

DTARPS-S 32 = TIC 24695044 = TOI-2195 = CTOI 24695044.01.   Disposition 'Planetary Candidate' with period P = 4.165693 d, transit depth 0.77\% and planet radius 9 R$_\oplus$. [NEA/TOI]   Planetary candidate from DIAmante analysis \citep{Montalto20}. [CTOI]

DTARPS-S 33 = TIC 25155310 = WASP-126 = TOI-114.  Spectroscopically 'Confirmed Planet' with period P = 3.2888 d, mass M = 89  M$_\oplus$ and radius R = 0.96 R$_J$ orbiting a G2 star. WASP-126 c has period P = 7.63 d. [NEA/TOI] Planetary candidate with period P = 3.288961 d, transit depth = 0.70\% and width 3.7 h, triage probability 99.6\% and vetting probability 89\% from deep learning analysis  \citep{Yu19} Eclipsing binary with P = 3.2887910 d. \citep{Prsa22} 

DTARPS-S 34 = TIC 26264354 = CTOI 26264354.01.   Planetary candidate from DIAmante analysis with period P = 3.24887 d, transit depth 0.77 mmag and planet radius 2.8 R$_\oplus$ \citep{Montalto20}. [CTOI]

DTARPS-S 35 = TIC 29430403 = CTOI 29430403.01.  Planetary candidate from DIAmante analysis with period P = 1.87756 d, transit depth 16.2 mmag and planet radius 13 R$_\oplus$ \citep{Montalto20}. [CTOI]

DTARPS-S 37 = TIC 31142436 = CTOI 31142436.01  Planetary candidate from DIAmante analysis with period P = 5.2714 d, transit depth 1.4 mmag and planet radius 3.0 R$_\oplus$ \citep{Montalto20}. [CTOI] Planetary candidate with period P = 5.271984 d, transit depth = 0.11\% and width 4.1 h, triage probability 45\% and vetting probability 25\% from deep learning analysis  \citep{Yu19}.

DTARPS-S 38 = TIC 32925763 = TOI-2352 = CTOI 32925763.01.   Disposition 'Planetary Candidate' with period P = 10.7846 d, transit depth 1.7 mmag and planet radius 8 R$_\oplus$. [NEA/TOI]  Planetary candidate from DIAmante analysis \citep{Montalto20}. [CTOI]

DTARPS-S 39 = TIC 32949757 = WASP-160A.  The planet is a transiting hot Saturn with period P = 3.7684952 d, mass $M_p = 0.3$ M$_J$ and radius $R_p = 1.1$ R$_J$  orbiting component K0 V WASP-160B of a near equal-mass visual binary with separation 29\arcsec\/ \citep{Lendl19}.  Planetary candidate with period P = 3.767558 d, transit depth = 1.1\% and width 2.9 h, triage probability 94\% and vetting probability 85\% from deep learning analysis  \citep{Yu19}. [SIMBAD] 

DTARPS-S 40 = TIC 33692729 = TOI-469.   Disposition 'Planetary Candidate' with period P = 13.6308024 d, transit depth 0.12\% and planet radius 3.5 R$_\oplus$. [NEA/TOI]

DTARPS-S 41 = TIC 34196883 = UCAC4 380-013011. Planetary candidate with period P = 1.617181 d, transit depth = 0.8\% and width 3.5 h, triage probability 99\% and vetting probability 36\% from deep learning analysis  \citep{Yu19}. [SIMBAD] 

DTARPS-S 42 = TIC 35703676.    Possible dilution  of transit depth by ZTF 399310100000336 with i=15.0 lying 8\arcsec\/ from the DTARPS-S candidate. 

DTARPS-S 43 = TIC 35857242 = WASP-138 = TOI-400.  Spectroscopically 'Confirmed Planet' with period P = 3.634433 d, mass M = 388 M$_\oplus$ and radius R = 1.09 R$_J$. [NEA/TOI] Eclipsing binary with P = 3.6344345 d. \citep{Prsa22}

DTARPS-S 46 = TIC 36734222 = WASP-43 = TOI-656.  Spectroscopically 'Confirmed Planet' with period P = 0.81347437 d, mass M = 651 M$_\oplus$ and radius R = 1.036 R$_J$  orbiting a K7 V  star. [NEA/TOI] Eclipsing binary with P = 0.8134735 d.  \citep{Prsa22}

DTARPS-S 47 = TIC 37718056 = WASP-158 = TOI-1908.  Spectroscopically 'Confirmed Planet' with period P = 3.656333 d, mass M = 886 M$_\oplus$ and radius R = 1.07 R$_J$  orbiting a F6 V  star. [NEA/TOI]

DTARPS-S 48 = TIC 38129218 = CTOI 38129218.01.   Planetary candidate from DIAmante analysis with period P = 2.40172 d, transit depth 3.2 mmag and planet radius 10 R$_\oplus$ \citep{Montalto20}. [CTOI]

DTARPS-S 49 = TIC 38696105 = TOI-281.   Disposition 'Planetary Candidate' with period P = 5.5764964 d, transit depth 0.07\% and planet radius 4 R$_\oplus$. [NEA/TOI] Planetary candidate with period P = 5.577113 d, transit depth = 0.06\% and width 2.4 h, triage probability 72\% and vetting probability 69\% from deep learning analysis  \citep{Yu19}.

DTARPS-S 52 = TIC 39018923 = TOI-2216.   Disposition 'Planetary Candidate' with period P = 5.68083 d, transit depth 0.21\% and planet radius  6 R$_\oplus$. [NEA/TOI]

DTARPS-S 53 = TIC 39414571 = TOI-2364 = CTOI 39414571.01 .   Disposition 'Planetary Candidate' with period P = 4.0197444 d, transit depth 0.97\% and planet radius 10 R$_\oplus$. [NEA/TOI]  Planetary candidate from DIAmante analysis \citep{Montalto20}. [CTOI] Confirmed giant planet with period 4.020 d and planet radius 0.77~R$_J$ \citep{Yee23}

DTARPS-S 54 = TIC 40083958 = TOI-851.   Disposition 'Planetary Candidate' with period P = 1.7018884 d.

DTARPS-S 55 = TIC 41995709 = TOI-4192.   Disposition 'Planetary Candidate' with period P = 4.2300182 d, transit depth 0.35\% and planet radius 17 R$_\oplus$. [NEA/TOI]

DTARPS-S 56 = TIC 43647325 = BD -06$^\circ$1077 = WASP-35 = TOI-423. Spectroscopically 'Confirmed Planet' with period P = 3.161575 d, mass M = 228 M$_\oplus$ and radius R = 1.32 R$_J$. [NEA/TOI]  Eclipsing binary with period 6.3231318 d \citep{Prsa22}.   ZTF 407204300018522 variable star 12\arcsec\/ from the TIC star at r=16.4-16.5 with flares to 16.2.     Possible dilution  of transit depth by ZTF 355114200027081 with g=15.8 lying 4\arcsec\/ from the DTARPS-S candidate.

DTARPS-S 57 = TIC 44737596 = HATS-75 = TOI-552. Spectroscopically 'Confirmed Planet' with period P = 2.7886556 d, mass M = 156 M$_\oplus$ and radius R = 0.88 R$_J$. [NEA/TOI]  Eclipsing binary with period 2.7886705 d. \citep{Prsa22}

DTARPS-S 58 = TIC 46096489 = WASP-16 = TOI-818.   Disposition 'Known Planet' with period P = 3.118536 d, transit depth 1.22\% and planet radius 12 R$_\oplus$. [NEA/TOI]  Spectroscopically 'Confirmed Planet' with period P = 3.1186068 d, mass M = 288 M$_\oplus$ and radius R = 1.22 R$_J$ . [NEA/TOI]

DTARPS-S 61 = TIC 49079670 = TOI-641.   Disposition 'Planetary Candidate' with period P = 1.8934942 d, transit depth 0.08\% and planet radius 3.2 R$_\oplus$. [NEA/TOI] Planetary candidate with period P = 1.891807 d, transit depth = 0.08\% and width 1.4 h, triage probability 89\% and vetting probability 81\% from deep learning analysis  \citep{Yu19}.

DTARPS-S 62 = TIC 50309953 = TOI-1109 = CTOI 50309953.01.   Disposition 'False Positive' with period P = 7.04083 d, transit depth 0.16\% and equivalent planet radius 8 R$_\oplus$. [NEA/TOI]  Planet candidate with period 7.04 d, transit depth 1.1 mmag and planet radius 5 R$_\oplus$  (Lipponen).  [CTOI] 

DTARPS-S 63 = TIC 50528326.     Possible dilution  of transit depth by ZTF 1401216300006538 with r=13.1 lying 10\arcsec\/ from the DTARPS-S candidate. 

DTARPS-S 64 = TIC 52216463 = HD 102197. Spectral type F8 V.  [SIMBAD]

DTARPS-S 65 = TIC 52369802.      Possible dilution  of transit depth by ZTF 1557204200008320 with r=16.0 lying 12\arcsec\/ and by ZTF 1557204200008312 with r=15.6 lying 14\arcsec\/ from the DTARPS-S candidate.

DTARPS-S 66 = TIC 52449797 = CTOI 52449797.01.   Planetary candidate from DIAmante analysis with period P = 2.44958 d, transit depth 0.9 mmag and planet radius 2.8 R$_\oplus$ \citep{Montalto20}  [CTOI]  Eclipsing binary with period 2.4511251 d \citep{Prsa22}.  TCF periodogram shows additional peaks suggesting possible multiplanet system. 

DTARPS-S 69 = TIC 53189332 = WASP`106 = TOI-660.   Spectroscopically 'Confirmed Planet' with period P = 9.28971 d, mass M = 541 M$_\oplus$ and radius R = 1.02 R$_J$  orbiting an F9 star. [NEA/TOI] Eclipsing binary with period 9.2908633 d \citep{Prsa22}. 

DTARPS-S 71 = TIC 53750200 = HATS-41 = TOI-1914.   Spectroscopically 'Confirmed Planet' with period P = 4.193649 d, mass M = 3082 M$_\oplus$ and radius R = 1.33 R$_J$ . [NEA/TOI]

DTARPS-S 72= TIC 54064834 = UCAC4 312-015385.  Planetary candidate with period P = 6.057196 d, transit depth = 0.3\% and width 2.5 h, triage probability 92\% and vetting probability 77\% from deep learning analysis  \citep{Yu19}.  [SIMBAD]

DTARPS-S 74 = TIC 55999432 = BD -07$^\circ$859 = ASAS J043811-0726.3 = 1RXS J043811.1-072621.  Photometric variable due to rotationally modulated starspots with amplitude 0.1 mag with T = 2.1 d rotational period \citep{Kiraga12}. Eclipsing binary with period 1.8555335 d \citep{Prsa22}.  X-ray source in the ROSAT Bright Survey.    The star appears in a study on super-flares (SIMBAD not yet ingested) \citep{Tu20}. [SIMBAD]

DTARPS-S 75 = TIC 59280552.     Possible dilution  of transit depth by ZTF 511209400013617 with r=14.8 lying 11\arcsec\/ from the DTARPS-S candidate.

DTARPS-S 77 = TIC 61117473 = TOI-3122.   Disposition 'Planetary Candidate' with period P = 6.182444 d, transit depth 1.04\% and planet radius 14 R$_\oplus$. [NEA/TOI]

DTARPS-S 79 = TIC 61404104 = TYC 5950-1237-1.  Planetary candidate with period P =  4.116773 d, transit depth = 0.13\% and width 3.6 h, triage probability 98\% and vetting probability 73\% from deep learning analysis  \citep{Yu19}.  [SIMBAD]

DTARPS-S 81 = TIC 61980227 = TOI-2316.   Disposition 'Ambiguous Planetary Candidate' with period P = 3.0958985 d, transit depth 1.04\% and planet radius 19 R$_\oplus$. [NEA/TOI]

DTARPS-S 83 = TIC 63720888 = TYC 6527-207-1.  High proper motion star [SIMBAD]  

DTARPS-S 84 = TIC 64070344 = CTOI 64070344.01.   Planetary candidate from DIAmante analysis with period P = 2.73716 d, transit depth 8.3 mmag and planet radius 14 R$_\oplus$ \citep{Montalto20}. [CTOI]

DTARPS-S 85 = TIC 64071894 = TYC 5862-52-1 = TOI-458.   Disposition 'Planetary Candidate' with period P = 17.5288347 d, transit depth 0.11\% and planet radius 2.9 R$_\oplus$. [NEA/TOI] High proper motion star [SIMBAD]

DTARPS-S 88 = TIC 67196573 = TOI-2499 = CTOI 67196573.01.   Disposition 'Planetary Candidate' with period P = 2.5574812 d, transit depth 0.19\% and planet radius 18 R$_\oplus$. [NEA/TOI] Planetary candidate with period P = 2.556024 d, transit depth = 0.33\% and width 1.1 h, triage probability 69\% and vetting probability 73\% from deep learning analysis  \citep{Yu19}.

DTARPS-S 89 = TIC 68007716  = TOI-2587.   Disposition 'Planetary Candidate' with period P = 5.4566563 d, transit depth 0.42\% and planet radius 11 R$_\oplus$. [NEA/TOI]  Confirmed giant planet with period 5.457 d and radius 1.08 R$_J$ \citep{Yee23}.   Possible dilution  of transit depth by ZTF 412210100031240 with r=13.6 lying 9\arcsec\/ from the DTARPS-S candidate.

DTARPS-S 90 = TIC 70914192 = TOI-427.  Disposition 'Ambiguous Planetary Candidate' with period P = 0.6090253 d, transit depth 0.42\% and planet radius 11 R$_\oplus$. [NEA/TOI]  Eclipsing binary with period 0.6095223 d \citep{Prsa22}. 

DTARPS-S 93 = TIC 72985822 = TOI-754.   Disposition 'Planetary Candidate' with period P = 3.8642753 d, transit depth 0.73\% and planet radius 14 R$_\oplus$. [NEA/TOI]

DTARPS-S 94 = TIC 73038411 = BD -01$^\circ$1919 = TOI-978.  Disposition 'Planetary Candidate' with period P = 13.1659202 d, transit depth 0.51\% and planet radius 11 R$_\oplus$.  [NEA/TOI] Planetary candidate with period P = 15.854 d and planet radius 12 R$_\oplus$ found in warm Jupiter survey \citep{Dong21}.  [SIMBAD]

DTARPS-S 99 = TIC 79172641 = CD -53$^\circ$8746.  

DTARPS-S 100 = TIC 79212406 = TYC 5962-3355-1.  

DTARPS-S 101 = TIC 83391903.  Eclipsing binary with period 2.4772474 s \citep{Prsa22}. 

DTARPS-S 102 = TIC 88571421 = TYC 6439-1132-1 = EBLM J0315-24.  Single-line spectroscopic binary, eclipsing binary with low mass companion with orbital period P = 3.190524 d with 5\% eccentricity \citep{Triaud17}.  The F5 primary has mass M = 1.3 M$_\odot$ and radius R = 1.3 R$_\odot$.  The secondary has mass M = 0.24 M$_\odot$.  [SIMBAD] Eclipsing binary with period 3.1905328 \citep{Prsa22}. 

DTARPS-S 103 = TIC 89020549 = CD -44$^\circ$14956 = TOI-132.  Spectroscopically 'Confirmed Planet' with period P = 2.1097019 d, mass M = 22 M$_\oplus$ and radius R = 0.305 R$_J$ orbiting a G8 V star. [NEA/TOI]

DTARPS-S 104 = TIC 92350408 = CTOI 92350408.01.   Planetary candidate from DIAmante analysis with period P = 2.95639 d, transit depth 13.2 mmag and planet radius 24 R$_\oplus$ \citep{Montalto20}. [CTOI] Likely False Positive due to large radius. [SPOC] Eclipsing binary with period 2.9574264 d \citep{Prsa22}. 

DTARPS-S 105 = TIC 92352621 = CD -34$^\circ$14724B = WASP-94B. Spectroscopically confirmed hot Jupiters orbiting each component of a wide binary system with separation 15\arcsec\/ \citep{Neveu14}.  The planet around F8 star WASP-94A has a retrograde orbit with period P=3.95019 d, mass M=0.45 M$_J$ and radius R = 1.7 R$_J$.  The planet around F9 star WASP-94B has period P = 2.00839 d, mass M = 196 M$_\oplus$.  The host stars have slightly different metallicities \citep{Teske16}.  [SIMBAD]

DTARPS-S 106 = TIC 92359850 = TOI-387.   Disposition 'False Positive' with period P = 4.155 d. [NEA/TOI]  Eclipsing binary with period 4.1499585 d \citep{Prsa22}. 

DTARPS-S 107 = TIC 92880568 = TYC 5924-1102-1 = TOI-958.   Disposition 'Ambiguous Planetary Candidate' with period P = 0.5875044 d, transit depth 0.17\% and planet radius 14 R$_\oplus$. [NEA/TOI] Planetary candidate with period P = 0.588447 d, transit depth = 0.35\% and width 1.4 h, triage probability 99.3\% and vetting probability 14\% from deep learning analysis  \citep{Yu19}.

DTARPS-S 109 = TIC 97156957 = TYC 7090-280-1.

DTARPS-S 110 = TIC 97165413 = TYC 7098-103-1.  

DTARPS-S 111 = TIC 97409519 = WASP-124 = TOI-113.   Spectroscopically 'Confirmed Planet' with period P = 3.37265 d, mass M = 190 M$_\oplus$ and radius R = 1.24 R$_J$ orbiting a F9 star. [NEA/TOI]

DTARPS-S 112 = TIC 97479438.     Possible dilution  of transit depth by ZTF 562202100006601 with r=15.6 lying 10\arcsec\/ from the DTARPS-S candidate.

DTARPS-S 113 = TIC 97861072 = TYC 729-700-1.     Possible dilution  of transit depth by ZTF 1557205400032288 with r=15.5 lying 8\arcsec\/ from the DTARPS-S candidate.

DTARPS-S 115 = TIC 98494121 = TYC 7436-1059-1 = TOI-2240.   Disposition 'False Positive' with period P = 3.06555 d, transit depth 0.29\% and equivalent planet radius 5 R$_\oplus$. [NEA/TOI]  High proper motion star [SIMBAD]

DTARPS-S 116 = TIC 100100827 = WASP-18 = HD 10069 = TOI-185.  Spectroscopically 'Confirmed Planet' WASP-18b has period P = 0.941452 d with mass M = 3300 M$_\oplus$ and radius R = 1.1 R$_J$.  A second planet WASP-18c has period P = 2.1558 d. [NEA/TOI] Eclipsing binary with period 0.9414549 d \citep{Prsa22}.    Transit depth differs in two TESS sectors.

DTARPS-S 117 = TIC 100589632 = TOI-2342 = CTOI 100589632.01 = TYC 7061-357-1.   Disposition 'False Positive' with period P = 3.9471567 d, transit depth 0.19\% and equivalent planet radius 4 R$_\oplus$. [NEA/TOI]  Planetary candidate from DIAmante analysis \citep{Montalto20}. [CTOI] Planetary candidate with period P = 3.946770 d, transit depth = 0.18\% and width 1.8 h, triage probability 80\% and vetting probability 69\% from deep learning analysis  \citep{Yu19}.

DTARPS-S 118 = TIC 101395259 = TOI-623.  Disposition 'False Positive' with period P = 15.51486 d, transit depth 0.62\% and equivalent planet radius 13 R$_\oplus$. [NEA/TOI]  Eclipsing binary with period 7.5746218 d \citep{Prsa22}.  

DTARPS-S 120 = UCAC2 13048233.  High proper motion star [SIMBAD]

DTARPS-S 121 = TIC 102192004 = WASP-174 = TOI-1919.  Spectroscopically 'Confirmed Planet' with period P = 4.2337 d, mass M = 413 M$_\oplus$ and radius R = 1.3 R$_J$ orbiting an F6 star. [NEA/TOI]

DTARPS-S 124 = TIC 111750317 = TYC 6390-110-1.  

DTARPS-S 126 = TIC 114749636 = WASP-147 = TOI-306.  Spectroscopically 'Confirmed Planet' with period P = 4.60273 d, mass M =  87 M$_\oplus$ and radius R = 1.1 R$_J$ orbiting a G4 star. [NEA/TOI]

DTARPS-S 127 = TIC 114987571 = TYC 7965-1050-1.

DTARPS-S 128 = TIC 115693377 = TYC 7528-481-1 = EBLM J0027-41.  Single-line spectroscopic binary, eclipsing binary with low mass companion with orbital period P = 4.9279889 d with 2\% eccentricity  \citep{Triaud17}.  The F8 primary has mass M = 1.2 M$_\odot$ and radius R = 1.2 R$_\odot$.  The secondary has mass M = 0.53 M$_\odot$. [SIMBAD] False Positive in the KELT survey \citep{Collins18}. Eclipsing binary with period 4.9279264 d \citep{Prsa22}. 

DTARPS-S 129 = TIC 116156517 = WASP-190 = TOI-338.  Spectroscopically 'Confirmed Planet' with period P = 5.367753 d, mass M = 318  M$_\oplus$ and radius R = 1.15 R$_J$ orbiting an F6 IV-V star. [NEA/TOI]

DTARPS-S 130 = TIC 117772549 = CTOI 117772549.01 = CD -33$^\circ$2972.  Planetary candidate from DIAmante analysis with period P = 7.87854 d, transit depth 0.5 mmag and planet radius 5 R$_\oplus$ \citep{Montalto20}. [CTOI]

DTARPS-S 132 = TIC 118956453 = WASP-170 = TOI-1925.  Spectroscopically 'Confirmed Planet' with period P = 2.34478022 d, mass M = 508  M$_\oplus$ and radius R = 1.096 R$_J$ orbiting an F6 IV-V star. [NEA/TOI]

DTARPS-S 133 = TIC 119292328 = TOI-512 = CD -38$^\circ$2650.   Disposition 'Planetary Candidate' with two planets.  TOI-512.01 has period P = 7.1887332 d, transit depth 0.03\% and planet radius 1.5 R$_\oplus$. [NEA/TOI] TOI 512.02 has period P = 20.2749388m trabsut depth 0.04\% and planet radius 1.7 R$_\oplus$. [NEA/TOI] High proper motion star [SIMBAD]

DTARPS-S 134 = TIC 120199940 = TYC 8753-82-1.  High proper motion star [SIMBAD]

DTARPS-S 135 = TIC 120247528 = TOI-1071 = HATS 21.   Disposition 'Known Planet' with period P = 3.553814 d, transit depth 1.00\% and planet radius 12 R$_\oplus$. [NEA/TOI]

DTARPS-S 136 = TIC 120610833 = WASP-45 = TOI-229.  Spectroscopically 'Confirmed Planet' with period P = 3.12609 d, mass M = 320  M$_\oplus$ and radius R = 1.15 R$_J$ orbiting an F6 IV-V star. [NEA/TOI]  Eclipsing binary with period 3.1261267 d \citep{Prsa22}.  

DTARPS-S 137 = TIC 122870280.     Possible dilution  of transit depth by ZTF 1457201200003550 with r=13.1 lying 10\arcsec\/ from the DTARPS-S candidate. 

DTARPS-S 138 = TIC 123627758 = TYC 5939-308-1.

DTARPS-S 139 = TIC 123702439 = TOI 499.  Disposition `Candidate Planet'  with period P = 8.5334865 d and planet radius 4.1 R$_\oplus$. [NEA/TOI]

DTARPS-S 140 = TIC 124206468 = TOI-2357 = CTOI 124206468.01 = TYC 7250-1125-1.  Disposition 'Planetary Candidate' with period P = 7.0955 d, transit depth 0.54\% and planet radius 12 R$_\oplus$. [NEA/TOI]   Planetary candidate from DIAmante analysis \citep{Montalto20}. [CTOI]

DTARPS-S 142 = TIC 128790976 = HD 189247 = TOI-1124.  Disposition 'Ambiguous Planetary Candidate' with period P = 3.51874 d, transit depth 0.58\% and planet radius 21 R$_\oplus$. [NEA/TOI]  Eclipsing binary with period 3.5187572 d \citep{Prsa22}. 

DTARPS-S 143 = TIC 134537478 = CD -42$\circ$3043 = KELT-14 = TOI-501.  Disposition 'Known Planet' with period P = 1.7100535 d, transit depth 1.27\% and planet radius 17 R$_\oplus$. [NEA/TOI]   Spectroscopically 'Confirmed Planet' with period P = 1.7100567 d, mass M = 400  M$_\oplus$ and radius R = 1.7 R$_J$ orbiting a G4 star. [NEA/TOI] 

DTARPS-S 145 = TIC 135050395 = CD -39$\circ$7570 = CTOI 135050395.01.   Planetary candidate from DIAmante analysis with period P = 6.76012 d, transit depth 23.1 mmag and planet radius 13 R$_\oplus$ \citep{Montalto20}. [CTOI] Eclipsing binary with period 6.7599877 d \cite{Prsa22}. 

DTARPS-S 146 = TIC 135145585 = CTOI 1315145585.01.   Planetary candidate from DIAmante analysis with period P = 1.51001 d, transit depth 4.5 mmag and planet radius 8 R$_\oplus$ \citep{Montalto20}. [CTOI]

DTARPS-S 147 = TIC 135163289 = CD -40$^\circ$7260.   

DTARPS-S 148 = TIC 135243705 = CTOI 135243705.01.  Planetary candidate from DIAmante analysis with period P = 2.65883 d, transit depth 9.5 mmag and planet radius 12 R$_\oplus$ \citep{Montalto20}. [CTOI]  Blended eclipsing binary in nearby 14~mag star. [SPOC]

DTARPS-S 149 = TIC 138727432 = TYC 7550-294-1 = TOI-853.   Disposition 'Planetary Candidate' with period P = 6.8646554 d, transit depth 0.51\% and planet radius 17 R$_\oplus$. [NEA/TOI]

DTARPS-S 150 = TIC 139285832 = TOI-332.   Disposition 'Planetary Candidate' with period P = 0.77685 d, transit depth 0.08\% and planet radius 3.2 R$_\oplus$. [NEA/TOI]  High proper motion star [SIMBAD]

DTARPS-S 151 = TIC 139733308 = HATS-5 = TOI-452.  Spectroscopically 'Confirmed Planet' with period P = 4.763387 d, mass M = 75  M$_\oplus$ and radius R = 0.912 R$_J$ orbiting a G star. [NEA/TOI]

DTARPS-S 152 = TIC 140067408 = CD -50$^\circ$13550. 

DTARPS-S 153 = TIC TIC 140444257 = TYC 142-544-1. Eclipsing binary with period 2.0845385 d \citep{Prsa22}. 

DTARPS-S 154 = TIC 140691463 = TOI-157.  Spectroscopically 'Confirmed Planet' with period P = 2.0845435 d, mass M = 375  M$_\oplus$ and radius R = 1.286 R$_J$ orbiting a G9 IV star. [NEA/TOI]  TCF preferred period is P = 2.08450 d.  Planetary candidate with period P = 2.084444 d, transit depth = 1.29\% and width 2.3 h, triage probability 99.2\% and vetting probability 60\% from deep learning analysis  \citep{Yu19}.  Eclipsing binary with period 2.0845385 d \citep{Prsa22}.  

DTARPS-S 155 = TIC 142305141 = CTOI 142305141.01.   Planetary candidate from DIAmante analysis with period P = 1.96047 d, transit depth 51.0 mmag and planet radius 14 R$_\oplus$ \citep{Montalto20}. [CTOI]

DTARPS-S 156 = TIC 142378043 = CD -30$\circ$9674 = TOI-1997.  Disposition 'Planetary Candidate' with period P = 1.8363262 d, transit depth 0.08\% and planet radius 5 R$_\oplus$. [NEA/TOI]

DTARPS-S 158 = TIC 142984508 = TYC 7028-394-1. 

DTARPS-S 159 = TIC 143257768 = HD 183845B = CTOI 143257768.01.   Planetary candidate from DIAmante analysis with Period 1.66241 d, transit depth 12.8 mmag and planet radius 18 R$_\oplus$ \citep{Montalto20}. [CTOI]  Double star. [SIMBAD]

DTARPS-S 160 = TIC 143858329 = TYC 8441-1369-1. 

DTARPS-S 162 = TIC 144164538 = CTOI 144164538.01.   Planetary candidate from DIAmante analysis with period P = 3.12849 d, transit depth 3.1 mmag and planet radius 1 R$_\oplus$ \citep{Montalto20}. [CTOI]

DTARPS-S 163 = TIC 144426921 = TYC 7779-146-1. Double-lined spectroscopic binary with period P = 3.105131 d and transit depth = 22 mmag in the KELT False Positive catalog \citep{Collins18}.  False Positives in DIAmante survey \citet{Montalto20}.  [SIMBAD]

DTARPS-S 164 = TIC 144700903 = TOI-532.  Spectroscopically 'Confirmed Planet' with period P = 2.3266508 d, mass M = 62  M$_\oplus$ and radius R = 0.519 R$_J$. [NEA/TOI]  Eclipsing binary with period 2.3237787 d \citep{Prsa22}. 

DTARPS-S 165 = TYC 7771-255-1. 

DTARPS-S 166 = TIC 145750719 = HATS-60 = TOI-303.  Spectroscopically 'Confirmed Planet' with period P = 10.695264 d, mass M = 13.8  M$_\oplus$ and radius R = 0.27 R$_J$. [NEA/TOI]

DTARPS-S 168 = TOI 146467675 = BD -20$\circ$999 = TOI-955.   Disposition 'False Positive' with period P = 1.63715 d, transit depth 0.08\% and equivalent planet radius 5 R$_\oplus$. [NEA/TOI]

DTARPS-S 169 = TIC 147115277 = HD 204019. High proper motion star.  [SIMBAD]

DTARPS-S 170 = TIC 147203645 = CD -44$^\circ$14526 = TOI-166.  Disposition 'Planetary Candidate' with period P = 5.0613599 d, transit depth 0.54\% and planet radius 10 R$_\oplus$. [NEA/TOI]  High proper motion star. [SIMBAD]

DTARPS-S 171 = TIC 150098860 = CD -61$^\circ$1276 = TOI-220.  Spectroscopically 'Confirmed Planet' with period P = 3.12609 d, mass M = 320  M$_\oplus$ and radius R = 1.15 R$_J$ orbiting an F6 IV-V star. [NEA/TOI] Planetary candidate with period P = 10.692789 d, transit depth = 0.06\% and width 2.5 h, triage probability 88\% and vetting probability 80\% from deep learning analysis  \citep{Yu19}. High proper motion star. [SIMBAD]

DTARPS-S 172 = TIC 150430484 = TOI-859.   Disposition 'Planetary Candidate' with period P = 51543182 d, transit depth 0.24\% and planet radius 10 R$_\oplus$. [NEA/TOI]  High proper motion star. [SIMBAD]

DTARPS-S 173 = TIC 150437346 = TOI-2220.   Disposition 'Planetary Candidate' with period P = 1.3926509 d, transit depth 0.72\% and planet radius 17 R$_\oplus$. [NEA/TOI] Planetary candidate with period P = 1.392659 d, transit depth = 0.71\% and width 2.1 h, triage probability 99.3\% and vetting probability 38\% from deep learning analysis  \citep{Yu19}.

DTARPS-S 174 = TIC 152147232 = TYC 7730-223-1 = TOI-759.   Disposition 'Ambiguous Planetary Candidate' with period P = 4.213316 d, transit depth 0.72\% and planet radius 13 R$_\oplus$. [NEA/TOI]  Spectroscopic binary. [SIMBAD]

DTARPS-S 175 = TIC 152226055 = CTOI 152226055.01.   Planetary candidate from DIAmante analysis with period P = 1.43115 d, transit depth 3.5 mmag and planet radius 5 R$_\oplus$ \citep{Montalto20}. [CTOI]  Blended eclipsing binary in 13~mag star 18\arcsec\/ from target. [SPOC]

DTARPS-S 176 = TIC 152799053 = TYC 8008-881-1. 

DTARPS-S 177 = TIC 153067709 = UCAC3 91-7106. High proper motion star. [SIMBAD]

DTARPS-S 178 = TIC 153095829 = TYC 6499-1918-1. 

DTARPS-S 179 = TIC 156075294 = TOI-299.   Disposition 'Planetary Candidate' with period P = 57216199 d, transit depth 0.54\% and planet radius 9 R$_\oplus$. [NEA/TOI]

DTARPS-S 180 = TIC 156241259 = TYC 723-1006-1.  Spectroscopic binary. [SIMBAD]

DTARPS-S 181 = TIC 156947629 = CTOI 156947629.01.   Planetary candidate from DIAmante analysis with period P = 5.25347 d, transit depth 2.1 mmag and planet radius 6 R$_\oplus$ \citep{Montalto20}. [CTOI]

DTARPS-S 183 = TIC 157230659 = NGTS-12.  Spectroscopically 'Confirmed Planet' with period P = 7.532806 d, mass M = 66  M$_\oplus$ and radius R = 1.048 R$_J$ orbiting a G star. [NEA/TOI]

DTARPS-S 184 = TIC 159951311 = WASP-139 = TOI-265.  Spectroscopically 'Confirmed Planet' with period P = 5.924262 d, mass M = 37  M$_\oplus$ and radius R = 0.8 R$_J$ orbiting a K0 star. [NEA/TOI]

DTARPS-S 185 = TIC 160148385 = WASP-96 = TOI-247.  Spectroscopically 'Confirmed Planet' with period P = 3.4252602 d, mass M = 155  M$_\oplus$ and radius R = 1.2 R$_J$ orbiting a G8 star. [NEA/TOI]  Eclipsing binary with period 6.8505104 d \citep{Prsa22}. 

DTARPS-S 187 = TIC 163260812 = TYC 8214-1633-1 = TOI-3098. `Candidate planet' with period P = 4.7332043 d and planet radius 17.1 R$_\oplus$.  [NEA/TOI]

DTARPS-S 188 = TIC 163360650 = TYC 8214-503-1. 

DTARPS-S 189 = TIC 165289668 = CD -37$^\circ$7549.

DTARPS-S 190 = TIC 166836920 = CD -50$^\circ$777 = WASP-99 = TOI-267.  Confirmed planet with period P = 5.75251 d, mass M = 883 M$_\oplus$ and radius R = 1.1 R$_J$ orbiting a F8 star. [NEA/TOI]

DTARPS-S 191 = TIC 167792080 = TYC 8912-693-1 = CTOI 167792080.01.  Planetary candidate with period P = 3.253503 d, transit depth 6.0 mmag and radius 15 R$_\oplus$ vetted with DAVE (Galgano) [CTOI] Eclipsing binary in 15~mag star 32\arcsec\/ from target. [SPOC]  Eclipsing binary with period 3.2532219 d \citep{Prsa22}. 

DTARPS-S 192 = TIC 168281028 = HATS-38  = CTOI 168281028.01.   Confirmed planet with period P = 4.375021 d, mass M = 23  M$_\oplus$ and radius R = 0.6 R$_J$. [NEA/TOI]  Planetary candidate from DIAmante analysis with period P = 4.37416 d, transit depth 3.5 mmag and planet radius 6 R$_\oplus$ \citep{Montalto20}. [CTOI]

DTARPS-S 193 = TIC 170102285 = WASP-23 = TOI-477.   Confirmed planet with period P = 2.94443 d, mass M = 279  M$_\oplus$ and radius R = 0.96 R$_J$ . [NEA/TOI] Planetary candidate with period P = 2.941959 d, transit depth = 2.03\% and width 2.6 h, triage probability 98\% and vetting probability 84\% from deep learning analysis  \citep{Yu19}.  Eclipsing binary with period 2.9444214 d \citep{Prsa22}. 

DTARPS-S 194 = TIC 172193428 = TYC 6520-198-1 = TOI-502.   Disposition 'Planetary Candidate' with period P = 2.9412694 d, transit depth 0.24\% and planet radius 9 R$_\oplus$. [NEA/TOI] Planetary candidate with period P = 2.939634 d, transit depth = 0.19\% and width 1.2 h, triage probability 89\% and vetting probability 80\% from deep learning analysis  \citep{Yu19}.  Eclipsing binary with period 2.9412329 d \citep{Prsa22}. 

DTARPS-S 195 = TIC 172409594 = TYC 6512-2125-1. Planetary candidate with period P = 0.341709 d, transit depth = 0.23\% and width 1.3 h, triage probability 92\% and vetting probability 18\% from deep learning analysis  \citep{Yu19}.  [SIMBAD]

DTARPS-S 196 = TIC 173640199 = TYC 7644-2302-1 = TOI-618.   Disposition 'Planetary Candidate' with period P = 4.4352884 d, transit depth 0.28\% and planet radius 8 R$_\oplus$. [NEA/TOI]

DTARPS-S 197 = TIC 176685457 = NGTS-9 = TOI-1935.   Confirmed planet with period P = 4.43527 d, mass M = 921  M$_\oplus$ and radius R = 1.07 R$_J$ orbiting a F8 star. [NEA/TOI]

DTARPS-S 199 = TIC 177079323 = TYC 9178-118-1 = TOI-2353 = CTOI 177079323.01.   Disposition 'Planetary Candidate' with period P = 5.7518756 d, transit depth 0.08\% and planet radius 6 R$_\oplus$. [NEA/TOI]  Planetary candidate from DIAmante analysis \citep{Montalto20}. [CTOI]

DTARPS-S 200 = TIC 177162886 = TOI-899.   Disposition 'Planetary Candidate' with period P = 12.8461782 d, transit depth 0.89\% and planet radius 12 R$_\oplus$. [NEA/TOI]

DTARPS-S 202 = TIC 177350401 = TYC 9186-659-1 = CTOI 177350401.01.   DIAmante planet candidate with period P = 12.07461 d, transit depth 4.3 mmag, planet radius 9 R$_\oplus$ \citep{Montalto20}. [CTOI]  Blended eclipsing binary in 14~map star 16\arcsec\/ from target.  [SPOC]  Eclipsing binary with period 12.0753454 d \citep{Prsa22}. 

DTARPS-S 204 = TYC 4852-2063-1. 

DTARPS-S 205 = TIC 179033846 = TOI-2675 = CTOI 179033846.01.   Disposition 'Planetary Candidate' with period P = 1.476869 d, transit depth 0.27\% and planet radius 13 R$_\oplus$. [NEA/TOI]  Planetary candidate from DIAmante analysis \citep{Montalto20}. [CTOI]

DTARPS-S 206 = TIC 180145006 = TOI-312.  Disposition 'Ambiguous Planetary Candidate' with period P = 3.5413787 d, transit depth 0.48\% and planet radius 10 R$_\oplus$. [NEA/TOI]

DTARPS-S 207 = TIC 180991313 = WASP 128 = TOI-2330 = CTOI 180991313.01.   Disposition 'Known Planet' with period P = 2.2088214 d, transit depth 0.78\% and planet radius 12 R$_\oplus$. [NEA/TOI] Planetary candidate from DIAmante analysis \citep{Montalto20}. [CTOI] 

DTARPS-S 208 = TIC 183120439 = TOI-169.   Confirmed planet with period P = 2.2554477 d, mass M = 251  M$_\oplus$ and radius R = 1.086 R$_J$ orbiting a G1 V star. [NEA/TOI]  

DTARPS-S 209 = TIC 183537452 = CD -40$^\circ$15273  = WASP-29 = TOI-192.  Spectroscopically 'Confirmed Planet' with period P = 3.92273 d, mass M = 73  M$_\oplus$ and radius R = 0.77 R$_J$. [NEA/TOI] High proper motion star [SIMBAD] 

DTARPS-S 210 = TIC 184240683 = WASP-5 = TOI-250.  Spectroscopically 'Confirmed Planet' with period P = 1.68243142  d, mass M = 498  M$_\oplus$ and radius R = 1.167 R$_J$. [NEA/TOI]

DTARPS-S 211 = TIC 186936449 = TYC 7146-1976-1  = TOI-2665 = CTOI 186936449.01.  Disposition 'Planetary Candidate' with period P = 2.8456575 d, transit depth 0.23\% and planet radius 7 R$_\oplus$. [NEA/TOI]  Planetary Candidate from convolutional neural network analysis \citep{Olmschenk21}.  [CTOI] 

DTARPS-S 212 = TIC 188570092 = HATS-72 = TOI-294.  Spectroscopically 'Confirmed Planet' with period P = 7.32829 d, mass M = 40  M$_\oplus$ and radius R = 0.722 R$_J$. [NEA/TOI] High proper motion star [SIMBAD]

DTARPS-S 213 = TIC 200324182 = HD 273029 = CTOI 200324182.01.   DIAmante planet candidate with period P = 1.29699 d, transit depth 1.9 mmag, planet radius 6 R$_\oplus$ \citep{Montalto20}. [CTOI] Planetary candidate with period P = 1.297000 d, transit depth = 0.19\% and width 1.5 h, triage probability 99.1\% and vetting probability 72\% from deep learning analysis  \citep{Yu19}.

DTARPS-S 214 = TIC 201177276 = TOI-2379 = CTOI 201177276.01.   Disposition 'Planetary Candidate' with period P = 5.4694164 d, transit depth 3.19\% and planet radius 11 R$_\oplus$. [NEA/TOI]  Planetary candidate from DIAmante analysis \citep{Montalto20}. [CTOI]

DTARPS-S 215 = TIC 201248411 = CD -55$^\circ$9423  = HIP 65 A = TOI-129.  Spectroscopically 'Confirmed Planet' with period P = 0.9809734  d, mass M = 1021  M$_\oplus$ and radius 2.03 R$_J$. [NEA/TOI] 

DTARPS-S 216 = TIC 201899356 = CTOI 201899356.01.   DIAmante planet candidate with period P = 3.10853 d, transit depth 11.1 mmag, planet radius 14 R$_\oplus$ \citep{Montalto20}. [CTOI]

DTARPS-S 217 = TIC 204376737 = WASP-6 = TOI-231.  Spectroscopically 'Confirmed Planet' with period P = 3.36100239  d, mass M = 153  M$_\oplus$ and radius R = 1.224 R$_J$. [NEA/TOI]

DTARPS-S 218 = TIC 204671232 = TOI-2361 = CTOI 204671232.01.  Disposition 'Planetary Candidate' with period P = 8.71546 d, transit depth 1.20\% and planet radius 12 R$_\oplus$. [NEA/TOI]  Planetary candidate from DIAmante analysis \citep{Montalto20}. [CTOI]

DTARPS-S 220 = TIC 209459275 = TOI-559.  Spectroscopically 'Confirmed Planet' with period P = 6.9839095  d, mass M = 1910  M$_\oplus$ and radius R = 1.091 R$_J$ orbiting a G V star. [NEA/TOI] High proper motion star [SIMBAD]

DTARPS-S 221 = TIC 219205407 = TYC 8114-653-1 = TOI-946.   Disposition 'False Positive' with period P = 6.1253092 d, transit depth 3.94\% and planet radius 22 R$_\oplus$. [NEA/TOI] Planetary candidate with period P = 6.125959 d, transit depth = 3.41\% and width 1.8 h, triage probability 98\% and vetting probability 18\% from deep learning analysis  \citep{Yu19}.  Spectroscopic binary [SIMBAD]

DTARPS-S 222 = TIC 219246401 = TYC 6091-94-1.  

DTARPS-S 223 = TIC 219388773 = TOI 399. Disposition `Candidate Planet' with period P = 1.571151 d and planet radius 3.5 R$_\oplus$. [NEA/TOI]

DTARPS-S 224 = TIC 219396153 = CTOI 219396153.01.   DIAmante planet candidate with period P = 10.24172 d, transit depth 10.2 mmag, planet radius 5 R$_\oplus$ \citep{Montalto20}. [CTOI]

DTARPS-S 225 = TIC 219421728 = HD 273953 = CTO 219421728.01.   DIAmante planet candidate with period P = 0.67113 d, transit depth 4.2 mmag, planet radius 10 R$_\oplus$ \citep{Montalto20}. [CTOI] Planetary candidate with period P = 0.671078 d, transit depth = 0.40\% and width 1.7 h, triage probability 99.6\% and vetting probability 29\% from deep learning analysis  \citep{Yu19}.

DTARPS-S 226 = TIC 219698950 = HATS-1 = TOI-766.   Disposition 'False Positive' with period P = 1.571151 d, transit depth 0.08\% and planet radius 3.5 R$_\oplus$. [NEA/TOI] Spectroscopically 'Confirmed Planet' with period P = 3.446459  d, mass M = 589 M$_\oplus$ and radius R = 1.3 R$_\odot$. [NEA/TOI]

DTARPS-S 227 = TIC 219701535 = TYC 6092-1130-1 = TOI-2399 = CTOI 219701535.01.   Disposition 'Planetary Candidate' with period P = 5.4681892 d, transit depth 0.43\% and planet radius 11 R$_\oplus$. [NEA/TOI] Planetary candidate from DIAmante analysis \citep{Montalto20}. [CTOI]

DTARPS-S 228 = TIC 219766989 = TOI-2354 = CTOI 219766989.01.   Disposition 'False Positive' with period P = 0.5198892 d, transit depth 0.27\% and planet radius 10 R$_\oplus$. [NEA/TOI] Planetary candidate from DIAmante analysis \citep{Montalto20}. [CTOI]

DTARPS-S 229 = TIC 219924802 = TYC 8775-952-1 = TOI-3282.  Disposition 'Ambiguous Planetary Candidate' with period P = 2.9478339 d, transit depth 0.50\% and planet radius 19 R$_\oplus$. [NEA/TOI]

DTARPS-S 230 = TIC 220068921 = TOI-2369 = CTOI 220068921.01. Disposition 'Planetary Candidate' with period P = 3.36376 d, transit depth 0.70\% and planet radius 10 R$_\oplus$. [NEA/TOI] Planetary candidate from DIAmante analysis \citep{Montalto20}. [CTOI]

DTARPS-S 231 = TIC 220076110 = TOI-2796 = CTOI 220076110.01. Disposition 'Planetary Candidate' with period P = 4.808498 d, transit depth 1.10\% and planet radius 15 R$_\oplus$. [NEA/TOI] Planetary candidate from DIAmante analysis \citep{Montalto20}. [CTOI]  Confirmed giant planet with period 4.808 d and planet radius 1.59 R$_J$ \citep{Yee23}. 

DTARPS-S 232 = TIC 220144363 = TOI-4199.  Disposition 'Planetary Candidate' with period P = 4.4466179 d, transit depth 0.79\% and planet radius 17 R$_\oplus$. [NEA/TOI]

DTARPS-S 233 = TIC 220396259 = CD -56$^\circ$952 = = TOI-379.  Disposition 'False Positive' with period P = 0.91677 d, transit depth 0.03\% and equivalent planet radius 3.1 R$_\oplus$. [NEA/TOI]

DTARPS-S 234 = TIC 220417892 = CTOI 220417892.01   DIAmante planet candidate with period P = 7.0339 d, transit depth 10.7 mag, planet radius 17 R$_\oplus$ \citep{Montalto20}. [CTOI]

DTARPS-S 235 = TIC 220459826 = TOI-872.  Disposition 'Planetary Candidate' with two planets.  TOI-872.01 has period P = 2.2395609 d, transit depth 0.11\% and planet radius 2.7 R$_\oplus$.  TOI-87202 has P = 4.9727684, transit depth 0.10\% and planet radius 2.6 R$_\oplus$. [NEA/TOI]  Planetary candidate with period P = 2.239526 d, transit depth = 0.10\% and width 1.3 h, triage probability 55\% and vetting probability 63\% from deep learning analysis  \citep{Yu19}.

DTARPS-S 236 = TIC 220518305 = TYC 8862-1459-1 = TOI-156.  Disposition 'False Positive' with period P = 14.152 d, transit depth 0.38\% and planet radius 11 R$_\oplus$. [NEA/TOI]

DTARPS-S 237 = TIC 220568520 = TYC 8863-203-1.  Single line spectroscopic binary with metal-rich primary star and a low-mass stellar companion in an eccentric orbit \citep{Mireles20}.  The primary has mass $M_\star$ = 1.03 M$_\odot$ and radius $R_\star$ = 1.01 R$_\odot$.   The secondary has orbital period P = 18.5577 d, eccentricity 10\%, radius $R_p$ = 1.25 R$_J$ and mass $M_p$ = 107 M$_J$.  [SIMBAD]

DTARPS-S 238 = TIC 229047362 = WASP-25 = TOI-767.  Disposition 'Known Planet' with period P = 3.764873 d, transit depth 2.22\% and planet radius 14 R$_\oplus$. [NEA/TOI] Spectroscopically 'Confirmed Planet' with period P = 3.764827  d, mass M = 190  M$_\oplus$ and radius R = 1.247 R$_J$. [NEA/TOI]

DTARPS-S 239 = TIC 229064630 = TYC 7511-114-1.

DTARPS-S 240 = TIC 229364030 = TYC 741-1429-1. 

DTARPS-S 243 = TIC 230980206 = TOI-2649 = CTOI 230980206.01.  Disposition 'Planetary Candidate' with period P = 3.375591 d, transit depth 1.16\% and planet radius 13 R$_\oplus$. [NEA/TOI] Planetary candidate from convolutional neural network analysis \citep{Olmschenk21}.  [CTOI] 

DTARPS-S 245 = TIC 230982885 = CD -56$\circ$324 = WASP-97 = TOI-195.  Spectroscopically 'Confirmed Planet' with period P = 2.07276  d, mass M = 432  M$_\oplus$ and radius R = 1.13 R$_J$ orbiting a G5 star. [NEA/TOI]

DTARPS-S 246 = TIC 231066944 = UCAC2 5487203.

DTARPS-S 247 = TIC 231274151 = TYC 8057-578-1.

DTARPS-S 248 = 	TIC 231930852 = CD -42$\circ$2405.

DTARPS-S 249 = TIC 231948825 = TYC 7624-259-1.

DTARPS-S 250 = TIC 233964642 = CD -64$\circ$69 = TOI-334.  Disposition 'Ambiguous Planetary Candidate' with period P = 17.9158854 d, transit depth 0.89\% and planet radius 21 R$_\oplus$. [NEA/TOI] Eclipsing binary with period 17.9157671 d \citep{Prsa22}. 

DTARPS-S 251 = TIC 234489976 = TOI-2308.  Disposition 'Planetary Candidate' with period P = 3.8686895 d, transit depth 0.80\% and planet radius 14 R$_\oplus$. [NEA/TOI]

DTARPS-S 252 = TIC 235015217 = CTOI 235015217.01.   DIAmante planet candidate with period P = 4.08793 d, transit depth 4.1 mmag, planet radius 13 R$_\oplus$ \citep{Montalto20}. [CTOI]

DTARPS-S 254 = TIC 235067594 = CTOI 235067592.01.   DIAmante planet candidate with period P = 8.29477 d, transit depth 4.7 mmag, planet radius 7 R$_\oplus$ \citep{Montalto20}. [CTOI] Planetary candidate with period P = 8.296909 d, transit depth = 0.37\% and width 3.0 h, triage probability 89\% and vetting probability 83\% from deep learning analysis  \citep{Yu19}. High proper motion star [SIMBAD]

DTARPS-S 256 = TIC 237885040 = TOI-2387.   Disposition 'Planetary Candidate' with period P = 5.6601098 d, transit depth 0.84\% and planet radius 12 R$_\oplus$. [NEA/TOI]

DTARPS-S 257 = TIC 238176110 = WASP-91 = TOI-116.  Spectroscopically 'Confirmed Planet' with period P = 2.798581 d, mass M = 425  M$_\oplus$ and radius R = 1.03 R$_J$ orbiting a K3 star. [NEA/TOI]

DTARPS-S 258 = TIC 242095191 = TYC 7276-1146-1.

DTARPS-S 259 = TIC 243641947 = TOI-3235.  Disposition 'Planetary Candidate' with period P = 2.5927363 d, transit depth 7.44\% and planet radius 10 R$_\oplus$. [NEA/TOI] High proper motion star. [SMBAD]

DTARPS-S 261 = TIC 247469203 = TYC 1298-163-1.

DTARPS-S 262 = TIC 248075138 = WASP-42 = TOI-769.  Spectroscopically 'Confirmed Planet' with period P = 4.9816872 d, mass M = 159 M$_\oplus$ and radius R = 1.063 R$_J$ . [NEA/TOI]  High proper motion star. [SIMBAD]

DTARPS-S 263 = TIC 248434289 = TYC 4759-1546-1.

DTARPS-S 264 = TIC 249022743 = CTOI 249022743.01.   DIAmante planet candidate with period P = 2.25601 d, transit depth 1.2 mmag, planet radius 4 R$_\oplus$ \citep{Montalto20}. [CTOI]

DTARPS-S 265 = TIC 250386181 = BD -05$^\circ$366 = TOI-390.  Disposition 'False Positive with period P = 1.470258 d, transit depth 0.26\% and planet radius 9 R$_\oplus$. [NEA/TOI] Eclipsing binary with period 1.4703473 d \citep{Prsa22}.   ZTF 400202200029768 variable star 5\arcsec\/ from TIC star with range r=15.4-17.4.  Possible blended contaminant,     Possible dilution  of transit depth by ZTF 1393210200029826 with r=15.6 lying 6\arcsec\/ from the DTARPS-S candidate.

DTARPS-S 266 = TIC 251848941 = TOI-178.   Confirmed planet with period P = 1.914558 d, mass M = 1.5  M$_\oplus$ and radius R = 0.103 R$_J$ orbiting a K star. TOI-178 c has period P = 3.23845 d, TOI-178 d has period P = 6.5577 d, TOI-178 e has period P = 9.955936 d, TOI-178 f has period P = 15.231915 d, TOI-178 g has period P = 2.07095 d.. [NEA/TOI] High proper motion star. [SIMBAD]

DTARPS-S 267 = TIC 254113311 = TOI-1130.  Spectroscopically 'Confirmed Planet' TOI-1130 b with period P = 4.066499 d, unknown mass and radius R = 0.326 R$_J$ orbiting a K7 star. TOI-1130 c has period P = 8.350381 d. [NEA/TOI]

DTARPS-S 269 = TIC 256158543 = TYC 9452-833-1 = TOI-3329 = CTOI 256158543.01.   Disposition 'Planetary Candidate' with period P = 5.0648607 d, transit depth 0.97\% and planet radius 13 R$_\oplus$. [NEA/TOI].  Planetary candidate from convolutional neural network analysis \citep{Olmschenk21}.  [CTOI] 

DTARPS-S 270 = TIC 257567854 = WASP-22 = TOI-403.  Spectroscopically 'Confirmed Planet' with period P = 3.53269  d, mass M = 178 M$_\oplus$ and radius R = 1.12 R$_J$ orbiting a G1 V star. [NEA/TOI]  Eclipsing binary with period 3.5334303 d \citep{Prsa22}. 

DTARPS-S 272 = TIC 260043723 = HD 42347 = TOI-217.  Disposition 'False Positive' with period P = 3.13879 d, transit depth 0.05\% and planet radius 5 R$_\oplus$. [NEA/TOI]

DTARPS-S 273 = TIC 260657291 = UCAC4 177-008238.  M0 V star at distance d = 83 pc in TESS Habitable Zone Star Catalog \citep{Kaltenegger21}. [SIMBAD]  Blended eclipsing binary on 17~mag star 17\arcsec\/ from target.  [SPOC] 

DTARPD 274 = TIC 261261490 = CTOI 261261490.01.  DIAmante planet candidate with period P = 3.539589 d, transit depth 0.7 mmag, planet radius 3.8 R$_\oplus$ \citep{Montalto20}. [CTOI] Eclipsing binary in 14~mag star 33\arcsec\/ from target. [SPOC] Eclipsing binary with period 113.0648644 d \citep{Prsa22}. 

DTARPS-S 276 = TIC 262330726 = TYC 103-1444-1.

DTARPS-S 277 = TIC 263003176 = HD 5278 = TOI-130.  Spectroscopically 'Confirmed Planet' HD 5278 b with period P = 14.339156 d, mass M = 7.8 M$_\oplus$ and radius R = 0.219 R$_J$ orbiting a F V star.  HD 5278 c has period P = 40.87 d. [NEA/TOI] High proper motion star [SIMBAD]

DTARPS-S 278 = TIC 264407995 = TYC 762-15-1. 

DTARPS-S 280 = TIC 264537668 = 1SWASP J052604.85+045135.4 = EBLM J0526+04. Eclipsing binary with period 4.0307037 d \citep{Prsa22}. 

DTARPS-S 283 = TIC 265077027 = TYC 205-2009-1.

DTARPS-S 285 = TIC 266980320 = HD 219666 = TOI-118.  Spectroscopically 'Confirmed Planet' with period P = 6.03607 d, mass M = 16.6 M$_\oplus$ and radius R = 0.42 R$_J$. [NEA/TOI] High proper motion star. [SIMBAD]

DTARPS-S 286 = TIC 268529943 = TYC 5850-383-1 = CTOI 268529943.01.   DIAmante planet candidate with period P = 4.30059 d, transit depth 21.6 mmag, planet radius 19 R$_\oplus$ \citep{Montalto20}. [CTOI] Eclipsing binary with period 4.3035024 d \citep{Prsa22}. 

DTARPS-S 287 = TIC 268644785 = CD -51$^\circ$2720 = KELT-15 = TOI-505.  Spectroscopically 'Confirmed Planet' with period P = 3.329441  d, mass M = 289 M$_\oplus$ and radius R = 1.443 R$_J$. [NEA/TOI]

DTARPS-S 288 = TIC 268766053 = WASP-53 = TOI-272.  Spectroscopically 'Confirmed Planet' WASP-53 b with period P = 3.3098443  d, mass M = 667 M$_\oplus$ and radius R = 1.07 R$_J$.  WASP-53 c has period P = 2840 d. [NEA/TOI] Eclipsing binary with period 6.6196826 d \citep{Prsa22}. 

DTARPS-S 290 = TIC 269558487 = BD -02$^\circ$146 = TOI-855.  Disposition 'Planetary Candidate' with period P = 1.8301278 d, transit depth 0.13\% and planet radius 5 R$_\oplus$. [NEA/TOI]

DTARPS-S 291 = TIC 270348758 = TYC 7282-1700-1.  Blended eclipsing binary with period P = 1.738217 d and transit depth 28 mmag in KELT False Positive catalog \citep{Collins18}.  [SIMBAD]

DTARPS-S 292 = TIC 270380593 = WASP-156 = TOI-465.  Spectroscopically 'Confirmed Planet' with period P = 3.836189  d, mass M = 40.68 M$_\oplus$ and radius R = 0.51 R$_J$. [NEA/TOI] Eclipsing binary with period 3.8769908 d \citep{Prsa22}. 

DTARPS-S 293 = TIC 270456887 = CTOI 270456887.01.   DIAmante planet candidate with period P = 2.26843 d, transit depth 7.8 mmag, planet radius 15.2 R$_\oplus$ \citep{Montalto20}. [CTOI]

DTARPS-S 294 = TIC 270468559 = HAT-P-42 = TOI-571.  Spectroscopically 'Confirmed Planet' with period P = 4.641878 d, mass M $\sim$ 300 M$_\oplus$ and radius R = 1.28 R$_J$. [NEA/TOI]

DTARPS-S 295 = TIC 270471727 = UCAC4 058-004726.

DTARPS-S 296 = TIC 271269442 = 1SWASP J075926.35+070728.4. False Positive (no radial velocity variations, \citet{Schanche19}).  ZTF  463213400031454 star 6\arcsec\/ from TIC star at r=17.0 with dips to 17.3.  Possible blended contaminant. 

DTARPS-S 297 = TIC 271893367 = TOI-150.  Spectroscopically 'Confirmed Planet' with period P = 5.857487 d, mass M = 798 M$_\oplus$ and radius R = 1.255 R$_J$  orbiting an F star. [NEA/TOI]  Eclipsing binary with period 5.8574622 d \citep{Prsa22}. 

DTARPS-S 298 = TIC 272430494 = UCAC2 937351.

DTARPS-S 300 = TIC 273695332 = TOI-2223. Disposition 'Planetary Candidate' with period P = 6.9320855 d, transit depth 0.69\% and planet radius 13 R$_\oplus$. [NEA/TOI]

DTARPS-S 301 = TIC 274017513 = UCAC3 82-4896. 

DTARPS-S 302 = TIC 277712294 = TYC 245-630-1 = EBLM J1013+01 = 1SWASP J101350.84+015928.1.   Single-line spectroscopic binary, eclipsing binary with low mass companion with orbital period P = 2.891182 day \citep{Schanche19}. The primary has mass M = 1.0 M$_\odot$, rotation velocity $v~sin(i) = 16$ km s$^{-1}$, metallicity $[Fe/H] = 0.3$ and age around 5 Gyr \citep{vonBoetticher19}. The secondary has mass M = 0.18 M$_odot$ and radius R = 0.15 R$_\odot$ producing a transit with depth 4\%.  The star appears in a study on super-flares (SIMBAD not yet ingested) \citep{Tu20}.  [SIMBAD]  Eclipsing binary with period 2.8922810 d \citep{Prsa22}. 

DTARPS-S 305 = TIC 280215684 = TYC 4803-981-1. 

DTARPS-S 306 = TIC 280614628 = TYC 179-562-1. ZTF 412214100006023 star 14\arcsec\/ from TIC star at r=18.0 with dips to 18.5. Possible blended contaminant.     Possible dilution  of transit depth by ZTF 412214100006142 with r=14.0 lying 12\arcsec\/ from the DTARPS-S candidate. 

DTARPS-S 307 = TIC 281143479.    Possible dilution  of transit depth by ZTF 460301200023076 with i=15.6 lying 10\arcsec\/ from the DTARPS-S candidate. 

DTARPS-S 308 = TIC 281408474 = HD 288842 = TOI-628.  Spectroscopically 'Confirmed Planet' with period P = 53.4095675 d, mass M = 2011 M$_\oplus$ and radius R = 1.06 R$_J$ . [NEA/TOI]

DTARPS-S 309 = TIC 282498590 = TOI-2492 = CTOI 282498590.01.  Disposition 'Ambiguous Planetary Candidate' with period P = 10.1011438 d, transit depth 1.31\% and planet radius 18 R$_\oplus$. [NEA/TOI] Planetary candidate from DIAmante analysis \citep{Montalto20}. [CTOI]

DTARPS-S 310 = TIC 282505014 = TYC 4809-264-1 = CoRoT 110769585. Spectral type F7 IV \citep{Guenther12}. [SIMBAD]

DTARPS-S 311 = TIC 282576340 = TYC 130-637-1 = TOI-2494.  Disposition 'Planetary Candidate' with two planets.  TOI-2494.01 has period P = 8.3761316 d, transit depth 0.90\% and planet radius 10 R$_\oplus$. TOI-2494.02 has P = 2.4087648 d, transit depth 0.06\% and planet radius 2.2 R$_\oplus$. [NEA/TOI] High proper motion star. [SIMBAD]

DTARPS-S 312 = TIC 283605976 = TOI-4679.  Disposition 'Planetary Candidate' with period P = 10.1011438 d, transit depth 1.31\% and planet radius 18 R$_\oplus$. [NEA/TOI]

DTARPS-S 313 = TIC 284160132 = TYC 701-1723-1. 

DTARPS-S 314 = TIC 286099128 = TOI-3422. Disposition 'Planetary Candidate' with period P = 3.7788872 d, transit depth 1.81\% and planet radius 14 R$_\oplus$. [NEA/TOI]

DTARPS-S 315 = TIC 287328202 = CTOI 287328202.01.   DIAmante planet candidate with period P = 3.8348 d, transit depth 1.1 mmag, planet radius 5.7 R$_\oplus$ \citep{Montalto20}. [CTOI]

DTARPS-S 316 = TIC 287721202 = TOI-3022. Disposition 'Planetary Candidate' with period P = 8.990904 d, transit depth 0.65\% and planet radius 10 R$_\oplus$. [NEA/TOI]

DTARPS-S 318 = TIC 2899889797 = BD -12$\circ$3697 = TOI-4178 = CTOI 2899889797.01.  Disposition 'Planetary Candidate' with period P = 2.2042197 d, transit depth 0.3\% and planet radius 14 R$_\oplus$. [NEA/TOI]  Planetary candidate from DIAmante analysis \citep{Montalto20}. [CTOI]

DTARPS-S 319 = TIC 290036495 = TOI-2351.  Disposition 'Planetary Candidate' with period P = 5.2524 d, transit depth 2.16\% and planet radius 19 R$_\oplus$. [NEA/TOI] 

DTARPS-S 321 = TIC 290777444 =  CTOI 290777444.01.   DIAmante planet candidate with period P = 6.10795 d, transit depth 14.8 mmag, planet radius 18 R$_\oplus$ \citep{Montalto20}. [CTOI]     Possible dilution  of transit depth by ZTF 461209100035300 with r=14.8 lying 9\arcsec\/ and by ZTF 461209100035304 with r=15.3 lying 11\arcsec\/ from the DTARPS-S candidate.

DTARPS-S 322 = TIC 292777663 = CTOI 292777663.01.   DIAmante planet candidate with period P = 4.60314 d, transit depth 1.1 mmag, planet radius 5 R$_\oplus$ \citep{Montalto20}. [CTOI]

DTARPS-S 324 = TIC 294301883 = WASP-55 = TOI-774.  Spectroscopically 'Confirmed Planet' with period P = 5.857487 d, mass M = 798 M$_\oplus$ and radius R = 1.255 R$_J$  orbiting an F star. [NEA/TOI]

DTARPS-S 325 = TIC 294691204 = TYC 9066-1809-1 = TOI-5018 = CTOI 294691204.01. Disposition 'Planetary Candidate' with period P = 6.5859565 d, transit depth 0.16\% and planet radius 6 R$_\oplus$. [NEA/TOI]  Planetary candidate from DIAmante analysis \citep{Montalto20}. [CTOI]

DTARPS-S 326 = TIC 294765664 = TYC 9289-1450-1.

DTARPS-S 327 = TIC 295176393 = CD -69$\circ$1709 = CTOI 295176393.01.   DIAmante planet candidate with period P = 27.2519 d, transit depth 5.3 mmag, planet radius 13 R$_\oplus$ \citep{Montalto20}. [CTOI]

DTARPS-S 328 = TIC 295541511 = TOI-1117. Disposition 'Planetary Candidate' with period P = 2.2264788 d, transit depth 0.06\% and planet radius = 2.7 R$_\oplus$. [NEA/TOI]

DTARPS-S 329 = TIC 299780329 = TYC 9357-866-1 = TOI-1869 = CTOI 299780329.01.  Disposition 'Planetary Candidate' with period P = 1.6024044 d, transit depth 0.14\% and planet radius 6 R$_\oplus$. [NEA/TOI]  Planetary candidate from eleanor analysis \citep{Feinstein19}. 

DTARPS-S 330 = TIC 300116105 = TYC = 7615-372-1 = CTOI 300116105.01.  DIAmante planet candidate with period P =  2.07693 d, transit depth 3.3 mmag, planet radius 7 R$_\oplus$ \citep{Montalto20}. [CTOI] Planetary candidate with period P = 2.075595 d, transit depth = 0.31\% and width 1.4 h, triage probability 94\% and vetting probability 65\% from deep learning analysis  \citep{Yu19}.

DTARPS-S 334 = TIC 300560295 = TYC 9179-1083-1 = CTOI 300560295.01.   Planetary candidate with period P = 10.28889 d, transit depth 0.42\%, planet radius 18 R$_\oplus$ (Lipponen).  [CTOI]  Eccentric eclipsing binary.  [SPOC] Eclipsing binary with period 20.5887760 d \citep{Prsa22}. 

DTARPS-S 335 = TIC 301343785 = TYC 4709-1048-1 = TOI-2347 = CTOI 301345785.01.  Disposition 'Ambiguous Planetary Candidate' with period P = 1.1736318 d, transit depth 0.15\% and planet radius 8 R$_\oplus$. [NEA/TOI]  Planetary candidate from DIAmante analysis \citep{Montalto20}. [CTOI]

DTARPS-S 336 = TIC 301980639 = CTOI 301980693.01.   Planetary candidate with period P = 4.782639 d, transit depth 24.1 mmag, planet radius 17 R$_\oplus$ (Lipponen). [CTOI] Eclipsoing binary with period 4.7830501 d \citep{Prsa22}. 

DTARPS-S 338 = TIC 302333151 = TYC 6946-123-1 = TOI-2315 = CTOI 302333151.01.  Disposition 'Ambiguous Planetary Candidate' with period P = 1.4408228 d, transit depth 0.45\% and planet radius 10 R$_\oplus$. [NEA/TOI]  Planetary candidate from DIAmante analysis \citep{Montalto20}. [CTOI]

DTARPS-S 339 = TIC 302495337 = TYC 1297-708-1 = CTOI 302495337.01.   DIAmante planet candidate with period P = 2.24986 d, transit depth 0.6 mmag, planet radius 3.7 R$_\oplus$ \citep{Montalto20}. [CTOI]     Possible dilution  of transit depth by ZTF 511216200000551 with r=13.8 lying 15\arcsec\/ from the DTARPS-S candidate. 

DTARPS-S 340 = TIC 302924206 = TOI-2362 = CTOI 302924206.01. Disposition 'Planetary Candidate' with period P = 6.9444047 d, transit depth 0.45\% and planet radius 10 R$_\oplus$. [NEA/TOI] Planetary candidate from DIAmante analysis \citep{Montalto20}. [CTOI] Double star. [SIMBAD]

DTARPS-S 341 = TIC 306337838 = TOI-2397 = CTOI 306337838.01.  Disposition 'Planetary Candidate' with period P = 6.00327 d, transit depth 0.49\% and planet radius 8 R$_\oplus$. [NEA/TOI] Planetary candidate from DIAmante analysis \citep{Montalto20}. [CTOI]

DTARPS-S 342 = TIC 306362738 = TYC 5936-2086-1 = WASP-49 = TOI-479 = CTOI 306362738.01.  Spectroscopically 'Confirmed Planet' with period P = 2.7817387 d, mass M = 127 M$_\oplus$ and radius R = 1.115 R$_J$  orbiting a G6 V star. [NEA/TOI] Eclipsing binary with period 5.5634778 d \citep{Prsa22}. 

DTARPS-S 343 = TIC 306580215 = TYC 8923-297-2. Eclipsing binary with period 3.4990050 d \citep{Prsa22}. 

DTARPS-S 344 = TIC 306603225 = TYC 745-1351-1. 

DTARPS-S 348 = TIC 307788096 = TYC 9194-1773-1.

DTARPS-S 349 = TIC 307943975 = TYC 741-1208-1. 

DTARPS-S 350 = TIC 308098254 = WASP-162 = TOI-1906.  Spectroscopically 'Confirmed Planet' with period P = 9.62468 d, mass M = 1652 M$_\oplus$ and radius R = 1 R$_J$  orbiting a K0  star. [NEA/TOI]

DTARPS-S 351 = TIC 309902656 = TYC 9282-1580-1. 

DTARPS-S 352 = TIC 309902656 = TYC 9282-1580-1. 

DTARPS-S 353 = TIC 311242780 = TYC 9443-2430-1. 

DTARPS-S 354 = TIC 312091232 = CTOI 312091232.01.   DIAmante planet candidate with period P = 4.34801 d, transit depth 4.7 mmag, planet radius 7 R$_\oplus$ \citep{Montalto20}. [CTOI]  Blended eclipsing binary on 13~mag star 21\arcsec\/ from target. [SPOC]

DTARPS-S 355 = TIC 314865962 = CPD -73$^\circ$228 = TOI-208.  Disposition 'Planetary Candidate' with period P = 22.4642328 d, transit depth 0.06\% and planet radius 2.5 R$_\oplus$. [NEA/TOI]

DTARPS-S 356 = TIC 316852947 = TYC 8807-759-1. 

DTARPS-S 357 = TIC 317548889 = HD 39688 = TOI-480.   Disposition 'Planetary Candidate' with period P = 6.8659582 d, transit depth 0.03\% and planet radius 2.8 R$_\oplus$. [NEA/TOI]  High proper motion star. [SIMBAD] 

DTARPS-S 358 = TIC 317860392 = TYC 172-13-1. 

DTARPS-S 360 = TIC 318756174 = CTOI 318756174.01.  Planet candidate with period 3.374443 d, transit depth 13.8 mmag and planet radius 14 R$_\oplus$  from convolutional neural network analysis \citep{Olmschenk21}.  [CTOI] 

DTARPS-S 361 = TIC 321011127 = TYC 7563-733-1 = TOI-860.  Disposition 'Planetary Candidate' with period P = 6.4535152 d, transit depth 1.12\% and planet radius 20 R$_\oplus$. [NEA/TOI]

DTARPS-S 362 = TIC 322807675 = TOI-2398 = CTOI 322807675.01.  Disposition 'Planetary Candidate' with period P = 19.4061437 d, transit depth 1.11\% and planet radius 9 R$_\oplus$. [NEA/TOI] Planetary candidate from DIAmante analysis \citep{Montalto20}. [CTOI]

DTARPS-S 364 = TIC 325680697 = HD 15301 = TOI-414.  Disposition 'False Positive' with period = 1.82488 d. [NEA/TOI]

DTARPS-S 365 = TIC 326092637 = CTOI 326092637.01.   DIAmante planet candidate with period P = 4.61824 d, transit depth 13.6 mmag, planet radius 18 R$_\oplus$ \citep{Montalto20}. [CTOI]

DTARPS-S 366 = TIC 327301957 = TYC 9316-731-1 = TOI-1074. Disposition 'Planetary Candidate' with period P = 13.9427597 d, transit depth 0.14\% and planet radius 2.8 R$_\oplus$. [NEA/TOI]

DTARPS-S 368 = TIC 327712163 = TYC 9324-987-1. 	

DTARPS-S 369 = TIC 327952677 = WASP 146 = TOI-295.  Disposition 'Known Planet' with period P =  3.3971091 d, transit depth 0.99\% and planet radius 14 R$_\oplus$. [NEA/TOI]

DTARPS-S 372 = TIC 332660150 = TYC 5315-221-1 = TOI-938.  Disposition 'Planetary Candidate' with period P = 8.8073507 d, transit depth 0.17\% and planet radius 4 R$_\oplus$. [NEA/TOI]

DTARPS-S 373 = TIC 333426440 = TYC 5390-2210-1.  Planetary candidate with period P = 2.423293 d, transit depth = 1.6\% and width 3.7 h, triage probability 99\% and vetting probability 16\% from deep learning analysis  \citep{Yu19}.  [SIMBAD]

DTARPS-S 375 = TIC 339672028 = CPD -57$^\circ$1712 = TOI-481.  Spectroscopically 'Confirmed Planet' with period P = 10.33111 d, mass M = 486 M$_\oplus$ and radius R = 0.99 R$_J$  orbiting a G star. [NEA/TOI]

DTARPS-S 376 = TIC 339733013 = TYC 8563-1174-1 = TOI-417.  Disposition 'False Positive' with period P = 5.62162 d, transit depth 0.06\% and planet radius 3.4 R$_\oplus$. [NEA/TOI]

DTARPS-S 377 = TIC 340320462 = TYC 8556-1958-1. 

DTARPS-S 378 = TIC 341105361.   Transit depth differs in two TESS  sectors.

DTARPS-S 379 = TIC 341411516 = TYC 8565-645-1 = CTOI 341411516.01.   DIAmante planet candidate with period P = 3.10911 d, transit depth 1.3 mmag, planet radius 2.7 R$_\oplus$ \citep{Montalto20}. [CTOI]

DTARPS-S 380 = TIC 343834897 = TYC 9298-1515-1. High proper motion star. [SIMBAD]

DTARPS-S 381 = TIC 343936388 = UCAC4 208-011547.   Warm Jupiter in TESS FFI images with period P = 15.144 d, radius 6.4 R$_\oplus$ and possible high eccentricity $\simeq 0.5$ orbiting a primary star with radius R = 0.86 R$\odot$  \citep{Dong21}.  A Transit Timing Variation signal may be present.  [SIMBAD]

DTARPS-S 384 = TIC 349521688 = TYC 8922-1135-1. 

DTARPS-S 385 = TIC 349575563 = TYC 8918-1023-1.  	

DTARPS-S 386 = TIC 350153977 = TOI-908.  Disposition 'Planetary Candidate' with period P = 3.1837681 d, transit depth 0.08\% and planet radius 3.1 R$_\oplus$. [NEA/TOI]

DTARPS-S 387 = TIC 350480660 = CTOI 365956062.01 = TYC 8527-329-1 = EBLM J0543-56.   Single-line spectroscopic binary, eclipsing binary with low mass companion with orbital period P = 4.463860 d  \citep{vonBoetticher19}.  The F8 primary has mass M = 1.3 M$_\odot$, radius R = 1.3 R$_\odot$, and age t = 1 Gyr.  The secondary has mass M = 0.16 M$_\odot$ and radius R = 0.19 R$_\odot$.  [SIMBAD] Planetary candidate vetted with DAVE \citep{Rao21}. [CTOI]  Eclipsiing binary with period 4.4638235 d \citep{Prsa22}.  

DTARPS-S 388 = TIC 355637190 = CD -55$^\circ$9400 = TOI-293.  Disposition 'Ambiguous Planetary Candidate' with period P = 0.80546 d, transit depth 0.31\% and planet radius 9 R$_\oplus$. [NEA/TOI]

DTARPS-S 389 = TIC 357026773 = TYC 8935-1712-1. High proper motion star. [SIMBAD]

DTARPS-S 390 = TIC 365956062 = TYC 9072-2452-1 = CTOI 365956062.01.   DIAmante planet candidate with period P = 1.80765 d, transit depth 1.2 mmag, planet radius 5 R$_\oplus$ \citep{Montalto20}. [CTOI]

DTARPS-S 392 = TIC 369376388 = TOI-2486 = CTOI 369376388.01.  Disposition 'Planetary Candidate' with period P = 1.5411563 d, transit depth 0.15\% and planet radius 3.7 R$_\oplus$. [NEA/TOI] Planetary candidate from DIAmante analysis \citep{Montalto20}. [CTOI] High proper motion star. [SIMBAD]

DTARPS-S 393 = TIC 369395383 = TYC 5850-2249-1.  	

DTARPS-S 394 = TIC 370138999 = TOI-3260 = CTOI 370138999.01.  Disposition 'Planetary Candidate' with period P = 11.4519176 d, transit depth 0.26\% and planet radius 8 R$_\oplus$. [NEA/TOI] Planetary candidate from DIAmante analysis \citep{Montalto20}. [CTOI]

DTARPS-S 395 = TIC 372909372 = TYC 8919-1482-1 = CTOI 372909372.01.   DIAmante planet candidate with period P = 9.01803 d, transit depth 1.1 mmag, planet radius 5 R$_\oplus$ \citep{Montalto20}. [CTOI]

DTARPS-S 397 = TIC 382068562 = TOI-924.  Disposition 'Planetary Candidate' with period P = 12.1273412 d, transit depth 1.93\% and planet radius 19 R$_\oplus$. [NEA/TOI]

DTARPS-S 398 = TIC 382188882 = CD -55$^\circ$1153 = TOI-276.  Disposition 'False Positive' with period P = 4.800523 d, transit depth 0.54\% and equivalent planet radius 15 R$_\oplus$. [NEA/TOI] Eclipsing binary with period 4.8000945 d \citep{Prsa22}. 

DTARPS-S 399 = TIC 382435735 = TYC 8919-191-1.  	

DTARPS-S 400 = TIC 382602147 = TOI-2384 = CTOI 382602147.01.  Disposition 'Planetary Candidate' with period P = 2.1356844 d, transit depth 2.88\% and planet radius 11 R$_\oplus$. [NEA/TOI] Planetary candidate from DIAmante analysis \citep{Montalto20}. [CTOI] Giant planet candidate around early-M dwarf with period 2.1357 d and planet radius 1.09 R$_J$ \citep{Gan23}. 

DTARPS-S 401 = TIC 382626661 = TOI-283.  Disposition 'Planetary Candidate' with period P = 17.6175219 d, transit depth 0.07\% and planet radius 2.3 R$_\oplus$. [NEA/TOI]

DTARPS-S 402 = TIC 384469512.   Transit depth differs in two TESS  sectors.

DTARPS-S 403 = TIC 384609312 = TYC 9455-1605-1.  	

DTARPS-S 404 = TIC 386066628 = TYC 5970-852-1. 

DTARPS-S 406 = TIC 388104525 = WASP-119 = TOI-112.  Spectroscopically 'Confirmed Planet' with period P = 2.49979 d, mass M = 391 M$_\oplus$ and radius R = 1.4 R$_J$  orbiting a G5  star. [NEA/TOI]

DTARPS-S 407 = TIC 388198242 = TOI-2224 = CTOI 388198242.01.  Disposition 'Planetary Candidate' with period P = 0.5055105 d, transit depth 0.23\% and planet radius 5 R$_\oplus$. [NEA/TOI]  Planetary candidate from DIAmante analysis \citep{Montalto20}. [CTOI]

DTARPS-S 408 = TIC 389520593 = TYC 7470-1016-1. High proper motion star. [SIMBAD]

DTARPS-S 409 = TIC 391949880 = HD 36481 = TOI-128.  Disposition 'Planetary Candidate' with period P = 4.9404681 d, transit depth 0.03\% and planet radius 2.2 R$_\oplus$. [NEA/TOI]

DTARPS-S 410 = TIC 393940766 = TOI-148.  Disposition 'Confirmed Planet' with period P = 4.8670689 d, transit depth 0.49\% and planet radius 8.2 R$_\oplus$. [NEA/TOI]

DTARPS-S 414 = TIC 394698182 = TOI-170.  Disposition 'Planetary Candidate' with period P = 1.0690773 d, transit depth 1.16\% and planet radius 14 R$_\oplus$. [NEA/TOI]

DTARPS-S 412 = TIC 394721720 = TYC 9494-273-1 = TOI-4378 = CTOI 394721720.01.  Disposition 'Planetary Candidate' with period P = 1.0690773 d, transit depth 0.30\% and planet radius 10 R$_\oplus$. [NEA/TOI]  Planetary candidate from DAVE vetting (Rao). 

DTARPS-S 414 = TIC 398943781 = CD -29$^\circ$9873 = WASP-41 = TOI-780.  Spectroscopically 'Confirmed Planet' with period P = 3052404 d, mass M = 299 M$_\oplus$ and radius R = 1.18 R$_J$ . [NEA/TOI]

DTARPS-S 415 = TIC 401125028 = TOI-2368 = CTOI 401125028.01.  Disposition 'Planetary Candidate' with period P = 5.1750108 d, transit depth 2.72\% and planet radius 15 R$_\oplus$. [NEA/TOI] Planetary candidate from DIAmante analysis \citep{Montalto20}. [CTOI]

DTARPS-S 416 = TIC 402026209 = WASP-4 = TOI-232.  Spectroscopically 'Confirmed Planet' with period P = 1.3382299 d, mass M = 397 M$_\oplus$ and radius R = 1.341 R$_J$  orbiting a G8  star. [NEA/TOI]

DTARPS-S 419 = TIC 402851922 = TYC 9428-52-1. 

DTARPS-S 420 = TIC 403135192 = CD -60$^\circ$7873 = TOI-2226.  Disposition 'Planetary Candidate' with period P = 0.9018226 d, transit depth 0.08\% and planet radius 4 R$_\oplus$. [NEA/TOI] High proper motion star. [SIMBAD]

DTARPS=S 421 = TIC 403154475 = TYC 8825-1703-1. 

DTARPS-S 422 = TIC 404467699 = BD -14$^\circ$324 = TOI-857.  Disposition 'Known Planet' with period P = 3.9080236 d, transit depth 0.81\% and planet radius 15 R$_\oplus$. [NEA/TOI]

DTARPS-S 423 = TIC 404518509 = HD 21520 = TOI-4320 = CTOI 404518509.01.  Disposition 'Planetary Candidate' with two planets.  TOI-4320.01 has period P = 703.6156148 d, transit depth 0.06\% and planet radius 2.9 R$_\oplus$.  TOI-4320.02 has period P = 46.4093732, transit depth 0.04\% and planet radius 2.2 R$_\oplus$]. [NEA/TOI]  A cool Jupiter with period $P \sim 27$ d,  transit depth 2.6\% and planet radius R = 2.9 R$_\oplus$ discovered photometrically in the Planet Hunters TESS survey \citep{Eisner21}.  [SIMBAD] 

DTARPS-S 424 = TIC 406241337 = TYC 8397-1196-1.

DTARPS-S 425 = TIC 407126408 = CD -80$^circ$565 = TOI-913.  Disposition 'Planetary Candidate' with period P = 11.094122 d, transit depth 0.11\% and planet radius 2.5 R$_\oplus$. [NEA/TOI]

DTARPS-S 426 = TIC 407127579 = TYC 9441-432-1 = TOI-5064.  Disposition 'Planetary Candidate' with period P = 3.5108018 d, transit depth 0.10\% and planet radius 5 R$_\oplus$. [NEA/TOI]

DTARPS-S 428 = TIC 408310006 = BD -20$^\circ$2976 = WASP-166 = TOI-576.  Spectroscopically 'Confirmed Planet' with period P = 5.44354 d, mass M = 32M$_\oplus$ and radius R = 0.63 R$_J$  orbiting a F9  star. [NEA/TOI]  Eclipsing binary with period 5.4435063d \citep{Prsa22}. 

DTARPS-S 430 = TIC 413162396 = TYC 6006-1522-1. Eclipsing binary with period 2.6729633 d \citep{Prsa22}	

DTARPS-S 431 = TIC 421455387 = TOI-578.  Disposition 'Ambiguous Planetary Candidate' with period P = 2.50876 d, transit depth 0.56\% and planet radius 19 R$_\oplus$. [NEA/TOI]

DTARPS-S 432 = TIC 423275733 = WASP-142 = TOI-578.  Spectroscopically 'Confirmed Planet' with period P = 2.052868 d, mass M = 266 M$_\oplus$ and radius R = 1.53 R$_J$  orbiting a F8  star. [NEA/TOI] Eclipsing binary with period 2.0521769 d \citep{Prsa22}. 

DTARPS-S 433 = TIC 425721385 = HD 180603 = TOI-1128.  Disposition 'False Positive' with period P = 13.5499 d, transit depth 0.12\% and planet radius 6 R$_\oplus$. [NEA/TOI] High proper motion star. [SIMBAD]

DTARPS-S 434 = TIC 428699140 = TOI-3082.  Disposition 'Planetary Candidate' with period P = 1.9268755 d, transit depth 0.24\% and planet radius 3.2 R$_\oplus$.  High proper motion star. [SIMBAD]

DTARPS-S 435 = TIC 429286485.  Double star. [SIMBAD]

DTARPS-S 436 = TIC 429304876 = HD 97260 = TOI-682.  Disposition 'Confirmed Planet' with period P = 6.8392339 d, transit depth 0.12\% and planet radius 3.8 R$_\oplus$. [NEA/TOI]

DTARPS-S 438 = TIC 434477274 = CTOI 434477274.01.   DIAmante planet candidate with period P = 2.89641 d, transit depth 6.3 mmag, planet radius 11 R$_\oplus$ \citep{Montalto20}. [CTOI]

DTARPS-S 439 = TIC 435868942 = TYC 9495-1074-1. High proper motion star. [SIMBAD]

DTARPS-S 441 = TIC 437242640  = CD -23$^\circ$9677 = WASP-34 = TOI-744.  Disposition 'Known Planet' with period P = 4.317685 d, transit depth 1.15\% and planet radius 14 R$_\oplus$. [NEA/TOI] Spectroscopically 'Confirmed Planet' with period P = 4.3176782 d, mass M = 187 M$_\oplus$ and radius R = 1.22 R$_J$. [NEA/TOI] Eclipsing binary with period 8.6353794 d \citep{Prsa22}. 

DTARPS-S 442 = TIC 437248515 = WASP-31 = TOI-683.  Disposition 'Known Planet' with period P = 3.4058883 d, transit depth 1.65\% and planet radius 16 R$_\oplus$. [NEA/TOI] Spectroscopically 'Confirmed Planet' with period P = 3.4059096 d, mass M = 151 M$_\oplus$ and radius R = 1.549 R$_J$. [NEA/TOI] Eclipsing binary with period 3.4058842 d \citep{Prsa22}. 

DTARPS-S 443 = TIC 438052638 = TYC 1315-225-1. 	

DTARPS-S 444 = TIC 439366537 = HD 202673 =  CTOI 439366537.01.   DIAmante planet candidate with period P = 4.31448 d, transit depth 1.0 mmag, planet radius 5 R$_\oplus$ \citep{Montalto20}. [CTOI]  Blended eclipsing binary in 11~mag star 22\arcsec\/ from target. [SPOC] Eclipsing binary with period 8.6289386 d \citep{Prsa22}.  Double star. [SIMBAD]

DTARPS-S 445 = TIC 440777904 = HAT-P-24 = TOI-511.  Spectroscopically 'Confirmed Planet' with period P = 3.3552479 d, mass M = 230 M$_\oplus$ and radius R = 1.242 R$_J$  orbiting a F8  star. [NEA/TOI] Eclipsing binary with period 3.3565037 d \citep{Prsa22}. 

DTARPS-S 446 = TIC 440872576 = TOI-3160.  Disposition 'Planetary Candidate' with period P = 3.9714761 d, transit depth 0.83\% and planet radius 15 R$_\oplus$. [NEA/TOI]

DTARPS-S 447 = TIC 440957227 = TYC 6760-962-1.  Single-line spectroscopic binary with period P = 3.418911 d and transit depth 20 mmag in the KELT False Positive catalog \citep{Collins18}.  [SIMBAD]

DTARPS-S 448 = TIC 441159680 = TOI-2367 = CTOI 441159680.01.  Disposition 'Planetary Candidate' with period P = 5.74984 d, transit depth 1.77\% and planet radius 14 R$_\oplus$. [NEA/TOI] Planetary candidate from DIAmante analysis \citep{Montalto20}. [CTOI]

DTARPS-S 450 = TIC 443618156 = 1SWASP J111644.43-015207.5 = EBLM J1116-01. Single-line spectroscopic binary, eclipsing binary with low mass companion with orbital period P = 7.375828 d with zero eccentricity  \citep{Triaud17}.  The G0 primary has mass M = 1.2 M$_\odot$ and radius R = 1.3 R$_\odot$.  The secondary has mass M = 0.21 M$_\odot$.  False Positive in the SuperWASP and DIAmante surveys \citep{Schanche19, Montalto20}.   [SIMBAD]

DTARPS-S 451 = TIC 453097786 = TYC 9188-1867-1.

DTARPS-S 452 = TIC 453100658 = TYC 9180-1606-1. Spectroscopic binary. [SIMBAD] 

DTARPS-S 453 = TIC 453123591 = TYC 7615-1413-1.  High proper motion star. [SIMBAD]

DTARPS-S 454 = TIC 455096220 = HAT-P-35 = TOI-489.  Spectroscopically 'Confirmed Planet' with period P = 3.646706 d, mass M = 335 M$_\oplus$ and radius R = 1.332 R$_J$. [NEA/TOI]

DTARPS-S 455 = TIC 455135327 = BD +06$^\circ$1909 = HAT-P-30 = TOI-490.  Spectroscopically 'Confirmed Planet' with period P = 2.810595 d, mass M = 225 M$_\oplus$ and radius R = 1.34 R$_J$. [NEA/TOI]

DTARPS-S 456 = TIC 458653596 = TYC 223-553-1.  Spectroscopic binary. [SIMBAD]

DTARPS-S 458 = TIC 469782185 = TOI-1123.  Disposition 'Ambiguous Planetary Candidate' with period P = 1.20928 d, transit depth 1.19\% and planet radius 15 R$_\oplus$. [NEA/TOI]

DTARPS-S 459 = TIC 469821652 = TYC 223-553-1 = TOI-3263.  Disposition 'Planetary Candidate' with period P = 5.6155837 d, transit depth 0.27\% and planet radius 8 R$_\oplus$. [NEA/TOI]

DTARPS-S 462 = TIC 738065944 = CTOI 738065944.01.   DIAmante planet candidate with period P = 7.75543 d, transit depth 7.7 mmag, planet radius 10 R$_\oplus$ \citep{Montalto20}. [NEA/CTOI]

\section{Notes on Galactic Plane List}
\label{sec:app_DTARPS-GP}

TIC 883943 = TYC 6002-1639-1.

TIC 1129033 = BD -07$\circ$436 = WASP-77 = TOI-398.  Spectroscopically 'Confirmed Planet' with period P = 1.360030 d, mass M = 560 M$_\oplus$ and radius R = 1.2 R$_J$ orbiting a G8 V star. [NEA/TOI]  Eclipsing binary with period 1.3600996	d \citep{Prsa22}. 

TIC 1605476 = TYC 4872-481-1. 

TIC 4616346 = TYC 5961-487-1. Planetary candidate with period P = 0.690059 d, transit depth = 0.12\% and width 1.8 h, triage probability 41\% and vetting probability 10\% from deep learning analysis  \citep{Yu19}. [SIMBAD] 

TIC 4784880 = TYC 5977-1721-1. High proper motion star. [SIMBAD]

TIC 5460908 = TYC 4815-637-1.

TIC 6432352 = TYC 5400-908-1.  Spectroscopic binary [SIMBAD]

TIC 7135184 = TYC 5397-3491-1. 

TIC 10827386 = TOI-2876.  Disposition 'Planetary Candidate' with period P = 62996307, transit depth 0.95\% and planet radius 9 R$_\oplus$. [NEA/TOI]]

TIC 13139556 = CD -25$\circ$16390. 

TIC 19346145 = TYC 5434-728-1.

TIC 19519368 = TOI-494.  Disposition 'Planetary Candidate' with period P = 1.7018884 d, transit depth 0.05\% and planet radius 1.7 R$_\oplus$. [NEA/TOI] High proper motion star [SIMBAD]

TIC 19937775.     Possible dilution  of transit depth by ZTF 1407203100035038 2\arcsec\/ from the DTARPS-S candidate  with magnitude 15.0. 

TIC 21184505 = BD +1246$\circ$1868 = EPIC 211424769.  Periodic variable discovered in the K2 ecliptic plane survey with P = 5.176243 d from a secondary object with radius $8 - 25$ R$_\oplus$ orbiting a mass M = 1.1 M$_\odot$ star.   Classified as a False Positive due to large radial velocity amplitude by \citet{Mayo18} but classified as a planet candidate by \citet{Kostov19}. [SIMBAD] Eclipsing binary with period 5.2919898 d \citep{Prsa22}. 

TIC 24773172.   Possible dilution  of transit depth by anonymous ZTF star 4\arcsec\/ from the DTARPS-S candidate with estimated magnitude 16. 

TIC 33521996 = HATS-6 = TOI-468.  Spectroscopically 'Confirmed Planet' with period P = 3.3452725 d, mass M = 100 M$_\oplus$ and radius R = 0.998 R$_J$  orbiting a M1 V  star [NEA/TOI] Eclipsing binary with period 3.3252654 d \citep{Prsa22}. Giant planet candidate orbiting early-M dwarf with period 3.3256 d and planet radius 1.00 R$_J$ \citep{Gan23}. 

TIC 34371411 = TYC 5371-981-1 = CTOI 3437411.01. Planet candidate with period 3.882755 d, transit depth 9.7 mmag and planet radius 15 R$_\oplus$  from convolutional neural network analysis \citep{Olmschenk21}.  [CTOI] Planetary candidate with period P = 3.881647 d, transit depth = 0.9\% and width 4.5 h, triage probability .99\% and vetting probability 82\% from deep learning analysis  \citep{Yu19}. [SIMBAD] 

TIC 36209863 = HD 295526 = CoRoT 102856619. Spectral type G4 V \citep{Guenther12}. [SIMBAD]     Possible dilution  of transit depth by anonymous ZTF star with g=15 lying 2\arcsec\/ from the DTARPS-S candidate.

TIC 36559570.    Possible dilution  of transit depth by ZTF 1455202200016625 with r=14.8 lying 10\arcsec, and by ZTF 410209200038811 lying 13\arcsec\/ from the DTARPS-S candidate.

TIC 36728161 = TYC 5288-201-1.

TIC 36916955.     Possible dilution  of transit depth by ZTF 410104100044112 with g=15.6 lying 4\arcsec\/ from the DTARPS-S candidate.

TIC 37167097 = CTOI 37167097.01.  Planet candidate with period 1.98 d and transit depth 0.9 mmag from Lightkurve analysis (Chiarello).  [CTOI] 

TIC 38088984 = TYC 5317-1668-1.

TIC 38696105 = TOI-281.   Disposition 'Planetary Candidate' with period P = 5.5764964 d, transit depth 0.07\% and planet radius 4 R$_\oplus$. [NEA/TOI] Planetary candidate with period P = 5.577113 d, transit depth = 0.06\% and width 2.4 h, triage probability 72\% and vetting probability 69\% from deep learning analysis  \citep{Yu19}.

TIC 48176862 = UCAC4 333-014584.  Planetary candidate with period P = 1.925622 d, transit depth = 2.0\% and width 4.2 h, triage probability 99\% and vetting probability 15\% from deep learning analysis  \citep{Yu19}. High proper motion star.  [SIMBAD]

TIC 50712784 = BD -11$^\circ$2350 = WASP-161 = TOI-1912.   Spectroscopically 'Confirmed Planet' with period P = 5.4060425 d, mass M = 791 M$_\oplus$ and radius R = 1.143 R$_J$ . [NEA/TOI]

TIC 52420398 = TOI-5427. Spectroscopic binary [SIMBAD]

TIC 52551181 = TYC 4802-1461-1

TIC 53588284 = TYC 4811-060-1.  ZTF 410202300051074  star  7\arcsec\/ from TIC star at r=17.3 mag with dips of 0.3-0.6 mag.  Possible blended contaminant. 

TIC 56096837 = TOI-2700 = CTOI 56096837.01.   Disposition 'Planetary Candidate' with period P = 4.6636858 d, transit depth 1.11\% and planet radius 13 R$_\oplus$. [NEA/TOI]Planet candidate from convolutional neural network analysis \citep{Olmschenk21}.  [CTOI]     Possible dilution  of transit depth by ZTF 1346213200001469 with r=15.8 lying 16\arcsec\/ from the DTARPS-S candidate. 

TIC 56127667 = TYC 5321-532-1. 
TIC 61755686.     Possible dilution  of transit depth by ZTF 1405216400024997 with r=13.9 lying 9\arcsec\/ from the DTARPS-S candidate.

TIC 64573956.     Possible dilution  of transit depth by ZTF 1405215100024359 with r=15.7 lying 8\arcsec\/ from the DTARPS-S candidate.

TIC 67772767 = TYC 4732-522-1 = TOI-939 = CTOI 67772767.01.   Disposition 'Planetary Candidate' with period P = 7.6107227 d, transit depth 0.64\% and planet radius 12 R$_\oplus$. [NEA/TOI]  Planetary candidate from DIAmante analysis \citep{Montalto20}. [CTOI]

TIC 68010197 = TYC 4831-2185-1. 

TIC 73191957 = TYC 4850-1897-1.

TIC 79137967.  Transit depth differs in two TESS  sectors.

TIC 79143083 =  TYC 6522-2373-1.  Planetary candidate with period P = 1.503826 d, transit depth = 0.32\% and width 3.1 h, triage probability 99\% and vetting probability 21\% from deep learning analysis  \citep{Yu19}.  [SIMBAD]

TIC 79431527 = TOI-2762.   Disposition 'Planetary Candidate' with period P = 3.9072351 d, transit depth 1.47\% and planet radius 14 R$_\oplus$. [NEA/TOI]

TIC 81591410 = TYC 5976-1697-1

TIC 93517731 = TYC 5397-2750-1.  	

TIC 94239926.  The TCF periodogram peak occurs at P=3.512516 d but the true period is more likely the 1/2 harmonic, P=1.756 d.  Transits are not perfectly aligned with either period.  

TIC 94986319 = BD -14$^\circ$1137 = TOI-421.  Spectroscopically confirmed  planet. TOI-421 b has period P = 5.19672 d, mass M = 7 M$_\oplus$ and radius R = 0.239 R$_J$ orbiting a G9 V star.  TOI-431 c has period P = 16.06819 d . [NEA/TOI] Eclipsing binary with period 16.0675528 d \citep{Prsa22}.

TIC 95057860 = TOI-4201.   Disposition 'Planetary Candidate' with period P = 3.5809733 d, transit depth 3.69\% and planet radius 12 R$_\oplus$. [NEA/TOI] Spectroscopically confirmed giant planet around M dwarf \citep{Hartman23}.  Giant planet candidate around early-M dwarf with period 3.5824 d and planet radius 1.05 R$_J$ \citep{Gan23}. 

TIC 95589845 = TOI-4794 = CTOI 95589845.01.  Disposition 'Planetary Candidate' with period P = 3.5658043 d, transit depth 3.69\% and planet radius 12 R$_\oplus$. [NEA/TOI]  Planet candidate with period 3.565656 d, transit depth 6.3 mmag and planet radius 15 R$_\oplus$  from convolutional neural network analysis \citep{Olmschenk21}.  [CTOI]

TIC 99493790 = TOI-162.   Disposition 'Ambiguous Planetary Candidate' with period P = 7.76686 d, transit depth 2.44\% and planet radius 17 R$_\oplus$. [NEA/TOI] High proper motion star. [SIMBAD] 

TIC 99935720 = CD -23$\circ$5158.

TIC 117979897 = WASP-141 = TOI-443.  Spectroscopically 'Confirmed Planet' with period P = 3.310651 d, mass M =  855 M$_\oplus$ and radius R = 1.2 R$_J$ orbiting a F9 star. [NEA/TOI] Eclipsing binary with period 3.3106801 d \citep{Prsa22}. 

TIC 120616194 = TYC 4812-4055-1. 

TIC 122981686.     Possible dilution  of transit depth by ZTF 413308300032702 with i=12.1 lying 13\arcsec\/ from the DTARPS-S candidate. 

TIC 12379043 = TOI-499.   Disposition 'Planetary Candidate' with period P = 8.533527 d, transit depth 0.17\% and planet radius 7 R$_\oplus$. [NEA/TOI]

TIC 124379043 = TOI-2803.   Disposition 'Planetary Candidate' with period P = 1.962292 d, transit depth 2.09\% and planet radius 17 R$_\oplus$. [NEA/TOI]  Confirmed as giant planet with period 1.962 d and planet radius 1.616 R$_J$ \citep{Yee23}. 

TIC 124498746.     Possible dilution  of transit depth by ZTF 1404203200050954 with r=15.8 lying 3\arcsec\/ from the DTARPS-S candidate.  Transit depth differs in two TESS  sectors.

TIC 125018207 = TYC 4826-1333-1.     Possible dilution  of transit depth by anonymous ZTF stars with r$\simeq$14 lying 3\arcsec, and by ZTF 1404207100022773 with r=15.7 lying 14\arcsec\/ from the DTARPS-S candidate. 

TIC 125201129.  ZTF 411203300034623 variable star 8\arcsec\/ from the TIC star  with range r=16.6-17.7 and possible dips.  Possible blended contaminants. 

TIC 125424836 = TOI-3250.   Disposition 'Planetary Candidate' with period P = 3.2800191 d, transit depth 0.84\% and planet radius 17 R$_\oplus$. [NEA/TOI]

TIC 139775416 = TYC 5987-2521-1.

TIC 139922016 = CD -22$^\circ$4953. 

TIC 140016207 = TYC 5984-423-1.  	

TIC 143350972 = HD 36152 = TOI-440.   Disposition 'False Positive' with period P = 1.0817 d, transit depth 0.03\% and equivalent planet radius 2.1 R$_\oplus$. [NEA/TOI] Eclipsing binary with period 1.0817620 d \citep{Prsa22}. 

TIC 143525808 = TYC 6005-1800-1.  Transit depth differs in two TESS  sectors.

TIC 148938758 = TYC 5968-1181-1. 

TIC 153610688 = TYC 143-2311-1. 

TIC 167032301 = TYC 739-835-1. 
 
TIC 168343381.     Possible dilution  of transit depth by ZTF 1456208400024340 with r=14.0 lying 8\arcsec\/ from the DTARPS-S candidate.

TIC 169226820 = TYC 4916-897-1. Secondary of wide binary with separation 40\arcsec\/ \citep{Andrews17}.  The DTARPS-S signal with P = 4.17188 d and transit depth 0.5\% appears to be blending from the G5 primary, BD -03$^\circ$2978, that has a  transiting planet WASP 127b lying in the Neptune desert.  It has orbital period P = 4.178070 d, transit depth 1\%, sub-Saturn mass ($M_p = 0.18$ M$_J$), and super-Jupiter radius ($R_p = 1.37$ R$_J$.  The atmosphere has water, clouds, and super-solar abundances of metals \citep{Palle17, Chen18, Skaf20}.  [SIMBAD]

TIC 170112990.     Possible dilution  of transit depth by ZTF 361209400014478 with r=12.5 lying 7\arcsec\/ from the DTARPS-S candidate.

TIC 175320880 = TYC 4903-874-1. 

TIC 177068644.     Possible dilution  of transit depth by ZTF 359213200011341 with r=13.9 lying 11\arcsec\/ from the DTARPS-S candidate.

TIC 177411679 = TYC 5385-1415-1. 

TIC 177722855 = TYC 5402-204-1 = TOI-971.   Disposition 'False Positive' with period P = 2.3903658 d, transit depth 0.09\% and planet radius 5 R$_\oplus$. [NEA/TOI]

TIC 178162579 = TOI-2842 = CTOI 178162579.01.   Disposition 'Planetary Candidate' with period P = 3.5514016 d, transit depth 0.85\% and planet radius 12 R$_\oplus$. [NEA/TOI]  Planetary candidate from convolutional neural network analysis \citep{Olmschenk21}.  [CTOI]  Confirmed giant planet with period 3.551 d and planet radius 1.146 R$_J$ \citep{Yee23}. 

TIC 178265008.  ZTF 411202400042939 variable star 6\arcsec\/ from TIC star at r$\simeq$15.5 mag with one dip to 15.9, and g$\simeq$16.3 with dips to 16.7.  Possible blended contaminant. 

TIC 178367144 = TYC 4851-299-1 = WASP-180 = TOI-966.  Spectroscopically 'Confirmed Planet' with period P = 3.409264 d, mass M = 286  M$_\oplus$ and radius R = 1.2 R$_J$ orbiting a F7 star. [NEA/TOI]

TIC 178580001 = TYC 4827-12-1.     Possible dilution  of transit depth by ZTF 411202400042939 with r=15.5 lying 6\arcsec\/ from the DTARPS-S candidate.

TIC 179159972 = TOI-965.   Disposition 'False Positive' with period P = 1.37101 d, transit depth 0.16\% and planet radius 5 R$_\oplus$. [NEA/TOI] High proper motion star.  [SIMBAD]

TIC 180145006 = TOI-312.  Disposition 'Ambiguous Planetary Candidate' with period P = 3.5413787 d, transit depth 0.48\% and planet radius 10 R$_\oplus$. [NEA/TOI]

TIC 187919451 = TYC 5400-1435-1. 

TIC 200545021 = TYC 126-1127-1. 	

TIC 206439308.  Eclipsing binary with period 5.5270087 d \citep{Prsa22}.

TIC 220294417 = TYC 746-671-1.  ZTF 513202300040813 star 9\arcsec\/ from the TIC star variable star at r=14.6 with occasional dips of 0.2 mag. Possible blended contaminant.     Possible dilution  of transit depth by ZTF 513202300017328 with r=15.6 lying 13\arcsec\/ from the DTARPS-S candidate.

TIC 229375359 = TYC 158-141-1.     Possible dilution  of transit depth by ZTF 1507206200040686 with r=14.4 lying 7\arcsec\/ from the DTARPS-S candidate.

TIC 231158418.  ZTF 461216400044713 star 6\arcsec\/ from TIC star, variable around 17 mag with powerful flares (13.9) and dips (17.5).  Possible blended contaminant. 

TIC 231387465 = TYC 6539-1365-1 = ASAS J073957-2247.6 = 1RXS J073957.1-224740	. Photometric variable due to rotationally modulated starspots with amplitude 0.03 mag \citep{Kiraga12}.  X-ray source in the ROSAT All Sky Survey.  High proper motion star. [SIMBAD]

TIC 232395760 = HD 291277  [SIMBAD]

TIC 234091431 = CoRoT 102768606.  Rotation period = 72 d \citep{Affer12}.   [SIMBAD]

TIC 234269325 = TYC 156-2033-1. 

TIC 234721262 = TYC 137-1530-1.    Possible dilution  of transit depth by ZTF 1454209200032757 with r=15.3 lying 7\arcsec\/ and by ZTF 1454209200032753 with r=15.5 lying 8\arcsec\/ from the DTARPS-S candidate.

TIC 234880209 = HD 259137. MARVELS survey shows a G2 V star with mass M = 1.0 M$_\odot$, R = 0.9 R$_\odot$ and [Fe/H] = -0.3 \citep{Grieves18}.  Spectroscopic binary.  [SIMBAD]

TIC 235117667 = TYC 160-1757-1. 

TIC 235166044.  ZTF 461215100039761 star 15\arcsec\/ from TIC star at r=16.2 with dips to 17.4. Possible blended contaminant. 

TIC 235183588.      Possible dilution  of transit depth by ZTF 513201400048716 with r=15.4 lying 5\arcsec\/ from the DTARPS-S candidate.

TIC 235276575 = TYC 747-763-1. 	

TIC 235500384.     Possible dilution  of transit depth by ZTF 461214100016409 with r=15.2 lying 9\arcsec\/ from the DTARPS-S candidate.

TIC 235507238 = TYC 752-611-1.

TIC 235548135.     Possible dilution  of transit depth by ZTF 514204400005818 with r=16.0 lying 5\arcsec\/ from the DTARPS-S candidate.

TIC 237566605.    Possible dilution  of transit depth by ZTF 461303100038845 with r=15.2 lying 12\arcsec\/ from the  DTARPS-S candidate.

TIC 237757429 = TYC 149-1204-1.  ZTF 461202400008213  star 12\arcsec\/ from TIC star with r=17.1 and dips to 18.5. Possible blended contaminant. 

TIC 256992368 = TYC 148-34-1.

TIC 262414864 = TYC 5977-2353-1. High proper motion stars. [SIMBAD]

TIC 264630054 = TYC 105-2208-1. 

TIC 265045591. ZTF 464201400031816 star 6\arcsec\/ from TIC star with r=17.1 and frequent dips to 17.4.  Possible blended contaminant. 

TIC 265446888 = TYC 140-1513-1. 	

TIC 265905545 = TYC 140-829-1. 

TIC 266009692 = TYC 140-447-1. ZTF 460207200012300  star 16\arcsec\/ from TIC star with long-term variability around r=15.7 and occasional dips to 16.6.  Possible blended contaminant. [SIMBAD] Spectroscopic binary. 

TIC 268587594 = HD 6217.  Spectral type G2-G3 V. [SIMBAD]

TIC 271098608.     Possible dilution  of transit depth by ZTF 1510208100027480 with r=13.0 lying 7\arcsec\/ from the DTARPS-S candidate. 

TIC 271374913 = TYC 180-2570-1 = 1SWASP J074943.21+005213.8. Planetary candidate with P = 1.267550762 day, transit depth = 0.64\% and width = 0.12 \citep{Schanche19}.  Blended star: TRAPPIST partial transit shows the eclipse is on the fainter star to the north.  [SIMBAD]

TIC 279942918.    Possible dilution  of transit depth by ZTF 412215100007035 with r=14.0 lying 12\arcsec\/ from the DTARPS-S candidate. 

TIC 283303192 = TOI-2741.  Disposition 'Planetary Candidate' with period P = 6.817089 d, transit depth 0.24\% and planet radius 6 R$_\oplus$. [NEA/TOI]    Possible dilution  of transit depth by ZTF 411216300042309 with r=15.2 lying 7\arcsec\/ from the DTARPS-S candidate. 

TIC 284859630 = TOI-2716 = CTOI 284859630.01. Disposition 'Planetary Candidate' with period P = 7.5420435 d, transit depth 0.77\% and planet radius 10 R$_\oplus$. [NEA/TOI] Planetary candidate from convolutional neural network analysis \citep{Olmschenk21}.  [CTOI]

TIC 286865921 = WASP-83 = TOI-773.  Spectroscopically 'Confirmed Planet' with period P = 4.4654291 d, mass M = 199 M$_\oplus$ and radius R = 1.335 R$_J$ . [NEA/TOI]

TIC 292068999 = TYC 174-496-1.  	

TIC 293847411 = TYC 137-1450-1.    Possible dilution  of transit depth by ZTF 460303100001701 with r=15.2 lying 5\arcsec\/ from the DTARPS-S candidate. 

TIC 296179330 = Ross 56 = Gliese 268.5 = NLTT 17658.  High-proper motion K5 star.  Long thought to be in the solar neighborhood with $d \simeq 20-35$ pc \citep{vanAltena95},  Gaia measures parallax = 2.3 mas ($d \simeq 430$ pc).  [SIMBAD]

TIC 307167913 = TYC 5406-1324-1. 

TIC 308051471 = BD -12$^\circ$3363. 

TIC 316877482 = TYC 5900-376-1. 

TIC 317022315 = UCAC4 370-009760.   Planetary candidate with period P = 2.226672 d, transit depth = 0.45\% and width 4.8 h, triage probability 98\% and vetting probability 28\% from deep learning analysis  \citep{Yu19}.  [SIMBAD]

TIC 318282352.    Possible dilution  of transit depth by ZTF 360211200008757 with r=12.6 lying 11\arcsec\/ from the DTARPS-S candidate.

TIC 318345342.     Possible dilution  of transit depth by ZTF 462202400034471 with r=15.5 lying 8\arcsec\/ from the DTARPS-S candidate.

TIC 318796593 = TOI=2886.   Disposition 'Planetary Candidate' with period P = 1.6020002 d, transit depth 2.15\% and planet radius 19 R$_\oplus$. [NEA/TOI]

TIC 318845924 = TYC 169-496-1. 

TIC 318899586.  ZTF 360115200036643 variable star 9\arcsec\/ from TIC star at g$\simeq$17.9 with dips to 18.4.  Possible blended contaminant. 

TIC 319014919 = TYC 186-2418-1. MARVELS survey shows a G6 V star with mass M = 1.0 M$_\odot$, R = 1.0 R$_\odot$ and [Fe/H] = 0.2, thin disk star with high proper motion \citep{Grieves18}. High proper motion star. [SIMBAD]

TIC 322512607 = TYC 5870-194-1. 

TIC 332911893 = TOI-4672.  Disposition 'Planetary Candidate' with period P = 6.6132132 d, transit depth 1.00\% and planet radius 15 R$_\oplus$. [NEA/TOI]     Possible dilution  of transit depth by ZTF 1457208200043405 with r=15.6 lying 4\arcsec\/ from the DTARPS-S candidate.

TIC 333259146 = TYC 5386-15-1.  	

TIC 336732616 = HATS-3 = TOI-103.  Spectroscopically 'Confirmed Planet' with period P = 3.547851 d, mass M = 340 M$_\oplus$ and radius R = 1.381 R$_J$  orbiting a F  star. [NEA/TOI]

TIC 340889095.   Transit depth differs in two TESS  sectors.

TIC 348244416 = TYC 5421-1222-1. 

TIC 348680748 = TYC 6013-1474-1. High proper motion star. [SIMBAD]

TIC 348756659 = TOI-4823 = CTOI 348756659.01.  Disposition 'Planetary Candidate' with period P = 3.7259874 d, transit depth 1.85\% and planet radius 15 R$_\oplus$. [NEA/TOI]  Planetary candidate from DIAmante analysis \citep{Montalto20}. [CTOI]

TIC 382323697 = TYC 5280-1246-1. Eclipsing binary with period 7.7742733	\citep{Prsa22}. 

TIC 382391899 = WASP-50 = TOI-391.  Spectroscopically 'Confirmed Planet' with period P = 1.9550959 d, mass M = 466 M$_\oplus$ and radius R = 1.153 R$_J$. [NEA/TOI]  Eclipsing binary with period 1.9550953 d \citep{Prsa22}. 

TIC 385267507 = TYC 6077-132-1. 

TIC 386275406 = TYC 6077-132-1.    Possible dilution  of transit depth by ZTF 1559201100022341 with r=15.7 lying 15\arcsec\/ from the DTARPS-S candidate.

TIC 386275406 = TYC 756-2020-1. 

TIC 404976695 = TYC 5983-1089-1. 

TIC 405116473 = TYC 5979-1449-1. High proper motion star. [SIMBAD]

TIC 409258019 = TYC 5965-412-1. 

TIC 409594381 = TOI-4214.  Disposition 'Planetary Candidate' with period P = 3.491383 d, transit depth 0.47\% and planet radius 11 R$_\oplus$. [NEA/TOI] High proper motion star. [SIMBAD]

TIC 409794137 = TOI-1478.  Spectroscopically 'Confirmed Planet' with period P = 10.180249 d, mass M = 270 M$_\oplus$ and radius R = 1.06 R$_J$. [NEA/TOI]

TIC 413166353 = TYC 6570-3121-1. 

TIC 415080840 = TYC 5988-2124-1. 

TIC 415559926 = TYC 1317-846-1.    Possible dilution  of transit depth by ZTF 1558212100025586 with r=15.6 lying 10\arcsec\/ from the DTARPS-S candidate.

TIC 427377458 = CTOI 427377458.01.  Planetary candidate with period P = 2.53 d, transit depth 0.05 from Lightkurve analysis. [CTOI]

TIC 432068814 = TYC 6554-2452-1. 

TIC 436243305.  Possible dilution  of transit depth by ZTF 1505207100007186 with r=15.2 lying 103\arcsec, by ZTF 1505207100007178 with r=15.0 lying 13\arcsec, and by ZTF 1505207100007124 with r=14.8 lying 14\arcsec\/ from the DTARPS-S candidate.

TIC 438338723 = UCAC4 532-026425 = EPIC 202086968.  Eclipsing binary discovered in K2 ecliptic plane survey with period P = 3.698, transit depth = 3.1\% and duration 3.5 h \citep{Kruse19}. The star is also in a visual binary system with separation 2\arcsec\/  \citep{Dressing19}.  [SIMBAD]

TIC 438429401 = TYC 1328-1268-1. High proper motion star. [SIMBAD]

TIC 443961200 = TYC 794-679-1. 

TIC 449050247 = BD +04$^\circ$701AB = TOI-951.  Disposition 'False Positive' with period P = 2.96845 d, transit depth 1.70\% and planet radius 20 R$_\oplus$. [NEA/TOI] Spectroscopic binary. [SIMBAD]

TIC 449050248.    Possible dilution  of transit depth by ZTF 1503204200026587 with r=15.9 lying 5\arcsec\/ from the DTARPS-S candidate.

TOI 468958331 = TYC 762-2164-1.    Possible dilution  of transit depth by ZTF 462216100010590 with r=13.6 lying 11\arcsec\/ from the DTARPS-S candidate.

\section{Catalog Comparison with Other Surveys} \label{sec:app_other}

Paper I (\S10 and Appendix) compared the performance of the DTARPS analysis and classifier to previous studies including the Confirmed Planets from the NASA Exoplanet Archive, the \TESS Objects of Interest (TOI) list, community TOI (cTOI) list, DIAmante candidate list (M20), and a variety of additional surveys.  Here, we provide a similar analysis for the DTARPS-S Candidate catalog after vetting procedures are applied.  We do not consider the DTARPS-S Galactic Plane list here because coverage of low Galactic latitudes in other surveys is uneven.  

For individual DTARPS-S Candidates, cross-listings to external surveys are given in Appendix \ref{sec:app_DTARPS}. DTARPS-S Candidates recover 44 objects from WASP and NGTS, 13 from HATS, 2 from KELT, and 1 from \TESS PlanetHunters.  Cross-listings include 13 from the HD/BD/CD catalogs of bright stars and 16 from the Tycho catalog.  The cross-listings also reveals a very small fraction ($\sim 1$\%) of known False Positives: five DTARPS-S Candidates are in the Low Mass Extreme Binary catalog \citep{Triaud17, vonBoetticher19}, and one is a spectroscopic binary discovered in the K2 survey \citep{Kruse19}. In addition, two are magnetically active X-ray emitting stars in the 1RXS catalog, and one is erroneously listed as a solar neighborhood star in the Gliese catalog. 

We now turn to the relationships between our DTARPS-S Candidate catalog and other planet searches treated as an ensemble. 

\subsection{DIAmante Candidates \label{sec:sens_DIA}}

The comparison with the DIAmante candidate exoplanets is important because the M20 and DTARPS studies are based on the same DIAmante data set though with different analysis methodology. The DTARPS-S Analysis List of 7,377 promising stars includes 213 M20 candidates (Paper I, \S10.1).  Of these, 158 (74\%) are DTARPS-S Candidates in Table \ref{tab:DTARPS_cands}. These include 56 Confirmed Planets on the NASA Exoplanet Archive, 91 planet candidates (46 of which had been promoted to the TOI list from the cTOI list), and 11 False Positives.  

The 55 M20 objects rejected by DTARPS vetting include 7 Confirmed Planets on the NASA Exoplanet Archive, 44 planetary candidates (20 of which had been promoted to the TOI list from the cTOI list), and 4 FPs from the TOI list. Of the 7 rejected Confirmed Planets, 2 are considered False Alarms (i.e. insufficient evidence for periodicity in the light curve), 1 as a potential photometric binary, and 4 as ephemeris matches.  The 44 planetary candidates not confirmed by DTARPS-S include 6 that failed centroid-crowding analysis, 15 False Alarms, 11 with even/odd transit differences, 6 with curvature in the light curve, 3 as possible photometric equal-mass binaries, and 3 for ephemeris matching.

\subsection{NASA Exoplanet Archive Confirmed Planets \label{sec:sens_KP}}

The Confirmed Planets from the NASA Exoplanet Archive and TOI lists represent the most reliable collection of transiting systems for analyzing the performance of the DTARPS method.  All the targets have been spectroscopically confirmed from previous transit surveys so the astrophysical False Positive should be negligible.

The DTARPS-S Analysis List includes 130 Confirmed Planets in the NASA Exoplanet Archive and TOI lists (Paper I, \S10.2-10.3), 82 are listed in Table \ref{tab:DTARPS_cands} indicating a 62\% sensitivity rate.  Thirty-seven Confirmed Planets were rejected by centroid-crowding analysis, 5 as False Alarms, 2 as potential photometric binary stellar hosts, and 4 for ephemeris matching.   

\begin{figure}[b]
    \centering
    \includegraphics[width=0.48\textwidth]{ 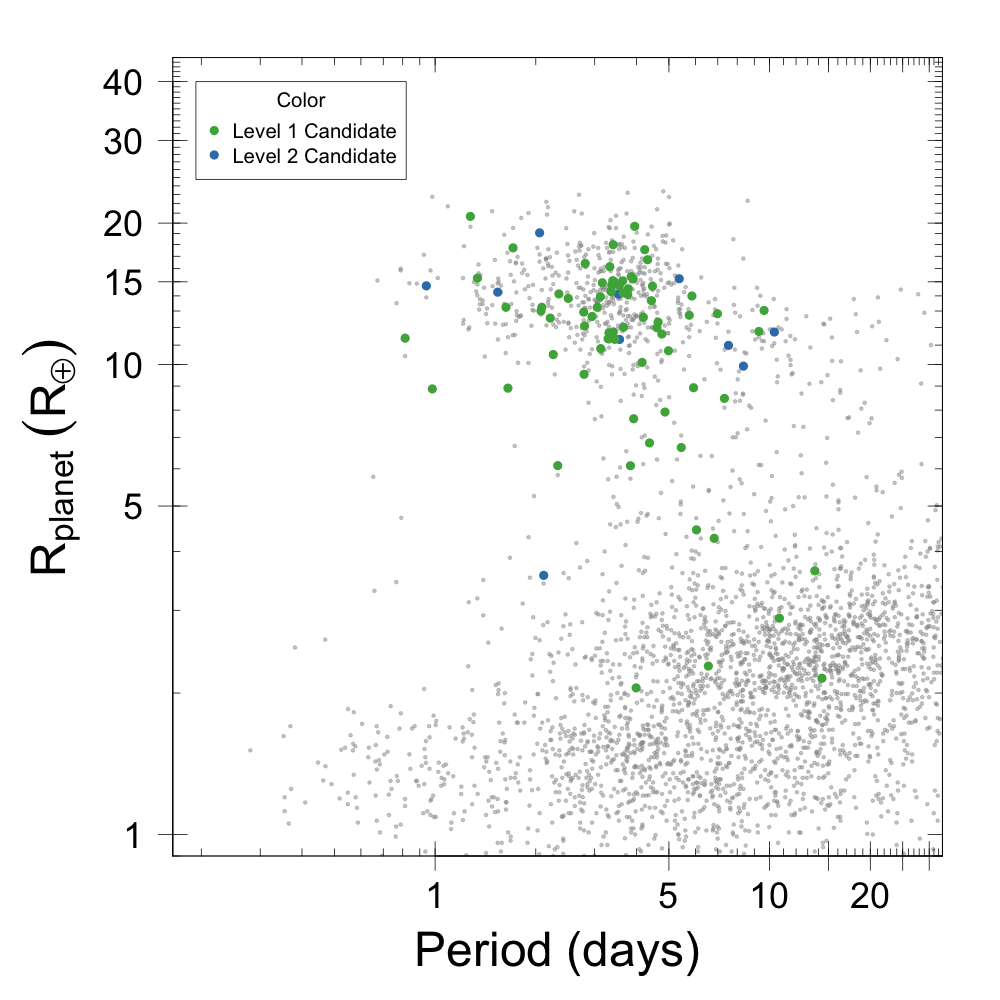}
    \includegraphics[width=0.48\textwidth]{ 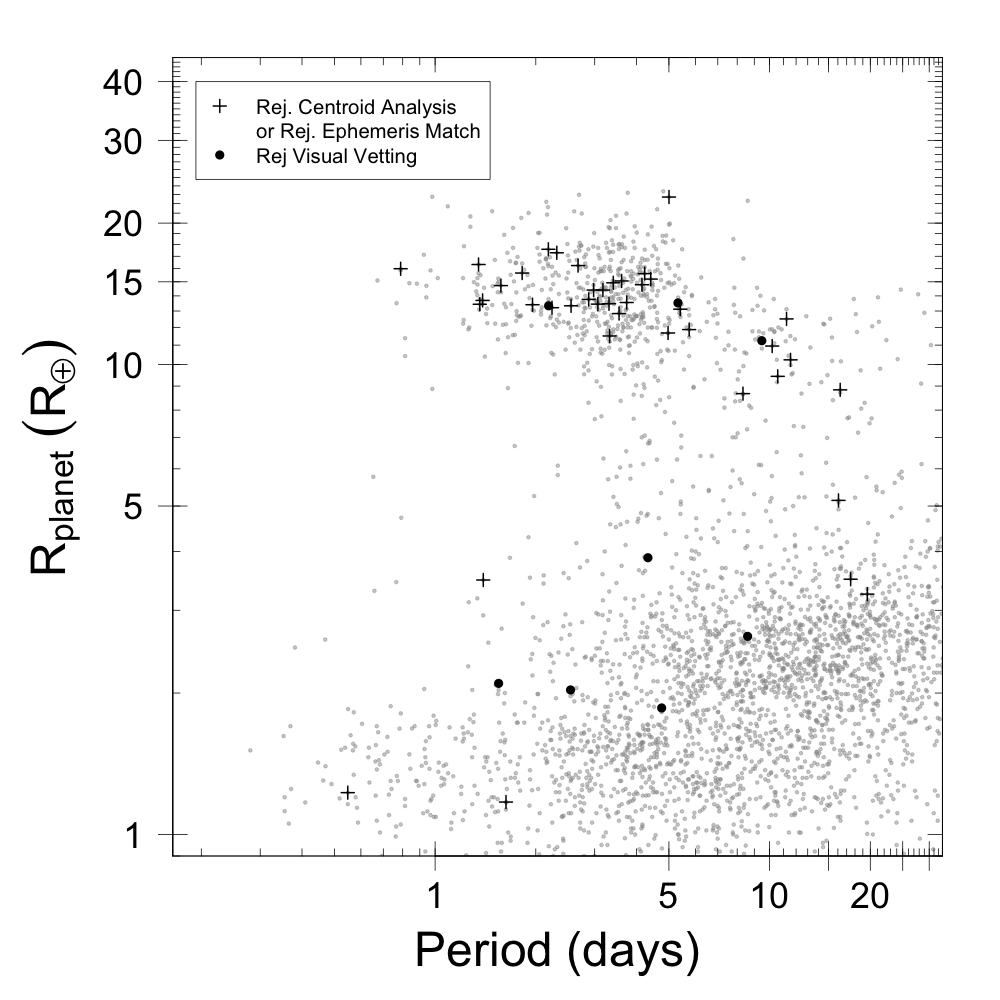}
    \caption{Radius-Period plots for Confirmed Planets in the DTARPS-S Analysis List (Paper I) that passed DTARPS vetting (left panel) and that were rejected by DTARPS vetting (right panel). All Confirmed Planets on the NASA Exoplanet Archive, mostly from the \Kepler survey, are plotted as small gray dots. The DTARPS-S candidate points are colored by DTARPS-S disposition (left panel). In the right panel, planets rejected by the DTARPS vetting process are plotted as pluses if the planet was rejected by either centroid-crowding analysis or ephemeris matching, or are plotted as circles if the planet was rejected by the visual vetting process.}
    \label{fig:cp_vet_comp}
\end{figure}

Most of the Confirmed Planets that were rejected did not pass centroid-crowding analysis (\S\ref{sec:centroid}) suggesting that these criteria may have been too strict (\S\ref{sec:eval_ca}).  Most centroid-crowding rejections (25 of 37) had been discovered by ground-based surveys with  much smaller pixel scales than the \TESS cameras making it easier to verify the source of the transit signal in crowded fields. Of the remaining 12 Confirmed planets rejected by centroid-crowding analysis, 1 was discovered in the K2 survey and the 11 were from two-minute cadence \TESS targets. If we exclude the Confirmed Planets rejected by centroid-crowding analysis or ephemeris matching in the sensitivity measurement, DTARPS-S has 87\% sensitivity towards Confirmed Planets using visual vetting tests. 

Figure \ref{fig:cp_vet_comp} shows the vetting sensitivity (left) and the vetting rejections (right) for the Confirmed Planets list. The DTARPS method is more sensitive to hot and warm Jovian planets ($R_p>$ 8 $R_{\oplus}$) with 64\% recovered than to super-Neptunes, Neptunes, or super-Earths ($R_p<$ 8 $R_{\oplus}$) with 52\% recovered. DTARPS does a poor job for any planets with longer periods ($P \gtrsim 12$ day).  In Figure \ref{fig:cp_vet_comp}-right, the vetting rejections have been split into two categories depending on whether the Confirmed Planet failed due to vetting restrictions based on the \TESS instrumentation (centroid-crowding analysis affected by the large pixel size or ephemeris matching affected by telescope optics) or vetting restrictions due to the light curve (visual vetting tests). The vetting rejections from \TESS instrumentation tend to exclude larger Jovian planets while the vetting rejections from visual vetting were biased against super-Earths and sub-Neptunes. The visual vetting bias is easily explained by weak transit signal strength in single-sector \TESS observations.  The centroid-crowding bias is discussed in   (\S\ref{sec:eval_ca}). 

\subsection{Previously Known False Positives \label{sec:sens_FP}}

The list of previously identified False Positives combines objects from the TOI list, cTOI list, \citet{Affer12}, \citet{Collins18}, \citet{Dressing19}, \citet{Eisner21}, \citet{Feinstein19}, \citet{Kostov19}, \citet{Kruse19}, \citet{Mayo18}, \citet{Olmschenk21}, \citet{Schanche19}, \citet{Tu20}, \citet{vonBoetticher19}, and \citet{Yu19}. The Random Forest classifier was effective in rejecting known False Positives, accepting 113 of 513 FPs in the full $\sim 1$ million star sample. The DTARPS vetting procedure was able to remove another 78 False Positives, leaving only 35 among the 462 DTARPS-S Candidates. The rejected False Positives include 48 removed by centroid-crowding analysis, 5 by ephemeris matches, and 25 by visual vetting.  

The False Positive list sources are heterogeneous with a wide range of information available. For the 35 False Positives in the DTARPS-S Candidate catalog,  17 are designated eclipsing binary systems, 5 as low-mass eclipsing binary system, 2 as blended eclipsing binary systems, 2 as single-line spectroscopic eclipsing binary systems, and 1 as a double-line spectroscopic eclipsing binary system (Appendix~\ref{sec:app_DTARPS}). 

\subsection{Eclipsing Binaries from Prsa et al. (2022)} \label{sec:prsa}

\citet{Prsa22} released a list of 4,584 eclipsing binaries from the \TESS Years 1 and 2 surveys of $\sim 200,000$ pre-selected targets with 2 minute cadences.  They were identified as EBs by various statistical methods including the \TESS science pipeline, and were validated by vetting with the \texttt{ICED LATTE} (Interlacing Code for Eclipsing binary Data validation Lightcurve Analysis Tool for Transiting Exoplanets) code along with other tools.  Comparing their list of 4,584 EBs with the dispositions in the NASA Exoplanet Archive \citep{NEA-CP} gives 448 overlaps with Confirmed Planets.  Among TOIs, the EB catalog has 85 spectroscopically Confirmed Planets, 111 Planet Candidates, and 63 previously identified False Positives. This indicates that roughly 10\% of the EB catalog may be contaminated by exoplanetary transits.

Comparing the \citet{Prsa22} catalog to our DTARPS-S Candidates Catalog gives 58 overlapping objects with 9 more in our DTARPS Galactic Plane List.  The dispositions for the 58 DTARPS-S Catalog overlapping objects are: 25 spectroscopically Confirmed Planets, 14 Planet Candidates, 13 previously identified False Positives and 6 newly identified DTARPS-S candidates.  The dispositions for the 9 DTARPS-S Galactic Plane list overlaps are: 5 spectroscopically Confirmed Planets, 1 Planet Candidates, 1 previously identified False Positive and 2 new DTARPS-S candidates. These cases are individually listed in Appendices A and B above. 

Sixty-seven likely EBs from the \citet{Prsa22} catalog among the 777 DTARPS candidates obtained here gives an estimated EB contamination rate of 8.6\%, assuming the EB catalog is complete and without contamination.  This is consistent with the calculation in Paper I (\S9.3) that the DTARPS-S `specificity' (fraction of known False Positives correctly identified by a classification procedure) ranges from 90\% to 97\% across the radius-period diagram. 

\end{document}